\pdfoutput=1
\RequirePackage{lineno}
\documentclass[11pt]{article}

\usepackage{epsfig}
\usepackage{hhline}
\usepackage{amsmath}
\usepackage{amssymb}
\usepackage{mathrsfs}

\DeclareMathOperator*{\argmin}{arg\,min}

\usepackage{ bbold }
\usepackage{color}
\usepackage{xspace}
\usepackage{paralist} 
\usepackage{caption}
\usepackage{bm} 
\usepackage{acronym}

\usepackage{hyperref} 
\hypersetup{colorlinks=true, urlcolor=blue}
\usepackage{cite} 
\hypersetup{
  citecolor=[rgb]{0.,0.,0.5},
  linkcolor=[rgb]{0.6,0.,0.0},
  urlcolor=[rgb]{0.,0.,0.5}
  }

\usepackage{booktabs} 

\usepackage{parskip}

\usepackage[utf8]{inputenc}
\newlength{\dinwidth}
\newlength{\dinmargin}
\newlength{\wmp}
\newlength{\wtz}
\newlength{\wee}
\newlength{\wat}

\newcommand{\chisq}{\ensuremath{\chi^{2}}}
\newcommand{\hchisq}{\ensuremath{\hat\chi^{2}}}
\newcommand{\ndf}{\ensuremath{n_{\rm dof}}}

\newcommand{\V}{\ensuremath{\mathbb{V}}}
\newcommand{\W}{\ensuremath{\mathbb{W}}}
\newcommand{\M}{\ensuremath{\tilde{M}}}
\newcommand{\Y}{\ensuremath{Y\M}}

\newcommand{\dt}{\ensuremath{\bm{d}}}
\newcommand{\mb}{\ensuremath{\bm{\bar{m}}}}
\newcommand{\mt}{\ensuremath{\bm{\tilde{m}}}}
\newcommand{\s}{\ensuremath{\bm{s}}}
\newcommand{\ahut}{\ensuremath{\bm{\hat{a}}}}
\newcommand{\mz}{\ensuremath{m_{\rm Z}}\xspace}
\newcommand{\as}{\ensuremath{\alpha_{\rm s}}\xspace}
\newcommand{\asmz}{\ensuremath{\as(\mz)}\xspace}
\newcommand{\ord}{\ensuremath{\mathcal{O}}\xspace}

\newcommand{\McT}{\ensuremath{      \begin{pmatrix}\bm{1}&\bm{\dot\alpha}\end{pmatrix}
      \left(
      \begin{pmatrix}\bm{1}&\bm{\dot\alpha}\end{pmatrix}
      \begin{pmatrix}\bm{1}&\bm{\dot\alpha}\end{pmatrix}^T
      \right)^{-1}
}}

\begin{document}
\pagestyle{empty}

\begin{titlepage}
\noindent
\begin{flushleft}
\end{flushleft}
\begin{flushright}
        {\href{https://doi.org/10.1140/epjc/s10052-022-10581-w}{Eur.\ Phys.\ J.\ C 82 (2022) 731}} \\
       {\tt MPP-2021-190 } \\
\end{flushright}

\noindent
\noindent

\vspace{2.5cm}
\begin{center}
\begin{LARGE}
{\bf
  The Linear Template Fit
}
\end{LARGE}

\vspace{1.5cm}
Daniel Britzger \\ 
\vspace{0.5cm}
\small
\textit{
Max-Planck-Institut f{\"u}r Physik, F{\"o}hringer Ring 6, 80805
Munich, Germany \\}
\settowidth{\wmp}{\texttt{mp}}
\settowidth{\wtz}{\texttt{tz}}
\settowidth{\wee}{\texttt{e}}
\settowidth{\wat}{\texttt{@}}
E-mail: \texttt{britz\hspace*{-\wtz}tzge\hspace*{-\wee}er@\hspace*{-\wat}@mp\hspace*{-\wmp}mpp.mp\hspace*{-\wmp}mpg.de\hspace*{-\wee}e}
\end{center}

\vspace{2.0cm}

\begin{abstract}
  \noindent
  The estimation of parameters from data is a common problem in many
  areas of the physical sciences, and frequently used algorithms rely
  on sets of simulated data which are fit to data.
  In this article, an analytic solution for simulation-based parameter
  estimation problems is presented.  
  The matrix formalism, termed the \textit{Linear Template Fit},
  calculates the best estimators for the parameters of interest.
  It combines a linear regression with the method of least squares.   
  The algorithm uses only predictions calculated for a few values of
  the parameters of interest, which have been made available prior to
  its execution. 
  The Linear Template Fit is particularly suited for performance critical applications and
  parameter estimation problems with computationally intense simulations,
  which are otherwise often limited in their usability for statistical inference.
  Equations for error propagation are discussed in detail and are given in closed analytic form.
  For the solution of problems with a nonlinear dependence on the
  parameters of interest, the \textit{Quadratic Template Fit} is
  introduced. 
  As an example application, a determination of the strong coupling
  constant from inclusive jet cross section data at the CERN Large Hadron Collider
  is studied and compared with previously published results.

\end{abstract}

\end{titlepage}

\clearpage
\setcounter{tocdepth}{1}
\tableofcontents
\clearpage

\pagestyle{plain} 

\section{Introduction}
The interpretation of collected research data is a frequent and
important step of the data analysis procedure in many scientific
fields.
A common task in this interpretation is the estimation of parameters
using predictions obtained from analytic calculations or simulation
programs, usually referred to as \textit{fitting}. 
Frequently used fitting algorithms rely on numerical methods and
utilize iterative function minimization 
algorithms~\cite{Fletcher1963ARC,Fletcher:1970,Goldstein1971OnDF,James:1975dr,Gill:99800,kingma2017adam,ceres-solver}.
The availability of more computational power and
the development of improved algorithms such as machine learning
techniques~\cite{LeCun:2015dt,Feickert:2021ajf} have led to more
comprehensive statistical methods that can be employed for the
estimation of parameters~\cite{Behnke:1517556,Baak:2014wma,Besjes:2015vns,Cranmer:2019eaq}.
At the same time, simulation programs have become more
complex and more computationally demanding, placing constraints on the
inference algorithms.
For example, in the field of particle physics, such simulations
provide fully differential predictions in perturbative Quantum Chromodynamics
and the Electroweak Theory (see
Refs.~\cite{Boughezal:2016wmq,Gehrmann-DeRidder:2016cdi,Currie:2016bfm,Czakon:2016ckf,Azzi:2019yne,Cepeda:2019klc,Kallweit:2019zez,Mazzitelli:2020jio,Heinrich:2020ybq}
for some recent advances)
and may take several hundreds up to a few thousand years on a single,
modern CPU core.
Sometimes, the simulation of the physical process is followed by a
detailed simulation of the experimental apparatus in order to provide a 
synthetic data as close as possible to the recorded raw data.
At the experiments at the CERN Large Hadron
Collider (LHC), the simulated data include the physics of
the proton--proton
collision~\cite{Gleisberg:2008ta,Sjostrand:2014zea,Bellm:2015jjp} and
the simulation of processes that take place subsequently in
the detector material~\cite{Agostinelli:2002hh}. This results in the simulation 
of about 100 million electronic channels~\cite{Outreach:1457044},
which are processed similarly to real data~\cite{Duckeck:2005rb}.
Consequently, these simulated data cannot be used in iterative
optimization algorithms because of their computational cost.
In addition, these predictions can be provided only
for a few selected values of the theory parameters of interest.

Besides the restrictions associated with computationally demanding simulations,
there exist other reasons why less complex
simulations cannot be used by certain optimization algorithms.
Numerical instabilities or statistical
fluctuations in the predictions
can result in fit instabilities, input quantities for the predictions
are sometimes only available for selected discrete values of the
parameters of interest (for example, parameterizations of the parton
distribution functions of the proton (PDFs) 
which are only available for a few  values of the strong coupling
constant $\alpha_s$~\cite{Gao:2017yyd,Buckley:2014ana}), 
or simply because there are
technical limitations when interfacing a simulation program to the
inference algorithm and the  parameter(s) of interest cannot
be made explicit to the minimization algorithm.
A further frequent limitation for simulation-based inference is
related to the intellectual property of the simulation program,
where the computer code is not publicly available, but the obtained
predictions are available for a given set of reference values.
In short, in many cases, the inference algorithm can only
use predictions that have been provided previously since a
recomputation by the inference algorithm is not possible.


%
Several variants of simulation-based template fits, or template-like fits, are nowadays
used in high energy
physics~\cite{Abulencia:2005aj,D0:2012kms,Malaescu:2012ts,CDF:2012gpf,ATLAS:2012aj,ATLAS:2015lek,ATLAS:2016muw,ATLAS:2017rzl,CMS:2017pcy,ATLAS:2018fwq,CMS:2019fak,Herwig:2019obz,ATLAS:2019ezb,Cranmer:2019eaq,Cranmer:2021oxr}. 
These make use of polynomial interpolation or regression between the
reference points or apply related techniques~\cite{Read:1999kh,Cranmer:2012sba,Baak:2014fta}.
Some algorithms use un-binned quantities, while others make use of
summary statistics.
Furthermore, different algorithms to find the best
estimator are employed, for example,
numerical methods
or iterative minimization algorithms are used
to find the extremum of a likelihood or \chisq-function.
In general, typical iterative minimization algorithms can be 
considered to be template fits, since every
iteration generates a new
template at a new reference point, which are then used to find the
extremum of the likelihood function.

In this article, the equations of the \emph{Linear Template Fit} (LTF)
are derived, which can be written in a closed matrix equation form.
The Linear Template Fit provides an analytic solution for the determination of the best
estimator in a least-squares problem under the assumption that the
predictions are available only for a finite set of values of the
parameters of interest.
The equations are obtained from a two-step marginalization of the
underlying statistical model, which assumes normal-distributed
uncertainties. The first step provides a linearized, but continuous, estimate of the
prediction\,\footnote{Usually,
 there exists prior knowledge about the value of the parameters of
 interest, for example from previous studies or theoretical
 considerations. The templates are then constructed in the vicinity of the expected best
estimator, such  that  the linearized model is a good approximation in general.},
and the second step provides the best estimator of the
fitting problem.
The Linear Template Fit is suited for a wide range of
parameter estimation problems, where the input data can be cross
section measurements, event counts, or other summary statistics like
histograms. 

This article is structured as follows.
After a brief review of the method of template fits in
Section~\ref{sec:tmplfits}, the equation
of the Linear Template Fit is derived in Section~\ref{sec:basicmodel} for a univariate problem.
While the Linear Template Fit turns out to be a simple matrix equation, to the
author’s knowledge it has not been published in its closed form before.
The emphasis of this article is on variations of the Linear Template
Fit, its applicability, and the propagation of uncertainties.
The  multivariate Linear Template Fit is discussed in
Section~\ref{sec:multivLTF}, and  the
Linear Template Fit with relative uncertainties is presented in
Section~\ref{sec:LogN}.
A detailed discussion on error treatment and propagation is given in
Section~\ref{sec:Errors}. 
A (detector) response matrix is inserted into the Linear Template Fit
in Section~\ref{sec:A}.
Several considerations for the applicability of the Linear Template Fit
are provided in Section~\ref{sec:considerations}, as well as a simple
algorithm for cross-checks and the relation to other algorithms.
While the prerequisite of the Linear Template Fit is the linearity of the
prediction in the parameters of interest, potential non-linear effects
are estimated and discussed 
in Section~\ref{sec:NonLinearModel}.
There, the Quadratic Template Fit is introduced, an algorithm with fast
convergence using second-degree polynomials for the 
parameter dependence of the model.
Three toy examples are discussed in
Sections~\ref{sec:example1},~\ref{sec:example2}
and~\ref{sec:example3}, and are also remarked upon occasionally in between.
A comprehensive and real example application of the Linear Template
Fit is given in Section~\ref{sec:CMSalphas}, where the value of the 
strong coupling constant \asmz\ is determined from inclusive jet cross
section data obtained by the CMS experiment at the LHC. The best estimators
are compared with previously published results, obtained with other inference algorithms.
Section~\ref{sec:summary} provides a summary.
Additional details are collected in the Appendix, where a table can be
found of the notation adopted in this article.

%
\section{Template fits}
\label{sec:tmplfits}
The objective function of a multivariate optimization problem 
assuming normal-distributed random variables is
a likelihood function calculated as the joint probability distribution of Gaussians
\begin{equation}
  \mathcal{L} =
  \prod_{i}^n\frac{1}{\sqrt{2\pi\sigma_i^2}}\exp\left(\frac{-(d_i-\lambda_i(\bm\alpha))^2}{2\sigma_i^2}\right)\,,
  \label{eq:L}
\end{equation}
where $d_i$ are the $i$ data values, $\sigma^2_i$ the variances,
and $\lambda_i$ is the value that is dependent on the model
parameters of interest $\bm{\alpha}$.
Gaussian probability distributions are often appropriate assumptions for real
data due to the central limit theorem~\cite{Cowan:1998ji,James:1019859}.
For numerical computation and optimization algorithms it is convenient to rewrite $\mathcal{L}$
in terms of a least-squares  equation using $\chisq=-2\log\mathcal{L}$
and omitting constant terms.
In matrix notation, and using a covariance matrix, the
objective function becomes\,\footnote{A table of the notation is provided in
  Appendix~\ref{sec:notation}. 
  It is assumed, that the hypothesized function $\lambda$ is obtained from
  the simulator and will be denoted as \emph{model},
  since it represents commonly a comprehensive calculation from a
  complicated theoretical \emph{model}.
}
\begin{equation}
  \chisq(\bm\alpha)= (\bm{d} - \bm{\lambda}(\bm{\alpha}))^{\text{T}} V^{-1} (\bm{d} - \bm{\lambda}(\bm{\alpha}))\,.
  \label{eq:chisqDetLev}
\end{equation}
The maximum likelihood estimators (MLE) of the parameters $\bm\alpha$ are found by minimizing \chisq
\begin{equation}
  \bm{\hat\alpha}=\argmin_{\bm\alpha}\chisq(\bm{d};\bm{\alpha})\,.
  \label{eq:minchi2lin}
\end{equation}
For this task, iterative function optimization algorithms are
commonly employed (see e.g.~Refs.~\cite{James:1975dr,Gill:99800,Nocedal:1638144,kingma2017adam}).  
The least squares method is then equivalent to the maximum likelihood
method in eq.~\eqref{eq:L}.

A common problem that physicists often face at this point is that
the model $\bm{\lambda}(\bm{\alpha})$ may be
computationally intense, time-consuming to calculate, or not be
available in a compatible implementation.
Hence,  the objective function \chisq\ cannot be evaluated
repeatedly and for arbitrary values of $\bm\alpha$,
as it would be required in gradient methods.
In contrast, it is often the case that the model is available for
a set of $j$ model parameters $\bm{\dot\alpha}_{(j)}$.
These predictions
$\bm{y}_{(j)}:=\bm{\lambda}(\bm{\dot\alpha}_{(j)})$
are denoted
as \emph{templates} for given \emph{reference values}
$\bm{\dot\alpha}_{(j)}$ in the following. 
Consequently, a template fit exploits only previously available
information and can be considered as a two-step algorithm:
\begin{enumerate}
\item
  In a first step, a (continuous) representation of the model in the
  parameter $\bm\alpha$ needs to be derived from the templates.
  This can be done by interpolation, regression or any other
  approximation of the true model
  with some reasonable function  
  $\bm{\mathbf{y}}(\bm{\alpha},{\bm{\hat\theta}})$, like
  \begin{equation}
    \bm{\lambda}(\bm{\alpha}) \approx
    \mathbf{y}(\bm{\alpha},{\bm{\hat\theta}})\,,
  \end{equation}
  where $\bm{\mathbf{y}}(\bm{\alpha},{\bm{\hat\theta}})$ is dependent
  on  $\bm\alpha$ and several parameters $\bm\theta$. 
  The values $\bm{\theta}$ are then determined from the templates,
  for example with a least-squares method and a suitable \chisq, similar to
  \begin{equation}
    \bm{\hat\theta} = \argmin_{\bm\theta}\chisq(\bm{y}_{(0)},\ldots,\bm{y}_{(j)};\bm{\theta})\,.
  \end{equation}
  Such a procedure can be done for each point $i$  separately, or for all points simultaneously.
\item
  In a second step, which is often decoupled from the first, 
  the best estimators of the parameter of interest are determined
  using the approximated true model from the first step.
  It can be expressed as
  \begin{equation}
    \bm{\hat\alpha}=\argmin_{\bm\alpha}\chisq(\bm{d};\bm{\alpha})\,,
  \end{equation}
  where the objective function could be a least-squares expression,
  like the one of eq.~\eqref{eq:chisqDetLev}, but using the
  approximated model $\mathbf{y}(\bm{\alpha},{\bm{\hat\theta}})$
  from the first step, like 
  \begin{equation}
    \chisq(\bm\alpha)= (\bm{d} - \mathbf{y}(\bm{\alpha},{\bm{\hat\theta}}) )^{\text{T}} V^{-1} (\bm{d} - \mathbf{y}(\bm{\alpha},{\bm{\hat\theta}}) )\,.
  \end{equation}
\end{enumerate}
Such a two-step algorithm appears to be cumbersome, and in the
following the equation of the Linear Template Fit is
introduced, which combines the two steps into a single equation, under
certain assumptions.

%
\section{The basic method of the Linear Template Fit}
\label{sec:basicmodel}
The basic methodology of the Linear Template Fit is
introduced, and
in order to simplify the discussion a
\emph{univariate} 
model $\bm\lambda(\alpha)$ is considered, where the parameter of
interest is denoted as $\alpha$.
It is assumed that the model is available for
several values ${\dot\alpha}_{(j)}$ of the parameter of interest,
and these $j$ predictions are denoted as the \emph{templates},
$\bm\lambda(\dot\alpha_{(j)})$.
These templates will be confronted with the vector of the data
$\bm{d}$ in order to obtain the best estimator $\hat\alpha$.

We consider a basic optimization problem based on Gaussian probability
distributions, and the objective function is written in
terms of a least-squares equation 
\begin{equation}
  \chisq(\alpha)= (\bm{d} - \bm{\lambda}({\alpha}))^{\text{T}} V^{-1} (\bm{d} - \bm{\lambda}({\alpha}))\,.
  \label{eq:chisq1}
\end{equation}
In a first step, the model $\bm\lambda(\alpha)$ is 
approximated linearly from the template distributions,
and in every entry $i$, $\lambda_i(\alpha)$ is approximated through
a linear function $\mathrm{y}_i(\alpha;\theta_0,\theta_1)$ like
\begin{equation}
  \lambda_i(\alpha) \approx \mathrm{y}_i(\alpha;\bm{\hat\theta})
  ~~~~\text{using}~~~~
  \mathrm{y}_i(\alpha;\bm{\theta}_{(i)}) := \theta^{(i)}_0 + \theta^{(i)}_1 \alpha\,.
  \label{eq:rmy}
\end{equation}
The best estimators of the function parameters, $\bm{\hat\theta}$, are obtained from the
templates by linear regression (``straight line fit,'' see
e.g.~Refs.~\cite{Blobel:1984cy,Barlow:213033,Seber2003LinearRA}), whose formalism also
follows the statistical concepts that were briefly outlined above.
Each of the templates is representative of a certain value of
$\alpha$, which will be denoted in the following as \emph{reference points} $\dot\alpha_j$,
and all reference values 
form the $j$-vector $\bm{\dot\alpha}$.
Next, an extended Vandermonde design matrix~\cite{seber2008matrix}, the \emph{regression matrix},  is constructed
from a unit column vector and $\dot{\bm\alpha}$:
\begin{equation}
M := \begin{pmatrix}\bm{1} &\bm{\dot\alpha} \end{pmatrix} =
\left(\begin{smallmatrix}
    1 & \dot\alpha_1 \\
    \vdots & \vdots \\
    1 & \dot\alpha_j  
\end{smallmatrix}\right)\,.
\label{eq:M}
\end{equation}
From the method of least squares the best-estimators of the polynomial
parameters for the  $i$th entry are~\cite{Blobel:437773} 
\begin{equation}
  {\bm{\hat\theta}}_{(i)}
  = \begin{pmatrix}\hat\theta_0\\\hat\theta_1\end{pmatrix}_{(i)} =
    M^+_{(i)}\begin{pmatrix} y_{(1),i} \\ \vdots \\ y_{(j),i} \end{pmatrix}
    ~~~{\rm using}~~~
    M^+_{(i)}=(M^{\text{T}}\mathcal{W}_{(i)}M)^{-1}M^{\text{T}}\mathcal{W}_{(i)}\,,
    \label{eq:theta}
\end{equation}
where the matrix $M^+_{(i)}$ is a $g$-inverse of a least squares
problem~\cite{RaoMitraBook,seber2008matrix},
 $y_{(j),i}$ denotes the $i$th entry of the $j$th template vector, 
and $\mathcal{W}_{(i)}$ is an inverse covariance matrix which
represents uncertainties in the templates, e.g.\
$\mathcal{W}_{(i),j,j}=\sigma_{y_{(j),i}}^{-2}$.
However, for the purpose of the Linear Template Fit, 
an important simplification is obtained since the special
case of an unweighted linear regression is applicable to a good approximation. 
The (inverse) covariance matrix $\mathcal{W}_{(i)}$ in this problem has two features: it is a
diagonal matrix, since the templates are generated independently, and secondly,
it has, to a very good approximation, equally sized diagonal elements,
$\mathcal{W}_{(i),j,j}\approx\mathcal{W}_{(i),j+1,j+1}$. 
This simplification is commonly well applicable, for example, 
if the model is obtained from an exact calculation or if Monte Carlo
event generators were employed and the templates all have a
similar statistical precision.
Consequently, the size of the uncertainties in the templates
$\sigma_{y_{(j),i}}$ factorizes from $\mathcal{W}_{(i)}$ and subsequently cancels
 in the calculation of $M^+_{(i)}$, eq.~\eqref{eq:theta}.
Thus, an unweighted polynomial regression is applicable and
the $g$-inverse simplifies to a left-side
Moore--Penrose pseudoinverse
matrix~\cite{Moore1920,penrose_1955,zbMATH03367052},
\begin{equation}
  M^+_{(i)}\simeq M^+= (M^{\text{T}}M)^{-1}M^{\text{T}}\,,
  \label{eq:Mc}
\end{equation}
with $M$ as defined in eq.~\eqref{eq:M}.
%
It is observed that the matrix $M^+$ is
\emph{universal}, so it is independent on the quantities
$y_{(j),i}$ and $\mathcal{W}_{(i)}$,
but is calculated only from the reference points $\bm{\dot\alpha}$.
Consequently, it is equally applicable in every  bin $i$. 

Henceforth, we decompose the matrix $M^+$
(eq.~\eqref{eq:Mc}), which is a 2$\times$$j$-matrix, into two row vectors, like
\begin{equation}
  M^+ =: 
  \begin{pmatrix}
        \bm{\bar{m}}^{\text{T}}\\
        \bm{\tilde{m}}^{\text{T}} \\
  \end{pmatrix} \,,~~~~~\text{so}~~~~~~
  \begin{matrix}
        \bm{\bar{m}}   = (M^+)^{\text{T}}\left(\begin{smallmatrix}1\\0\end{smallmatrix}\right) \\
        \bm{\tilde{m}} = (M^+)^{\text{T}}\left(\begin{smallmatrix}0\\1\end{smallmatrix}\right) 
  \end{matrix}\,.
    \label{eq:mbar}
\end{equation}
This introduces the $j$-column-vectors $\bm{\bar{m}}$ and
$\bm{\tilde{m}}$.

We further introduce the \emph{template matrix} $Y$, a $i\times j$ matrix,
which is constructed from the column-vectors of the template
distributions, and can be written as
\begin{equation}
  Y :=
  \begin{pmatrix}
    Y_{1,1} & \hdots & Y_{1,j} \\
    \vdots & \ddots & \vdots \\
    Y_{i,1} & \hdots & Y_{i,j} \\
  \end{pmatrix}
  =
  \bigg(\begin{matrix}
    \bm{y}_{(\dot\alpha_1)} &
    \hdots &
    \bm{y}_{(\dot\alpha_j)} 
  \end{matrix}
  \bigg)
  =
  \begin{pmatrix}
    y_{(1),1} & \hdots & y_{(j),1} \\
    \vdots & \vdots & \vdots \\
    y_{(1),i} & \hdots & y_{(j),i}
  \end{pmatrix}\,.
  \label{eq:Y}
\end{equation}
Hence, substituting the best estimators $\bm{\hat\theta}$ 
(eq.~\eqref{eq:theta}) into the linearized model
$\mathbf{{y}}(\alpha,\bm{\hat\theta})$ (eq.~\eqref{eq:rmy}),
and using $\bm{\bar{m}}$ and $\bm{\tilde{m}}$ (eq.\,\eqref{eq:mbar}) and $Y$ (eq.\,\eqref{eq:Y}), 
the model can be expressed as a matrix equation\,\footnote{Note that eq.~\eqref{eq:ymt} represents already a very
  useful (linearly approximated) representation of the simulation, since it
  transforms a discrete representation of the model into a continuous
  function.
  This equation may become useful for several other purposes, or optimization
  algorithms, whenever only templates are available, but a (quickly
  calculable) continuous function is required.
  Higher-degree polynomials beyond
  the linear approximation can also be used for that purpose, so 
  $\bm{\lambda}^{\text{T}}(\alpha)\simeq
  \begin{pmatrix}\alpha^0&\ldots&\alpha^n\end{pmatrix}
    \mathcal{M}^+Y^{\text{T}}$, using a $n$-degree regression matrix $\mathcal{M}$
    (eq.~\eqref{eq:multtheta}). 
}: 
\begin{equation}
  \bm{\lambda}(\alpha)\approx \mathbf{{y}}(\alpha,\bm{\hat\theta})
  = Y\bm{\bar{m}} + Y\bm{\tilde{m}}\alpha\,.
  \label{eq:ymt}
\end{equation}
It is seen that the polynomial parameters $\bm{\theta}$ are no longer explicit in this
expression.
Consequently, the objective function (eq.~\eqref{eq:chisq1}) becomes a linear least-squares expression
\begin{equation}
  \chisq(\alpha) = (\dt - Y\mb - Y\mt \alpha)^{\text{T}} W (\dt - Y\mb - Y\mt\alpha)\,,
  \label{eq:chisqDetLev2}
\end{equation}
where $W$ is the inverse covariance matrix, $W=V^{-1}$.

In the second step of the derivation of the Linear Template Fit the
best estimator $\hat\alpha$ is determined.
Due to the linearity of $\chisq$ in $\alpha$, the best-fit value
of $\alpha$ is defined by the stationary point of
\chisq~\cite{Blobel:437773,James:1019859,Cowan:PDG:2020}, and the 
best linear unbiased estimator of the parameter of interest is
\begin{equation}
  \hat\alpha
  = \frac{(Y\mt)^{\text{T}}W }{(Y\mt)^{\text{T}}WY\mt} (\dt-Y\mb)\,.
  \label{eq:mtDetLev}
\end{equation}
When introducing the $g$-inverse of least squares, a $1\times i$ matrix,
\begin{equation}
  F := \frac{(Y\mt)^{\text{T}}W}{ (Y\mt)^{\text{T}}WY\mt }\,,
  \label{eq:F}
\end{equation}
the master formula of the \emph{Linear Template Fit} is found to be
\begin{equation}
  \hat\alpha=F ( \dt - Y\mb)\,,
  \label{eq:master}
\end{equation}
where eqs.~\eqref{eq:mbar},~\eqref{eq:Y}, and \eqref{eq:F} were used, and $\dt$
  denotes the vector of the data.

Since the distribution of the data is assumed to be normal-distributed, from
the linearity of the least squares, it follows that the estimates are also
normal-distributed.
From error propagation, the variance of the best
estimator is 
\begin{equation}
  \sigma_{\hat\alpha}^2= FVF^{\text{T}} = (F^{\text{T}}WF)^{-1}\,.
  \label{eq:errorprop}
\end{equation}
Given the case that the approximation in eq.~\eqref{eq:ymt} holds,
the estimator $\hat\alpha$ represents a best and unbiased
estimator according to the
Gau\ss--Markov-theorem~\cite{Blobel:437773,James:1019859}, and also the 
variances are the smallest among all possible
estimators.
Since $\bm{d}$ does not enter the calculation of the variances,
eq.~\eqref{eq:errorprop}, the uncertainties are equivalent to the
expected uncertainties from the Asimov~\cite{Cowan:2010js} data  $\bm{\lambda}(\hat\alpha)$.

When using the right expression of eq.~\eqref{eq:mbar}, it can be directly
seen that the best estimator is indeed obtained from a single matrix
equation using only the template quantities ($Y$, $\bm{\dot\alpha}$)
and the data details ($\bm{d}$, $W$) and the Linear Template Fit
can alternatively be written as
\begin{align}
  \hat\alpha=
  \left(
  \left(
  \Upsilon\left(\begin{smallmatrix}0\\1\end{smallmatrix}\right)  \right)^{\text{T}} 
    W
    \Upsilon\left(\begin{smallmatrix}0\\1\end{smallmatrix}\right) 
      \right)^{-1} 
      \left( \Upsilon\left(\begin{smallmatrix}0\\1\end{smallmatrix}\right) \right)^{\text{T}} 
        W
        \left(\dt - \Upsilon\left(\begin{smallmatrix}1\\0\end{smallmatrix}\right) \right)
          ~~~\text{using}\\
          \Upsilon=Y(M^+)^{\text{T}}=Y\McT\,.~~~~~~~~
\end{align}
In contrast, when the (commonly well justified) approximation of the
unweighted regression in each bin is not applicable (see left side of
eq.~\eqref{eq:Mc}), then the bin-dependent regression matrices
$M_{(i)}^+$  remain explicit.
Such a case may be present when a Monte Carlo event generator is used
for the generation of the templates and the statistical uncertainties
have to be considered, and one sample was generated with higher
statistical precision than the others, for instance, since it also serves for
further purposes like unfolding.
However, when using such a weighted regression, the best estimator may
still be expressed using eq.~\eqref{eq:mtDetLev}, but the elements of
the two vectors $Y\mb$ and $Y\mt$ would need to be calculated
separately and become
  \begin{equation}
      (Y\mb)_i = \left(\begin{smallmatrix}1\\0\end{smallmatrix}\right)^{\text{T}} M^+_{(i)}\bm{x}_{(i)} ~~~~\text{and}~~~~
      (Y\mt)_i = \left(\begin{smallmatrix}0\\1\end{smallmatrix}\right)^{\text{T}} M^+_{(i)}\bm{x}_{(i)} \,,
        \label{eq:wgtreg}
  \end{equation}
  where the vectors $\bm{x}_{(i)}$  denote the $i$th row of the
  template matrix, $\bm{x}_{(i)}=Y_i^{\text{T}}$, and the matrices
  $M^+_{(i)}$ may include uncertainties in the templates
  $\bm{x}_{(i)}$ through $\mathcal{W}_{(i)}$ (see
  eq.~\eqref{eq:theta}, right). This case, however, will not be
  considered any further in that article.
  
In the following sections, an example application of the Linear
Template Fit is presented, and subsequently, generalizations of the
Linear Template Fit are discussed, and formulae for error propagation
are presented.

\section{Example 1: the univariate Linear Template Fit}
\label{sec:example1}
An example application is constructed from a random number generator
within the \texttt{ROOT} analysis framework~\cite{Brun:1997pa}.
The physics model is a normal-distribution with a standard
deviation of $6.0$.
The mean value, denoted as $\alpha$, is subject to the inference.
Similarly to particle physics, a counting experiment is simulated, and
the (pseudo-)data are generated from 500 \emph{events} at a \emph{true} mean
value of $\alpha=170.2$, while limited acceptance restricts the measurement to
values larger than 169.
Some measurement distortions are
simulated by using a standard deviation of $6.2$ for the pseudo-data.
Limited acceptance of the experimental setup
is simulated by considering only one side of the normal distribution,
and altogether 14 ``bins'' with unity width are ``measured''.
Let us assume that from a previous measurement it is known that the physics parameter of
interest has a value of about $\alpha\approx170.5\pm1.0$.
Therefore, seven templates in a range from $\dot\alpha=169.0$ to
$172.0$ in steps of $0.5$ are generated, and each template is
generated using 40,000 random numbers.
From eq.~\eqref{eq:errorprop} it is then found that the pseudo-data will be
able to determine the value of $\alpha$ approximately with an
uncertainty of $\pm0.5$, and subsequently, additional templates at values of
$\dot\alpha=169.0$, 169.5, 170.5, 171 and 171.5 are generated.

\begin{figure}[!tbp]
     \begin{center}
      \begin{minipage}[c]{0.48\textwidth}
        \includegraphics[width=0.98\textwidth]{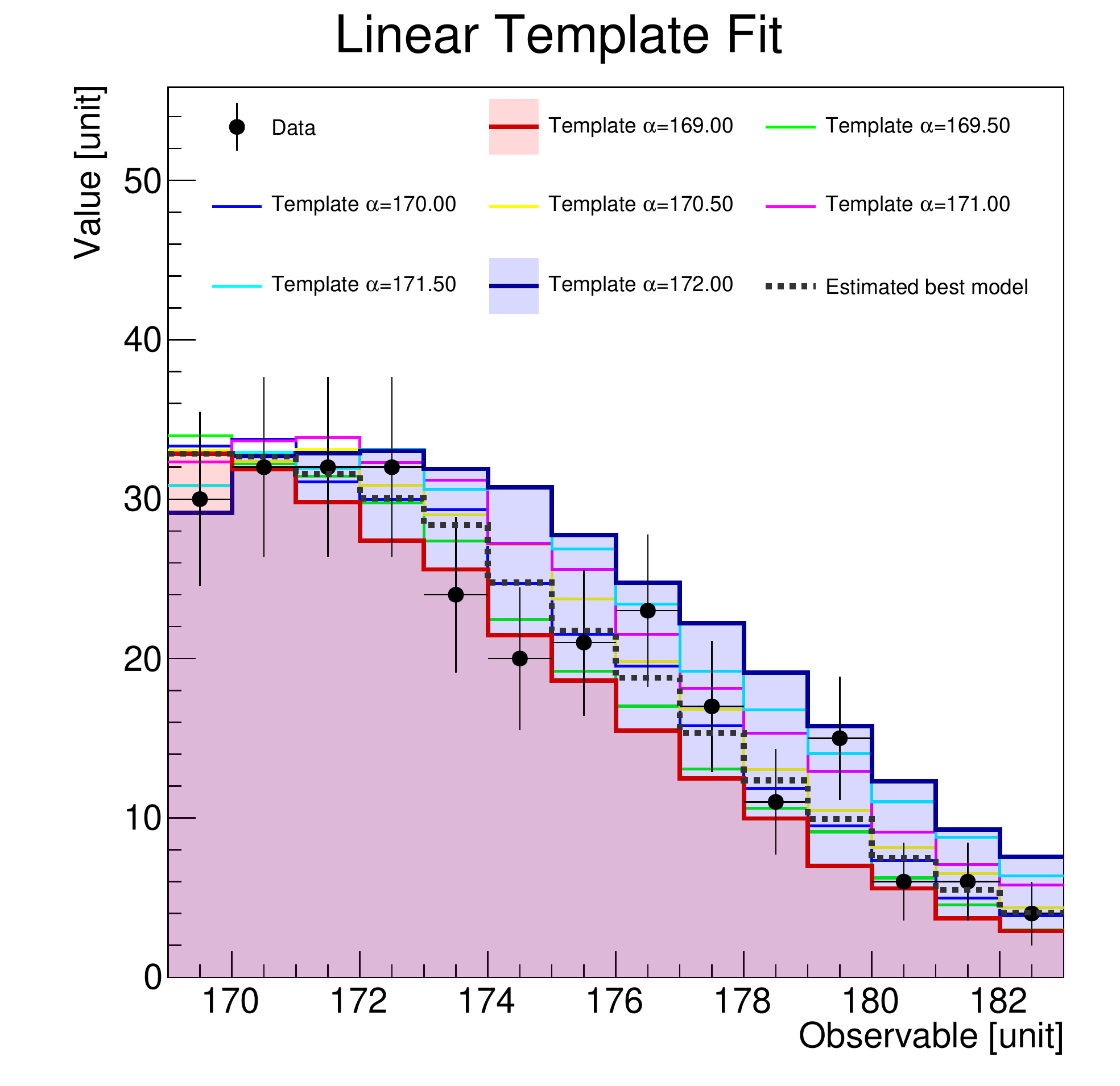}
      \end{minipage}
      \begin{minipage}[c]{0.48\textwidth}
        \includegraphics[width=0.32\textwidth]{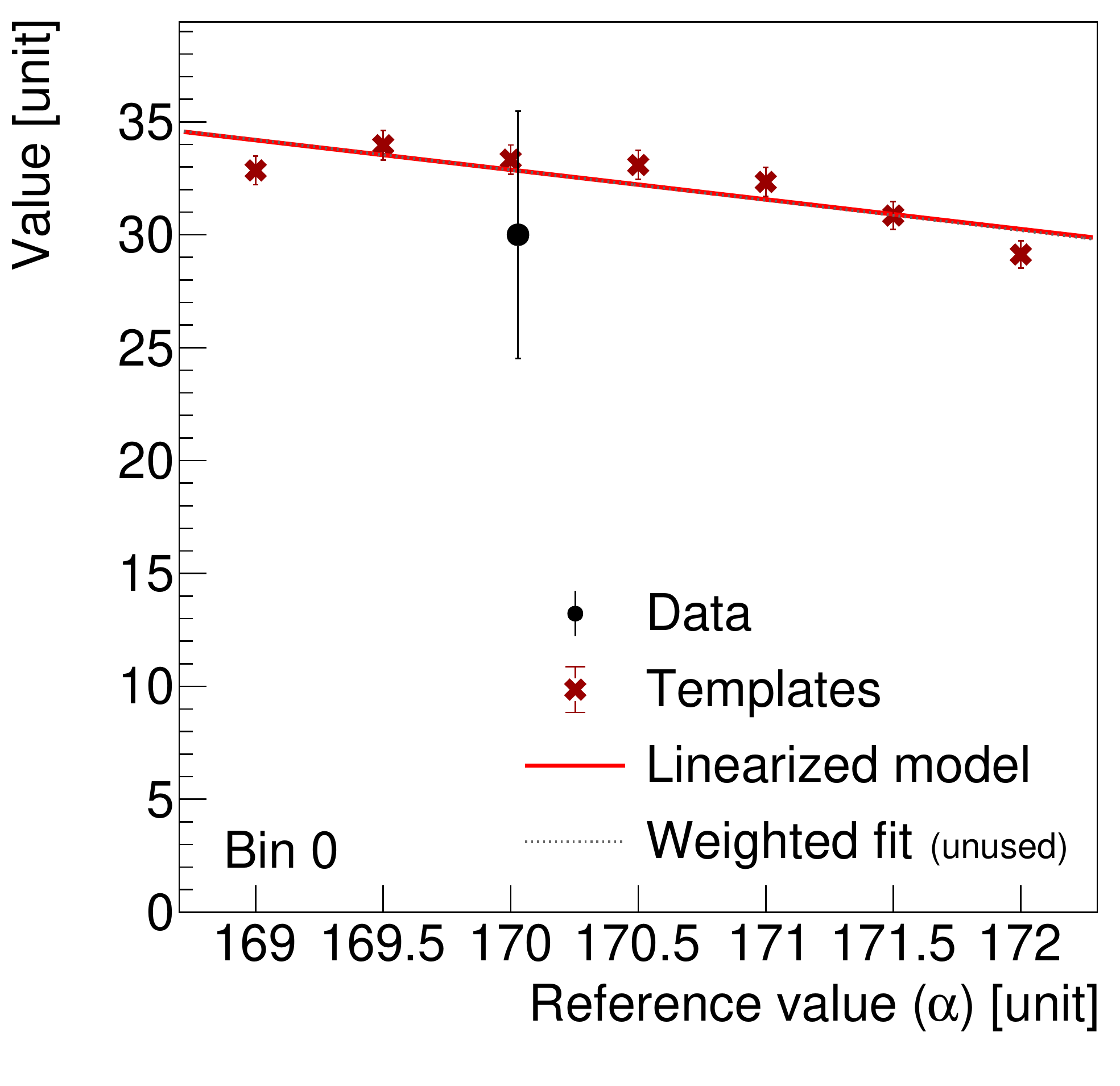}
        \includegraphics[width=0.32\textwidth]{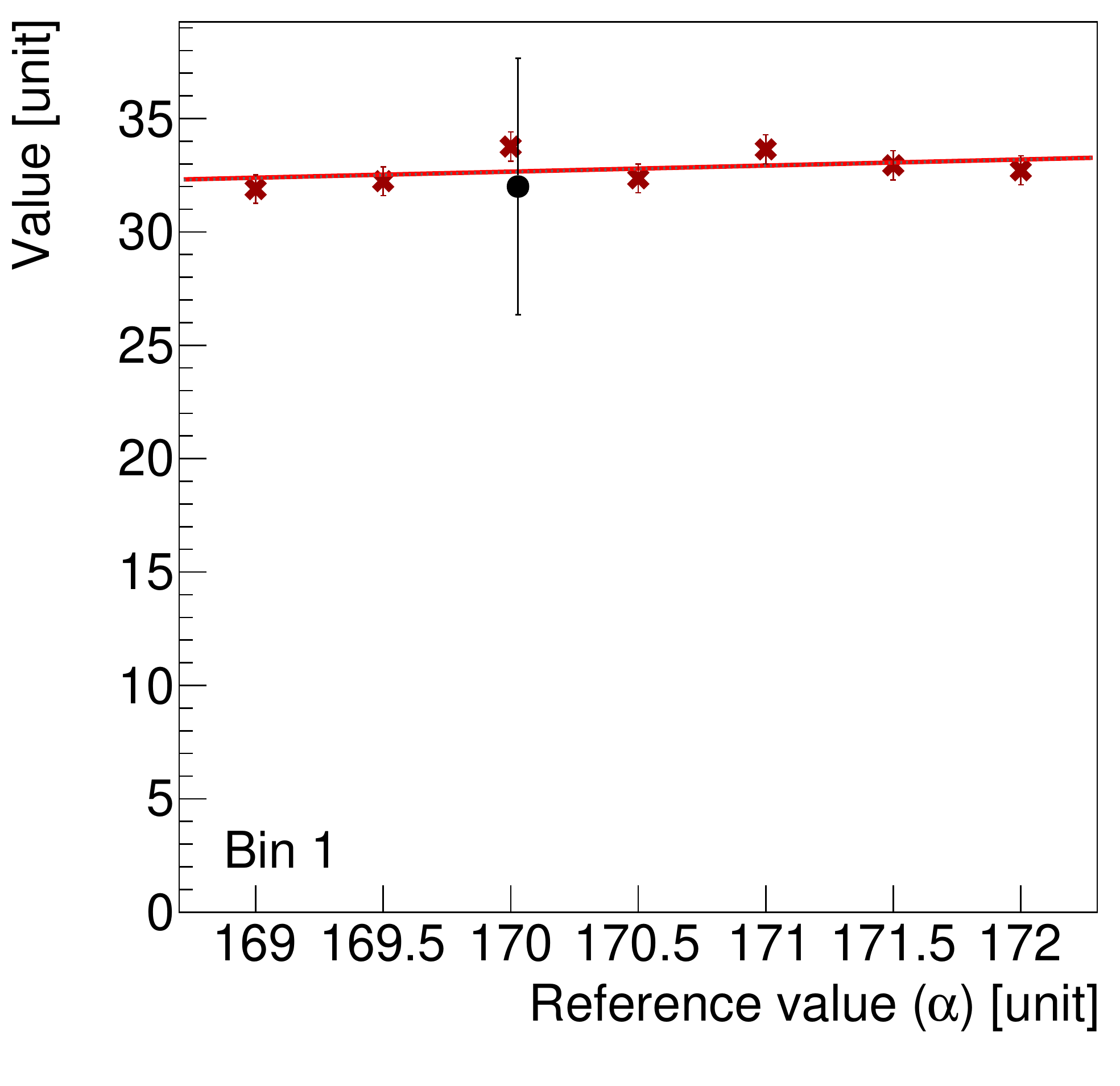}
        \includegraphics[width=0.32\textwidth]{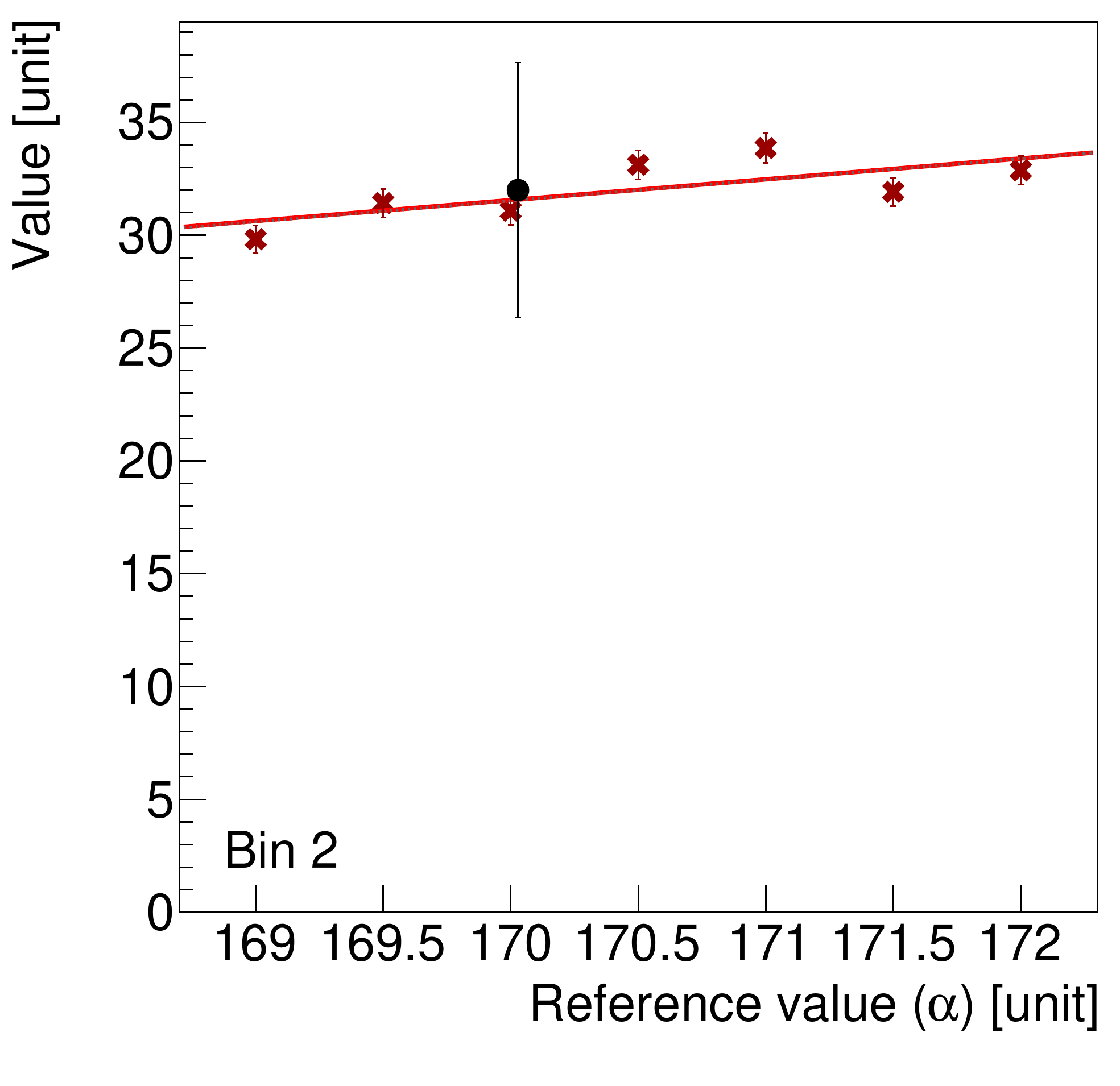}
        \includegraphics[width=0.32\textwidth]{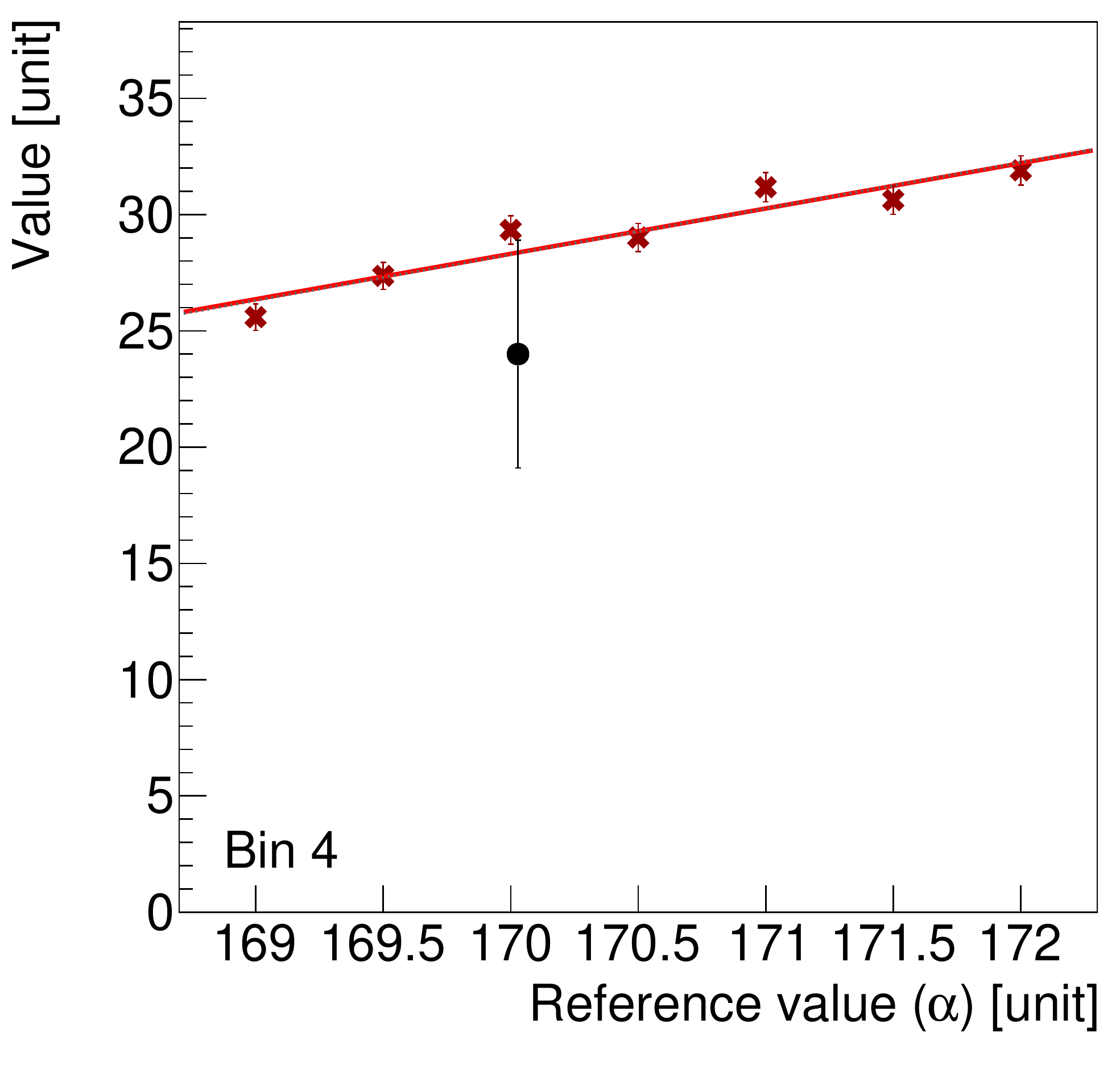}
        \includegraphics[width=0.32\textwidth]{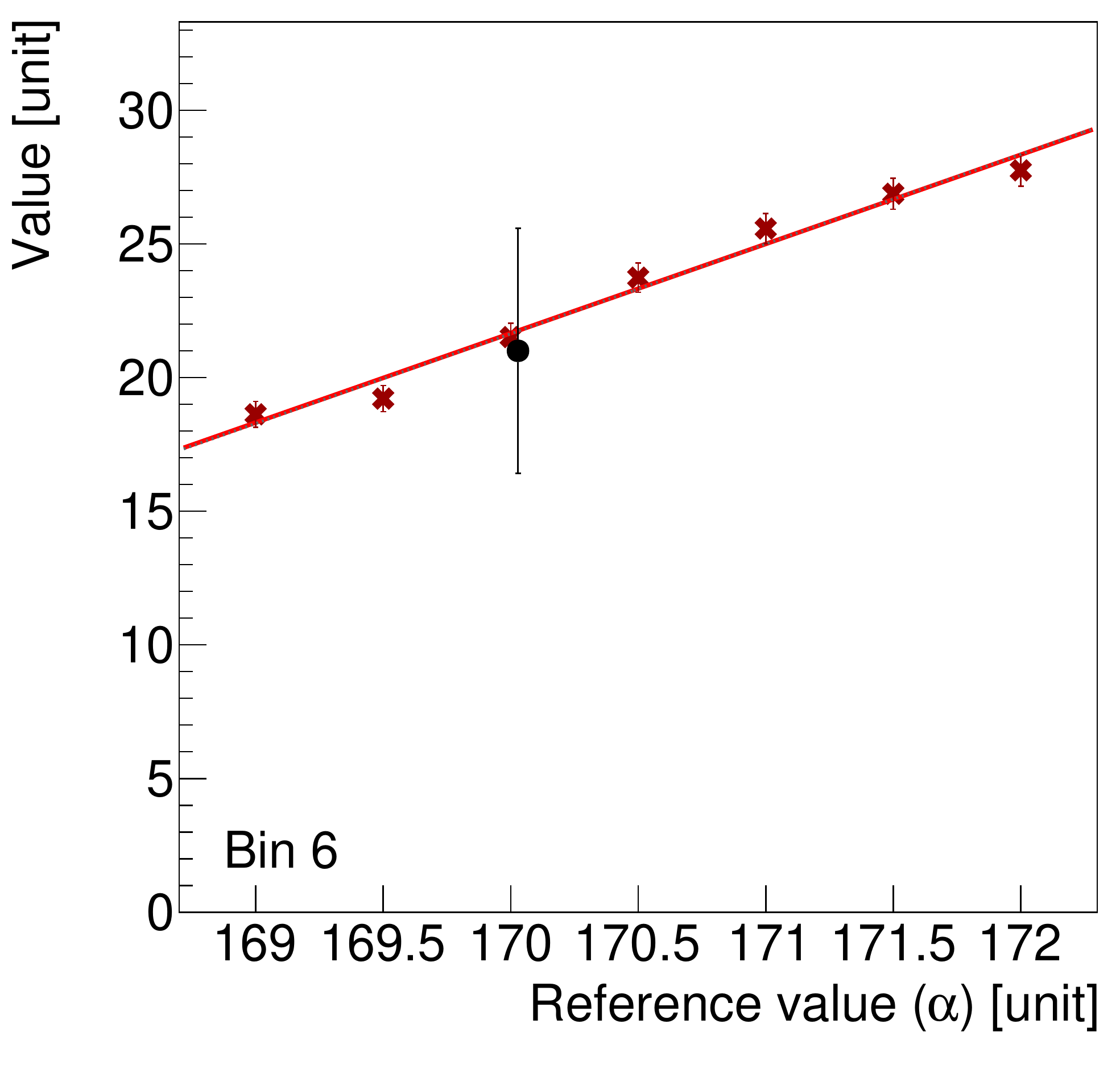}
        \includegraphics[width=0.32\textwidth]{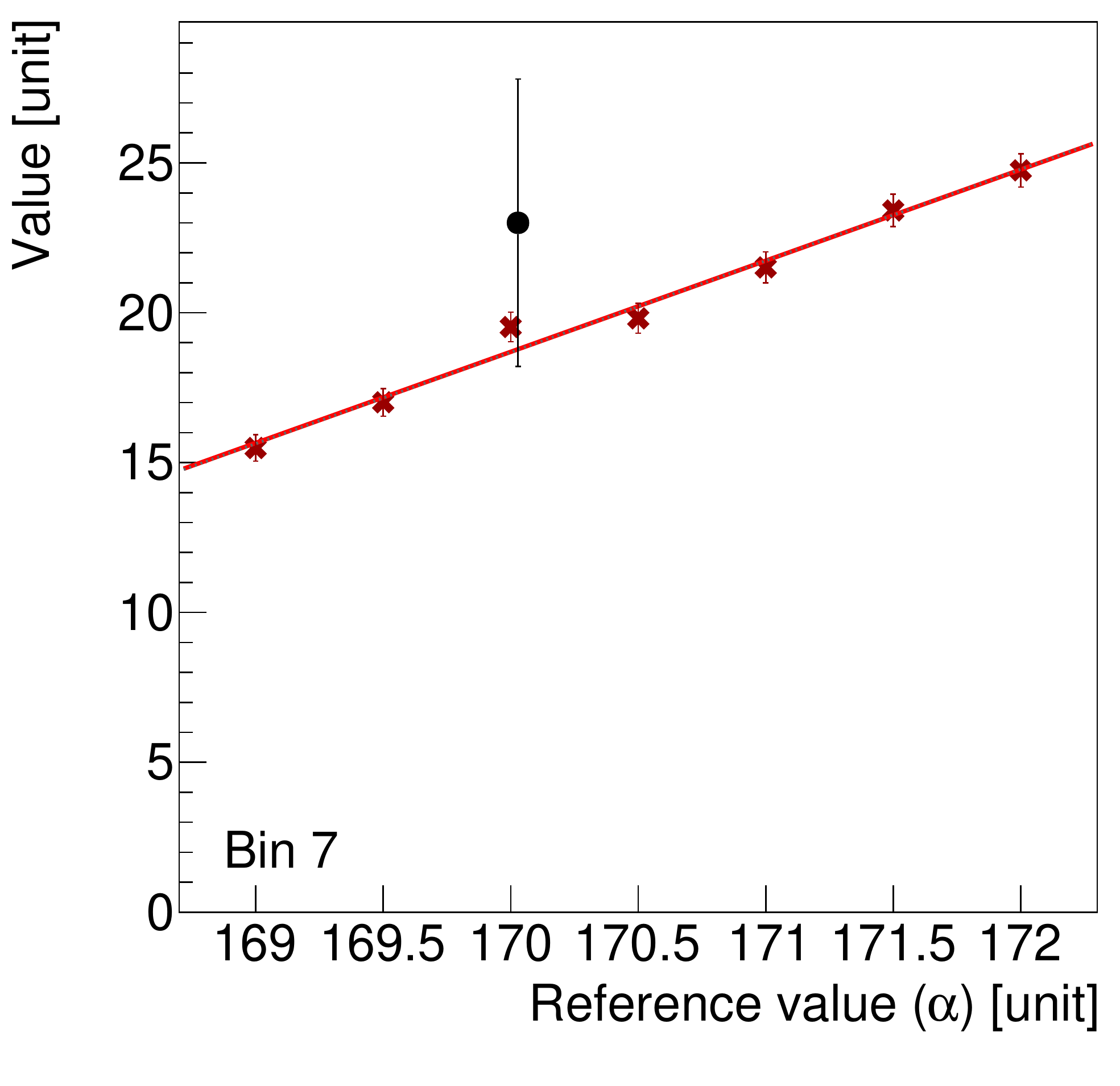}
        \includegraphics[width=0.32\textwidth]{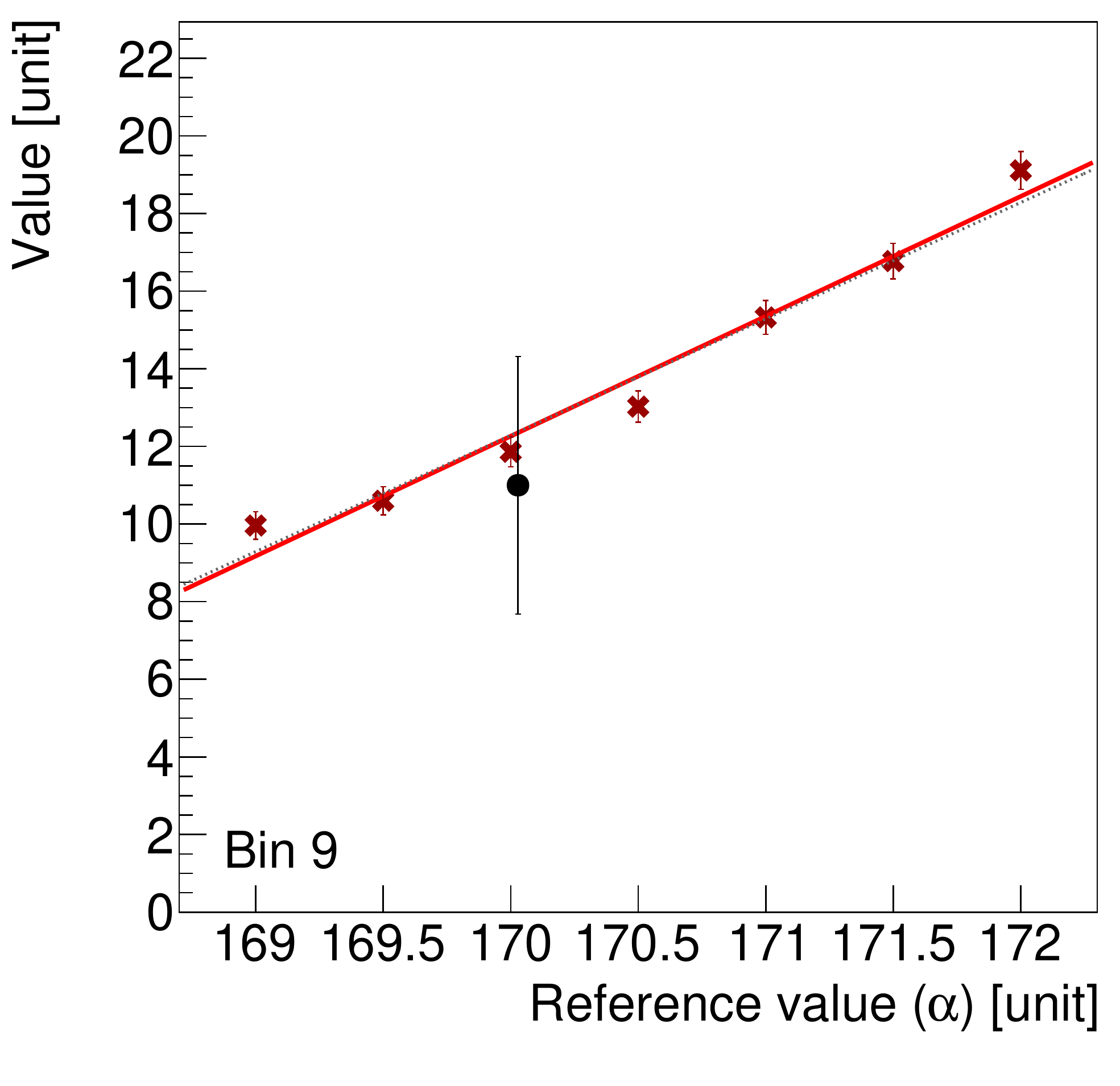}
        \includegraphics[width=0.32\textwidth]{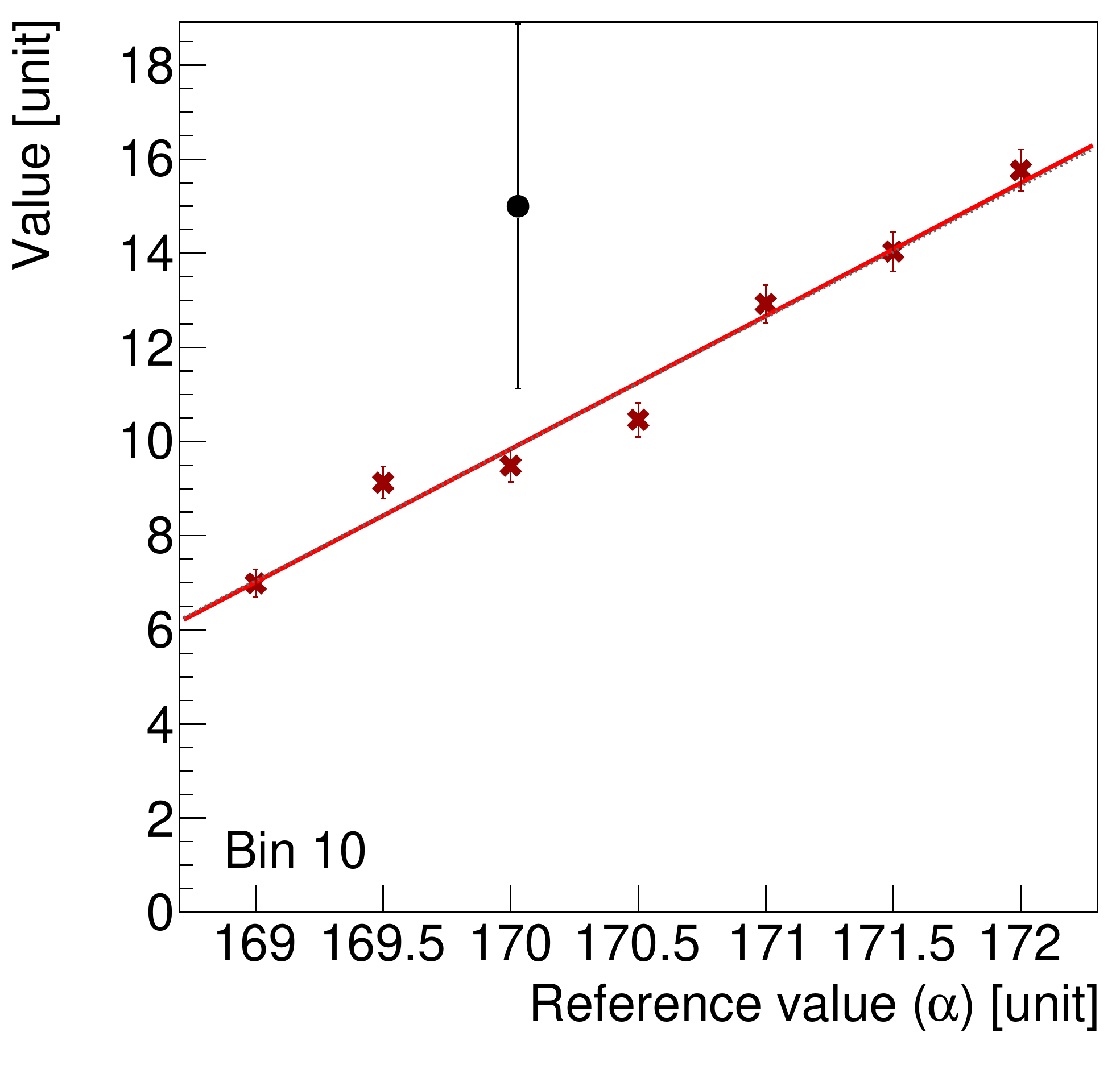}
        \includegraphics[width=0.32\textwidth]{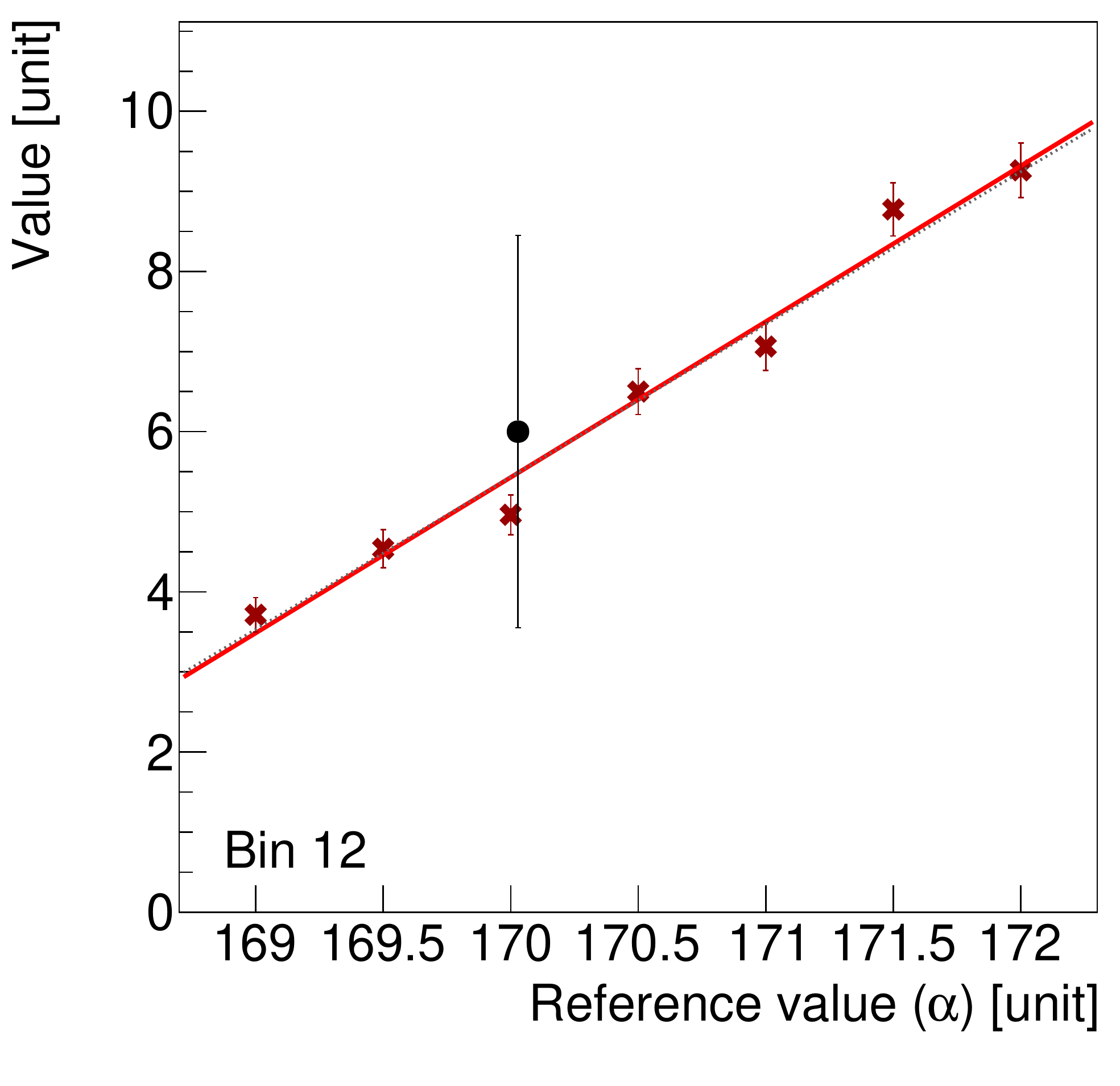}
      \end{minipage}
   \end{center}
\caption{
  Two views of the template matrix $Y$ of an example application of the
  Linear Template Fit.
  Left: the template distributions $Y$ as a function of the observable.
  The full circles display a pseudo-data set with statistical
  uncertainties, and the dotted line indicates the estimated model
  after the fit, ${\mathbf{y}}(\hat{\alpha})$.
  Right: the template distributions $Y$ of an example application of the
  Linear Template Fit, but as a function of the reference values of
  the parameter of interest $\alpha$. Only
  nine selected `bins' $i$ are displayed. The red line shows the
  linearized model $\hat{\mathrm{y}}_i(\alpha)$, eq.~\eqref{eq:ymt},
  in the respective bin, and the full circles are again the pseudo-data.
  The dotted line indicates a weighted linear regression, which is not
  used in the Linear Template Fit.
}
\label{fig:example1}
\end{figure}

The Linear Template Fit using these seven templates is illustrated in
figure~\ref{fig:example1}.
The Linear template Fit from eq.~\eqref{eq:master} reports a value of
$\hat\alpha=170.03 \pm 0.41$, which is fully consistent within the
statistical uncertainty with the
generated true value of 170.2, although a somewhat different standard
deviation was used.
In figure~\ref{fig:example1} (right), the result from a hypothesized
weighted linear regression is shown (cf.\ eq.~\eqref{eq:wgtreg}), but it is
almost indistinguishable 
from the unweighted linear regression as is used in the Linear
Template Fit.
This is because the generation of the individual templates
always employs the same methodology, as already argued above.
Furthermore, the estimated best model is displayed in
figure~\ref{fig:example1} (left), which is defined from
eq.~\eqref{eq:ymt} when substituting the best estimator,
${\hat{\mathbf{y}}}={\mathbf{y}}(\hat{\alpha})$.
This results in $\chisq=5.6$ for 14 data points and one free parameter.
Other seeds for the random number generator of course result in different values.
It should be noted that this example is purposefully constructed with large
(statistical) uncertainties in order to obtain a visually clearer
presentation in the figures, although in some bins the assumption of
normal-distributed random variables is then a rather poor 
approximation to the Poisson distribution.

\section{The multivariate Linear Template Fit}
\label{sec:multivLTF}
Phenomenological models often depend on multiple parameters, and thus it is a common
task to determine multiple parameters at a time. 
Such $k$ parameters of interest are referred to as $\alpha_1$, \dots, $\alpha_k$,
or simply $\bm\alpha$,
and the best estimators are denoted as $\hat{\alpha}_1$, \dots,
$\hat{\alpha}_k$, or $\bm{\hat\alpha}$.
The 
linear representation of the model is a hyperplane
$\rm{y}_i(\bm\alpha)$ in each of the $i$ data points, defined as
\begin{equation}
  \bm{\lambda}(\bm{\alpha}) \approx
    \mathbf{y}(\bm{\alpha};{\bm{\hat\theta}}) = \hat\theta_0 + \hat\theta_{(1,1)} \alpha_1 + \dots + \hat\theta_{(1,k)} \alpha_k\,,
   \label{eq:linapprox}
\end{equation}
where a constant $\theta_0$ and the first-degree
parameters $\theta_{(1,k)}$ are considered.
Since higher-degree terms or interference terms are not included, the
fit parameters need to be (sufficiently) independent or they have been
made orthogonal by applying a variable transformation beforehand.

In the multivariate case, each template $\bm{y}_{(j)}$ is representative
of a reference point $\bm{\dot\alpha}_{(j)}$ in the $k$-dimensional space.
The regressor matrix $M$ is constructed as a $j\times(1+k)$
design matrix:
\begin{equation}
  M :=
  \left(\begin{matrix}
    1 & \dot\alpha_{(1),1} & \dots & \dot\alpha_{(1),k} \\
    \vdots&  \vdots         & \vdots      &   \vdots       \\
    1 & \dot\alpha_{(j),1} & \dots & \dot\alpha_{(j),k} \\
  \end{matrix}\right)\,.
  \label{eq:MnD}
\end{equation}
As in the univariate case, the pseudoinverse $M^+$ is calculated from eq.~\eqref{eq:Mc},
and also the same considerations for the justification
of the \emph{unweighted} multiple regression are applicable.
Instead of a single vector $\mt$, there is now a
vector for each regression parameter $\theta_k$.
Therefore, the  $j\times k$ matrix $\tilde{M}$ is introduced, which is defined by
decomposing $M^+$ like
\begin{equation}
  M^+ =:
  \begin{pmatrix}
        \bm{\bar{m}}^{\text{T}}\\
        \tilde{M}^{\text{T}} \\
  \end{pmatrix}\,.
  \label{eq:Mbar}
\end{equation}
Hence, the best estimator for the linearized multivariate model
$\mathbf{y}(\bm{\alpha};{\bm{\theta}})$ becomes  
\begin{equation}
  \bm{\lambda}(\bm\alpha) \approx
  {\hat{\mathbf{y}}}(\bm\alpha) = Y\mb + Y\M\bm\alpha\,,
  \label{eq:ymt2}
\end{equation}
where the $\theta$-parameters are again no longer explicit.

When using the linear approximation of the model,
$\bm\lambda(\bm\alpha)\approx\hat{\mathbf{y}}(\bm\alpha)$, the
\chisq~function (eq.~\eqref{eq:chisqDetLev})
for the  multivariate Linear Template Fit becomes
\begin{align}
  \chisq 
         &\simeq \left( \dt -\hat{\mathbf{y}}(\bm\alpha)\right)^{\text{T}} W \left( \dt - \hat{\mathbf{y}}(\bm\alpha)\right)\\
         &= \left( \dt - Y\mb - Y\M\bm{\alpha}\right)^{\text{T}} W \left( \dt - Y\mb - Y\M\bm{\alpha}\right)\\
         &= \left( \dt - {\textstyle\sum}_l\hat\epsilon_l \s_{(l)} - \hat{\mathbf{y}}(\bm\alpha)\right)^{\text{T}} \mathbb{V}^{-1} \left(\dt - {\textstyle\sum}_l\hat\epsilon_l \s_{(l)} - \hat{\mathbf{y}}(\bm\alpha)\right) + {\textstyle\sum}_l \hat\epsilon_l^2\,.
  \label{eq:chisqNuisance}
\end{align}
The last equation introduces an equivalent expression in terms of
nuisance parameters~\cite{Heinrich:2007zza}.
The factors $\hat\epsilon_l$ are related to uncertainties with
full bin-to-bin correlations when writing the covariance matrix as 
\begin{equation}
  V
  =\mathbb{V}+V_\textrm{corr} 
  ~~~\text{ using}~~~
  V_\textrm{corr}=\textstyle\sum_l \s_{(l)}^{ }\s^{\text{T}}_{(l)}\,,
  \label{eq:Vsum}
\end{equation}
where the sum $l$ runs over all uncertainties with full
bin-to-bin correlations and the
vectors $\s_l$ denote the individual systematic uncertainties (also
called \emph{shifts}), while the matrix $\mathbb{V}$ includes all
other uncertainty components.   
It is common practice that the systematic shifts are calculated
from relative uncertainties and multiplied with the measured
data. Implications of this practice are discussed in section~\ref{sec:LogN}
and~\ref{sec:errorrescaling}, and care must be taken that the result
does not become
biased~\cite{Lyons:1989gh,Lincoln:1993yi,DAgostini:1993arp,Takeuchi:1995xe,D'Agostini:642515,Ball:2009qv};
 a common technique to avoid that bias is discussed in section~\ref{sec:errorrescaling}.

Equation~\eqref{eq:chisqNuisance} is again a linear least 
squares expression and the best linear unbiased estimators for the parameters
$\bm{\hat\alpha}$ and the nuisance parameters $\bm{\hat\epsilon}$
are obtained from the stationary point.
Hence, the best estimators from the \emph{multivariate Linear Template Fit} become
\begin{equation}
  \ahut 
    = \begin{pmatrix}\bm{\hat\alpha}\\\bm{\hat\epsilon}\end{pmatrix}
    = \mathcal{F} ( \dt-Y\mb)\,,
        \label{eq:full2}
\end{equation}
where $\ahut$ was introduced and
the shorthand notations $\mathcal{F}$ (a $g$-inverse of least squares)
and $\mathcal{D}$ are calculated as
\begin{equation}
  \mathcal{F}
  :=
  \mathcal{D}^{-1}   \begin{pmatrix}\Y & ~S\end{pmatrix}^{\text{T}} \W
    :=
  \left(
  \begin{pmatrix}\Y & ~S\end{pmatrix}^{\text{T}}
    \W
    \begin{pmatrix}\Y & ~S\end{pmatrix}
      + \begin{pmatrix} {0}_k & 0 \\ 0 & {1}_l\end{pmatrix}
        \right)^{-1}
        \begin{pmatrix}\Y & ~S\end{pmatrix}^{\text{T}} \W\,,
        \label{eq:full0}
\end{equation}
using $\W=\mathbb{V}^{-1}$.
The matrix with the fully bin-to-bin correlated uncertainties $S$ is composed
from the column-vectors of the systematic shifts $\s_{(l)}$ as
\begin{equation}
  S := \begin{pmatrix}\s_{(1)} & \hdots & \s_{(l)} \end{pmatrix}\,,
  \label{eq:S}
\end{equation}
and the symbol $\begin{pmatrix}\Y & ~S\end{pmatrix}$ denotes a composed
  $i\times(k+l)$ matrix
  from  $\Y$
  and $S$, while
  ${0}_{k}$ and ${1}_{l}$ denote a $k\times k$ or $l\times l$
  zero or unit-matrix, respectively.  
The matrix $\mathcal{D}$ 
is commonly a full-ranked symmetric matrix
and thus invertible.
A brief discussion about a more efficient numerical calculation of
$\mathcal{D}^{-1}$ is given in~\ref{sec:numerics}. 
If all uncertainty components are represented through a single
covariance matrix $V$, cf.\ left side of eq.~\eqref{eq:Vsum}, the
multivariate Linear Template Fit simplifies to
\begin{equation}
  \bm{\hat\alpha}
    =
    \left((Y\M)^{\text{T}}WY\M\right)^{-1}(Y\M)^{\text{T}} W ( \dt-Y\mb)\,.
    \label{eq:full}
\end{equation}

\section{Example 2: the multivariate Linear Template Fit}
\label{sec:example2}
\begin{figure}[!thbp]
  \begin{center}
     \begin{minipage}[c]{0.48\textwidth}
       \includegraphics[width=0.98\textwidth]{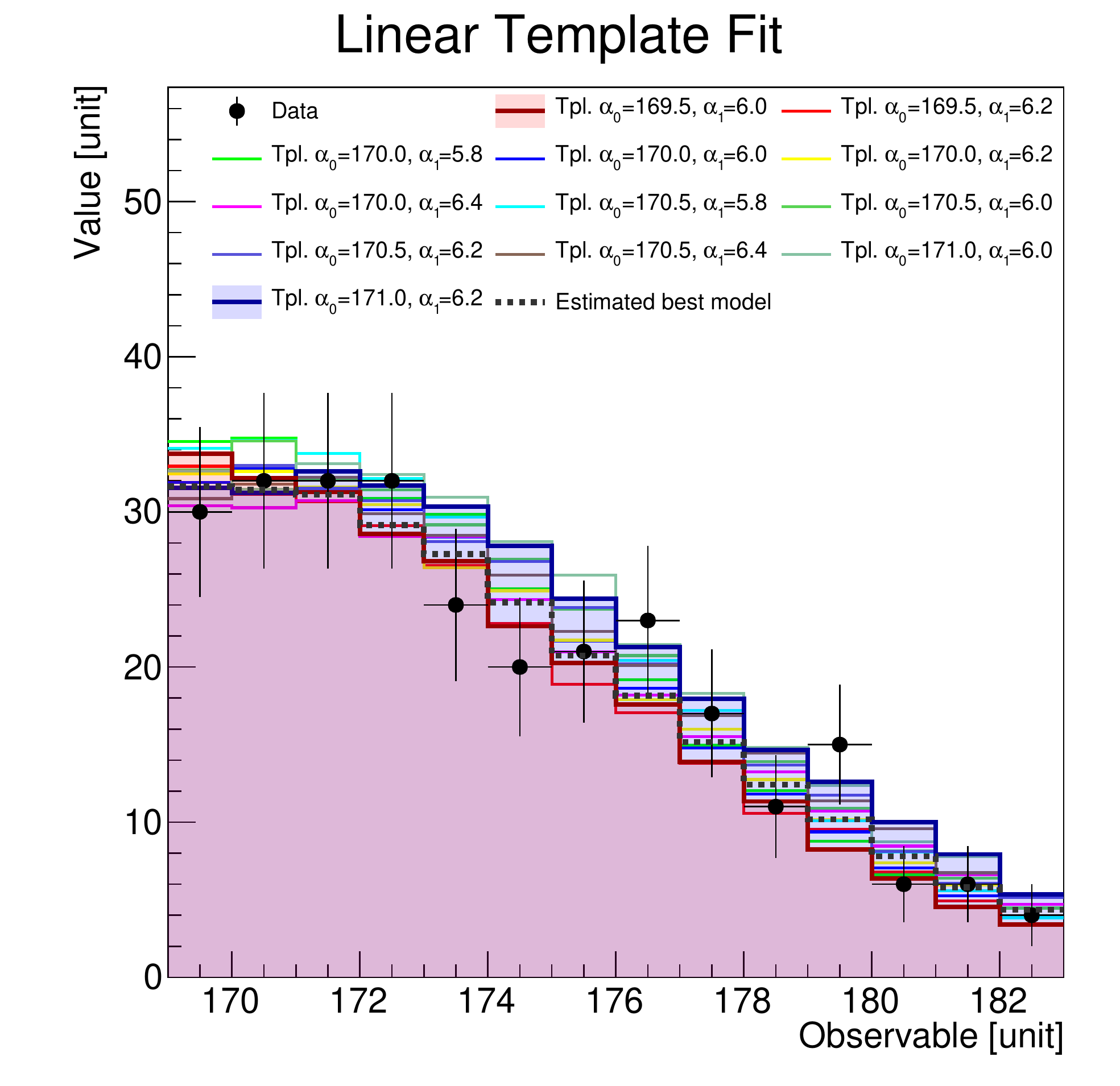}
     \end{minipage}
     \begin{minipage}[c]{0.48\textwidth}
       \includegraphics[width=0.49\textwidth]{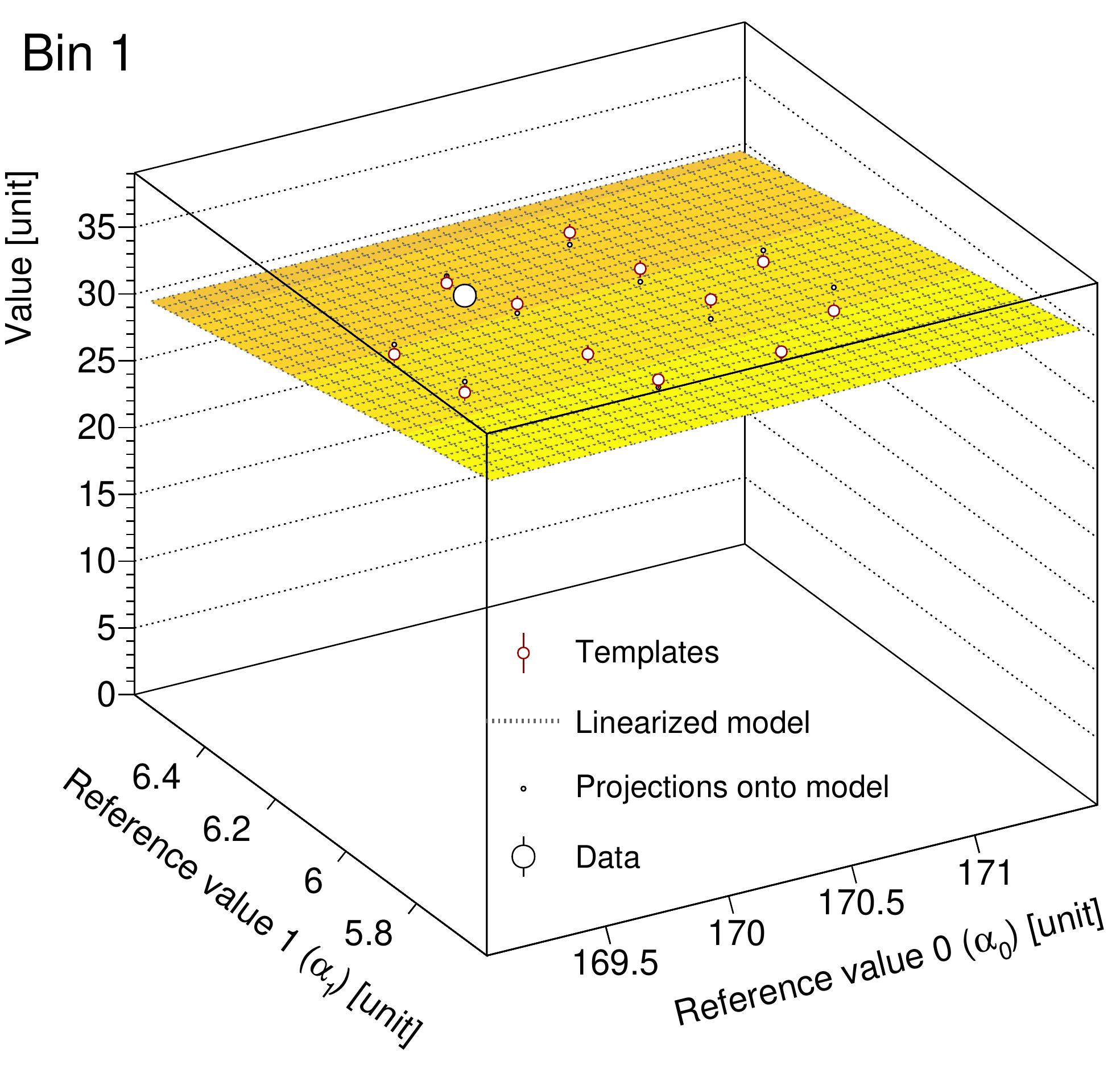}
       \includegraphics[width=0.49\textwidth]{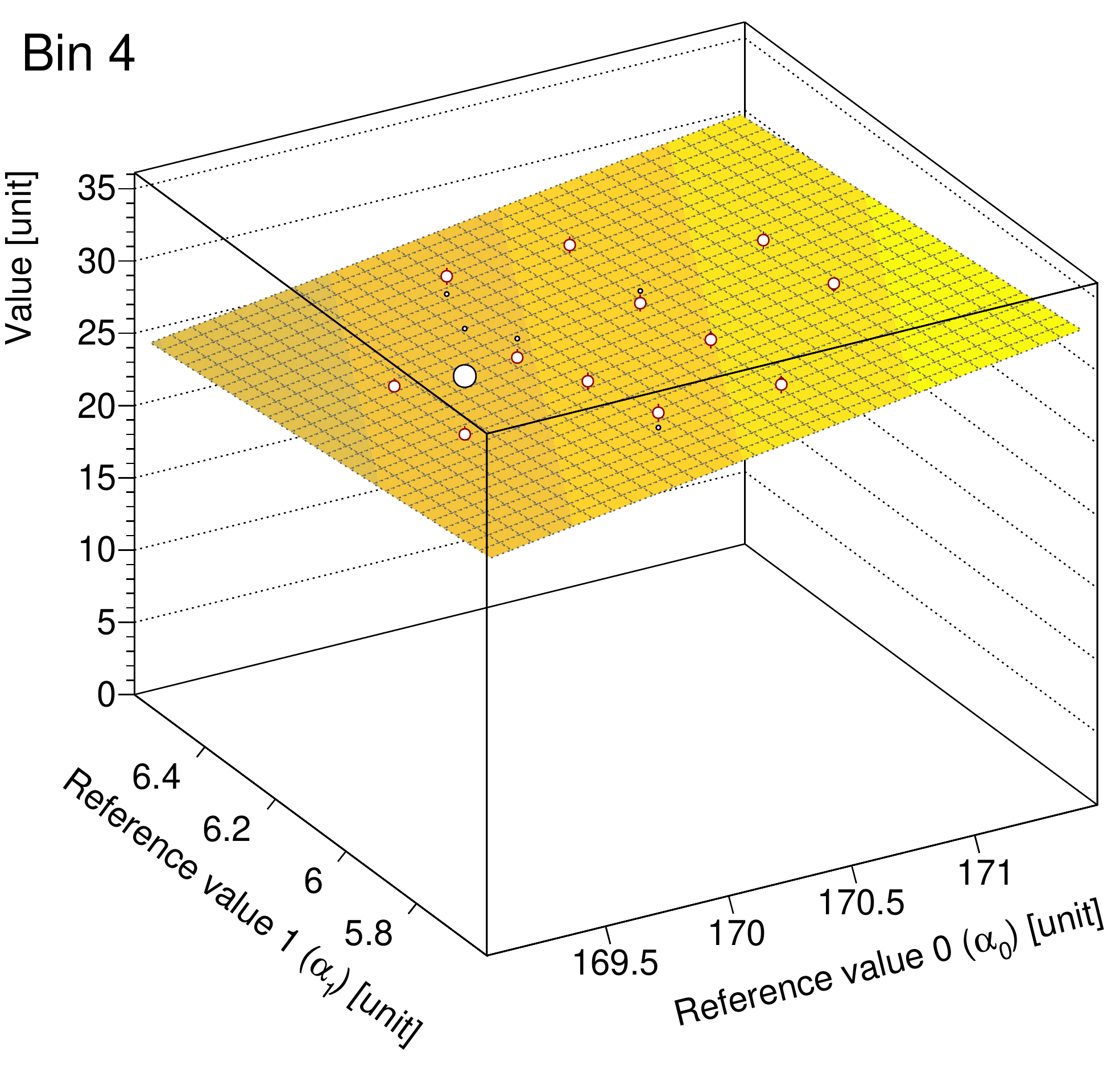}
       \includegraphics[width=0.49\textwidth]{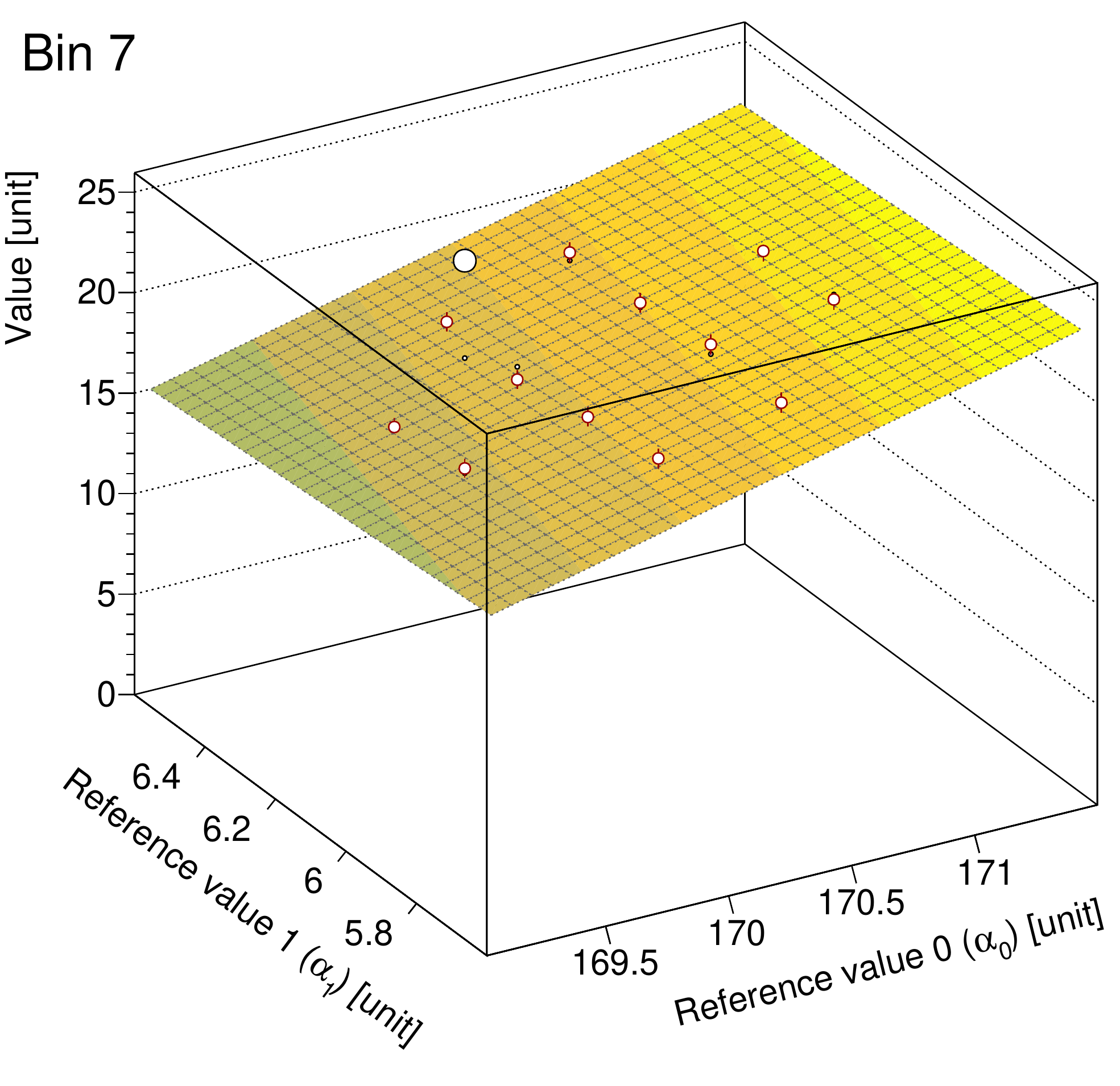}
       \includegraphics[width=0.49\textwidth]{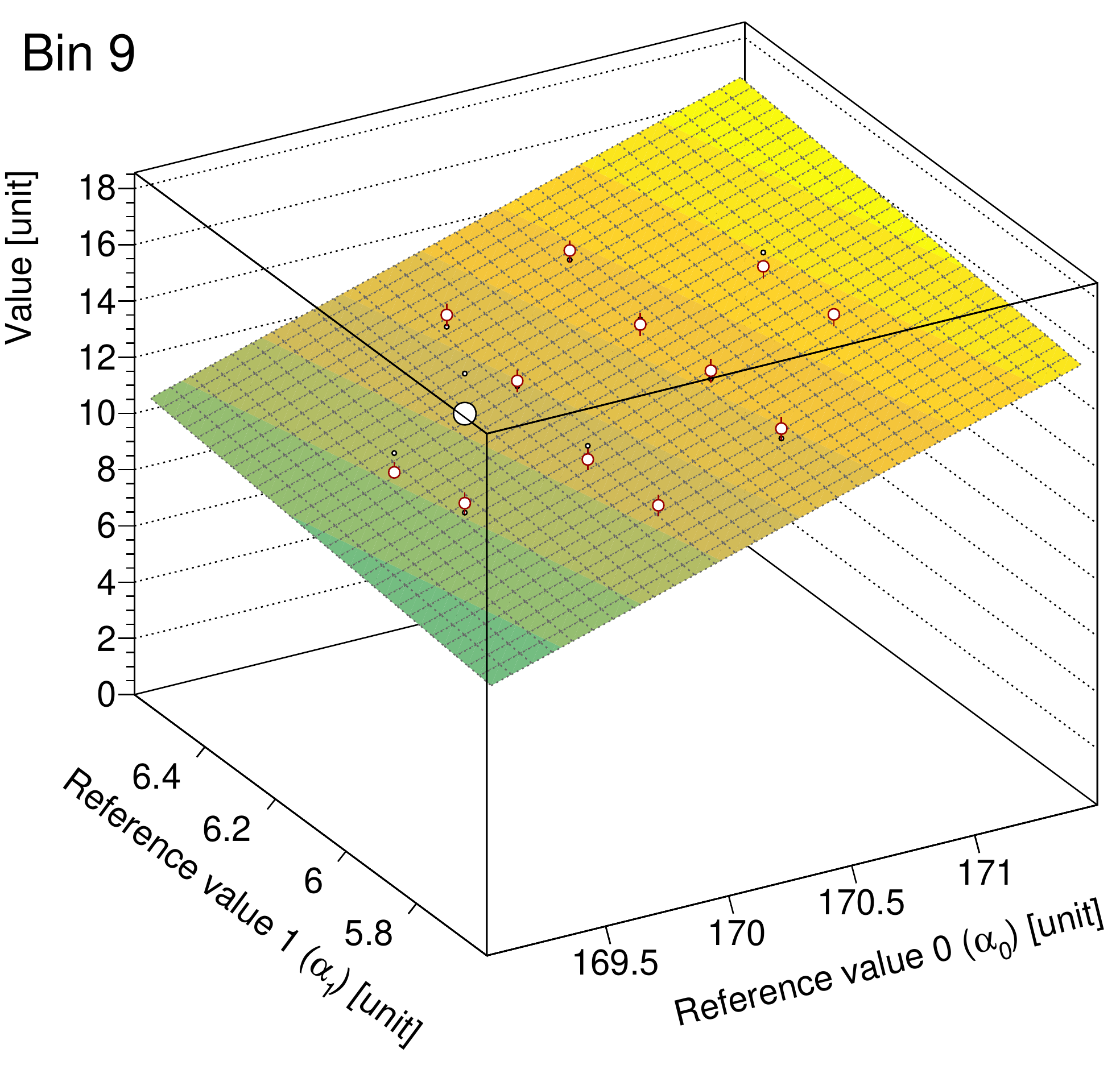}
     \end{minipage}
  \end{center}
\caption{
  Two views of the template matrix $Y$ of an example application of
  the multivariate Linear Template Fit.
  Left:
  the template distributions $Y$ as a function the
  observable. More details are given in the caption of figure~\ref{fig:example1}.
  Right:
  the template distributions $Y$ (open medium-sized circles) as a
  function of the two reference   values for four selected ``bins'' $i$.
  The plane displays the linearized model
  $\hat{\mathrm{y}}_i(\bm{\alpha})$, eq.~\eqref{eq:ymt2}, 
  in the respective bin.
  The large open circles are again the pseudo-data, and small circles
  indicate the projection onto the plane $\hat{\mathrm{y}}$.
}
\label{fig:example2}
\end{figure}

In Example 1, only the mean value of the
model (which is a normal-distribution) was determined, although
the pseudo-data had a slightly different standard deviation than the
model.
In the following example, a multivariate Linear Template Fit is
performed and the same pseudo-data as in Example 1 are used, but both
values of the model will be determined: the mean value and the standard
deviation of the Gaussian.
Therefore, templates are generated for some selected values for the
mean (between values of 169.5 and 171) and
the standard deviation (between values of 5.8 and 6.4) of a Gaussian,
and again 40,000 events are used for each template.
The four ``extreme'' variations of the two parameters are omitted.
The multivariate Linear Template Fit using all these templates is illustrated in
figure~\ref{fig:example2}.

The best estimators from the multivariate Linear Template Fit
(eq.~\eqref{eq:full}) are  $169.91\pm 0.46$ for the mean and
$6.28\pm0.38$ for the standard deviation.
Their correlation coefficient is found to be $-0.3$.
Within the
statistical uncertainties, both values are in very good agreement with the
simulated input values of $170.2$ and $6.2$, respectively.
Thus, this example illustrates the application of the multivariate Linear
Template Fit with two free parameters, where a visual representation
of the linearized model is still possible. However, any number of free
parameters is in principle possible, and for an $n$-parameter fit, the minimum
number of linearly independent templates is just $n+1$.

\section{The Linear Template Fit with relative uncertainties}
\label{sec:LogN}
\begin{figure*}[t!b]
\begin{center}
      \includegraphics[width=0.49\textwidth,trim={0 0 30 0 },clip]{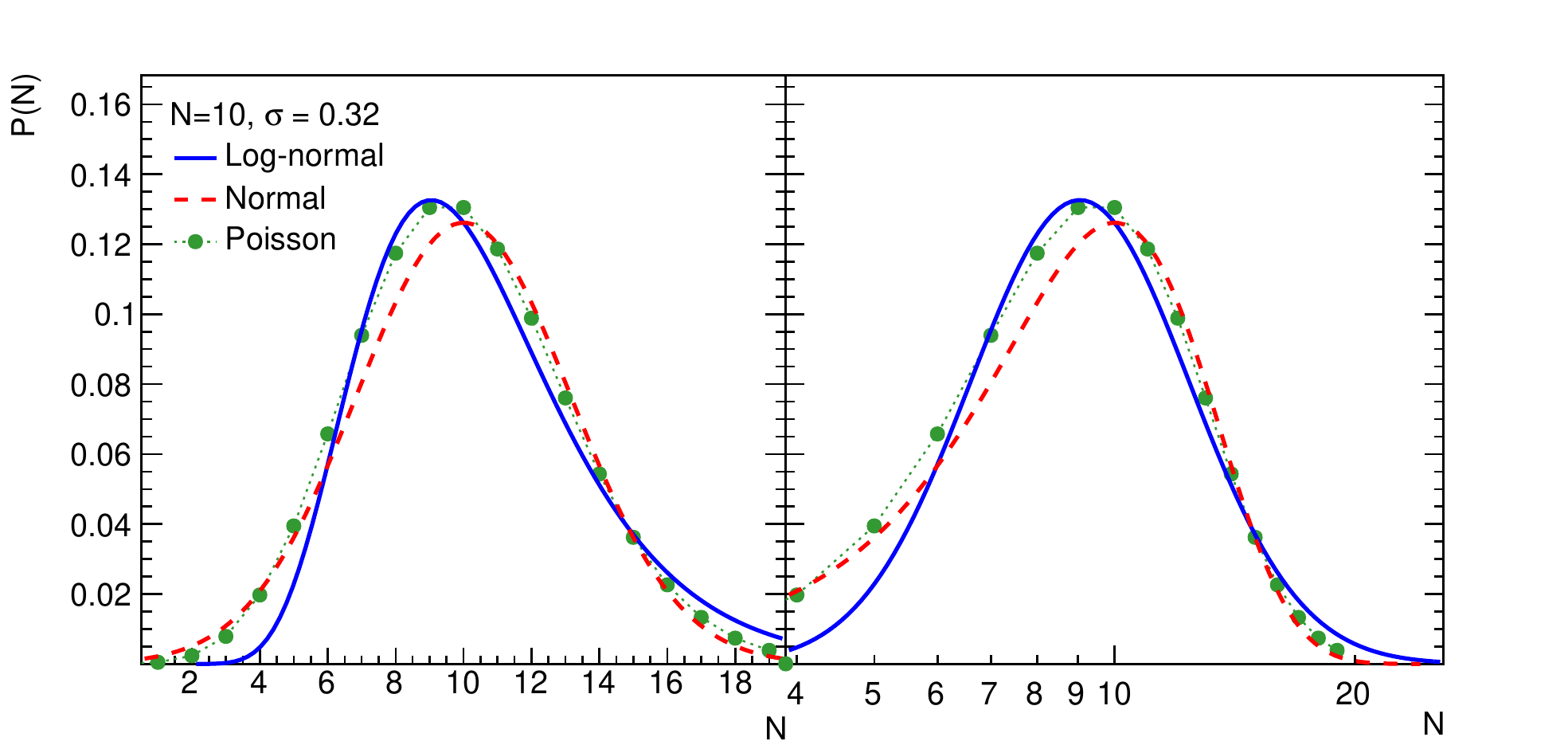}
      \includegraphics[width=0.49\textwidth,trim={0 0 30 0 },clip]{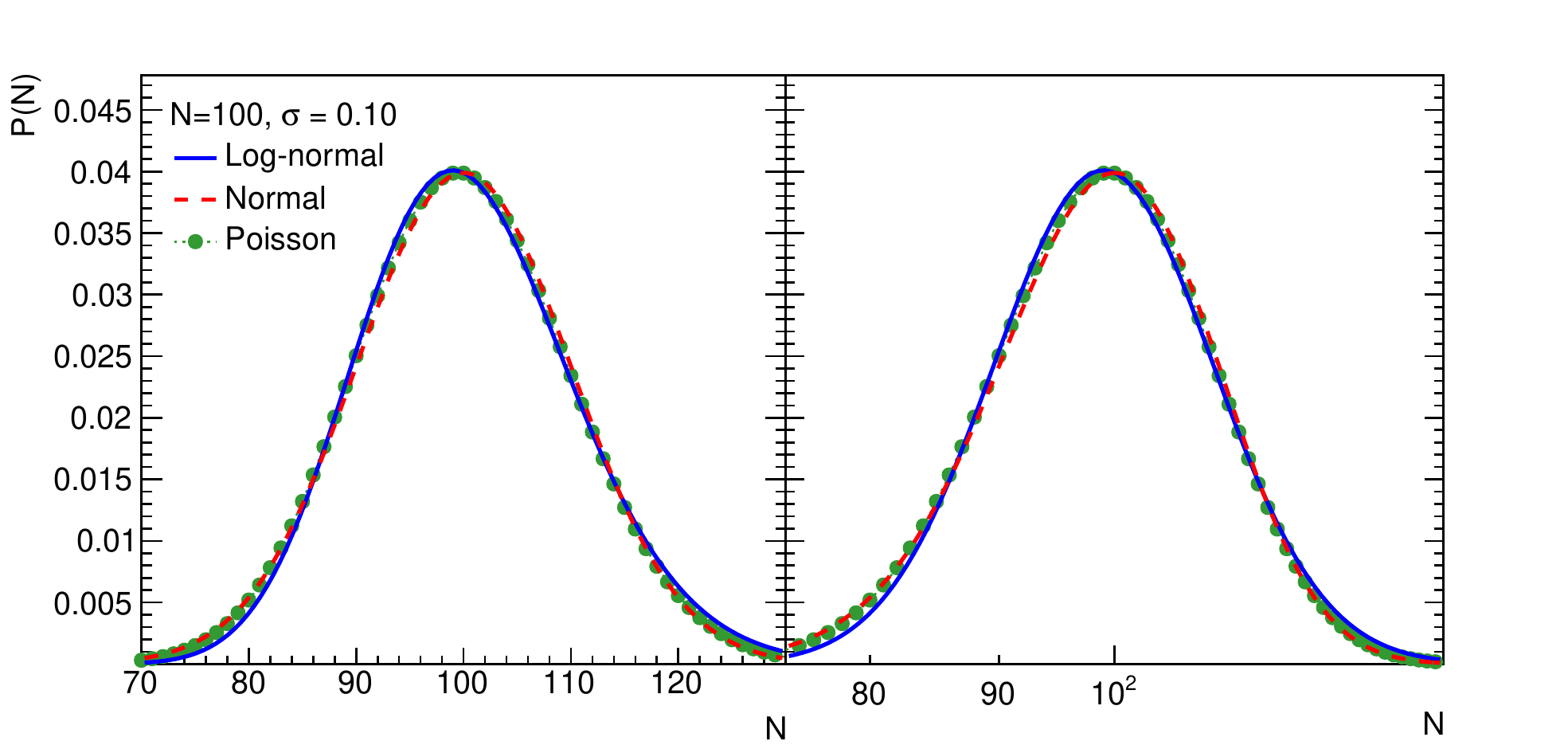}
\end{center}
\caption{
  The log-normal, normal, and Poisson probability distribution
  functions with mean values of 10 (left) and 100 (right)
  on a linear and logarithmic axis.
  These values are commonly referred to as relative uncertainties of
  32\,\% and 10\,\%, respectively.  For larger uncertainties, the
  log-normal distribution provides a 
  good approximation of the Poisson distribution around the peak. The
  log-normal and the Poisson distributions are strictly positive.
  For large $N$, all distributions become similar.
}
\label{fig:lognormal}
\end{figure*}

In this section we present the equations for the Linear Template Fit
when the estimators obey a log-normal distribution. 
This could be the case for data when the determination of a variable
is affected by a number of multiplicative factors that are subject to
uncertainties. Consequently, the variable follows a log-normal 
distribution due to the central limit
theorem~\cite{Cowan:1998ji,James:1019859}.
Also, when the value of an observed variable is a random proportion of the
previous observation, it follows a log-normal
distribution~\cite{James:1019859}. 
An example would be the measurement of the electron energy in a
calorimeter which is affected by a number of fractional energy losses
and corrections~\cite{Behnke:1517556}.
Another example would be measurements that are dominated by systematic
multiplicative uncertainties: a prominent multiplicative error is due
to uncertainties of the luminosity measurement in 
particle collider experiments which results in a relative uncertainty.
Contrary to additive errors, multiplicative errors cannot change the
sign of the variable, and a positively defined observable always remains
positive, which is an important prerequisite for several physical quantities
such as cross sections. 
A comparison of the log-normal, the normal, and the Poisson probability
distribution function for two selected values of their mean value are
displayed in figure~\ref{fig:lognormal}.

The likelihood of a set of $i$ measurements is a joint function of
log-normal probability distribution functions
\begin{equation}
  \mathcal{L} = \prod_{i}^N\frac{1}{\sqrt{2\pi\varsigma_i^2d_i^2}}\exp\left(\frac{-(\log{d_i}-\mu_i)^2}{2\varsigma_i^2}\right)\,.
  \label{eq:Llog}
\end{equation}
When writing $\mu_i=\log m_i$~\cite{Forbes:1271091,Andreev:2014wwa}, it
can be directly
seen from the residuals that the ratios
$d_i/m_i$ enter the likelihood calculation instead of their differences.
The variance $\varsigma$ denotes a relative quantity, thus it
represents \emph{relative uncertainties}. 
In fact, the log-normal distribution is equivalent to
considering random variables with normal-distributed relative uncertainties. 
For small relative errors, $\varsigma_i\lesssim20\,\%$, which is a
common case in particle physics, the log-normal,
the normal, and the Poisson distributions become similar.
For larger relative errors, the log-normal distribution 
provides some approximation of the Poisson distribution, and for
practical purposes one benefits from its strictly positiveness, in
contrast to the often used normal distribution.

Similarly to the case for normal-distributed data,
we choose as a first-order approximation of the model in every bin
\begin{equation}
  \mu_i(\bm\alpha)
  \approx  \theta^{(i)}_0 + \theta^{(i)}_{(1,1)} \alpha_1 +\hdots +
  \theta^{(i)}_{(1,k)} \alpha_k\,,
  \label{eq:logapprox}
\end{equation}
and the best estimators for the $\theta$ parameters are again
determined from the templates; consequently,
the best estimator of the approximated model is
\begin{equation}
  \bm{\mu}(\bm{\alpha})
  \approx
  \bm{\hat{{\mu}}}(\bm\alpha) 
  = \log(Y) \mb  + \log(Y) \M\bm{\alpha}\,.
  \label{eq:ymtlog}
\end{equation}
The term $\log(Y)$ denotes a coefficient-wise
application of the logarithm to the elements of the template matrix
$Y$, and it is used to match the logarithmized data ($\log{d_i}$) in
the likelihood.
Eq.~\eqref{eq:ymtlog} is an equally valid first-order approximation of the
model as compared with the default linear approximation in
eq.~\eqref{eq:linapprox}.

\begin{figure}[tb!]
\begin{center}
       \includegraphics[width=0.24\textwidth,trim={0 0 10 0 },clip]{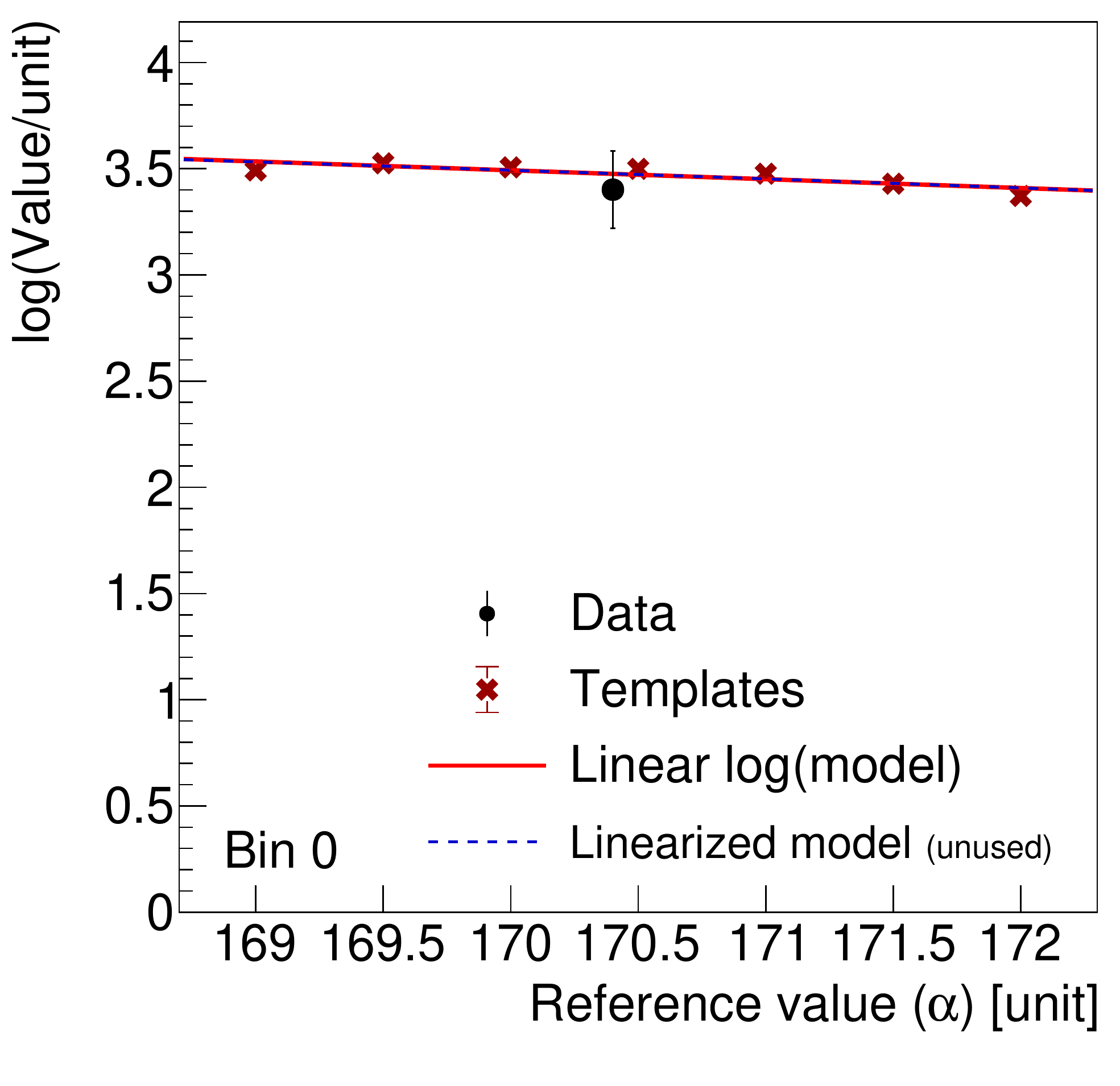}
       \includegraphics[width=0.24\textwidth,trim={0 0 10 0 },clip]{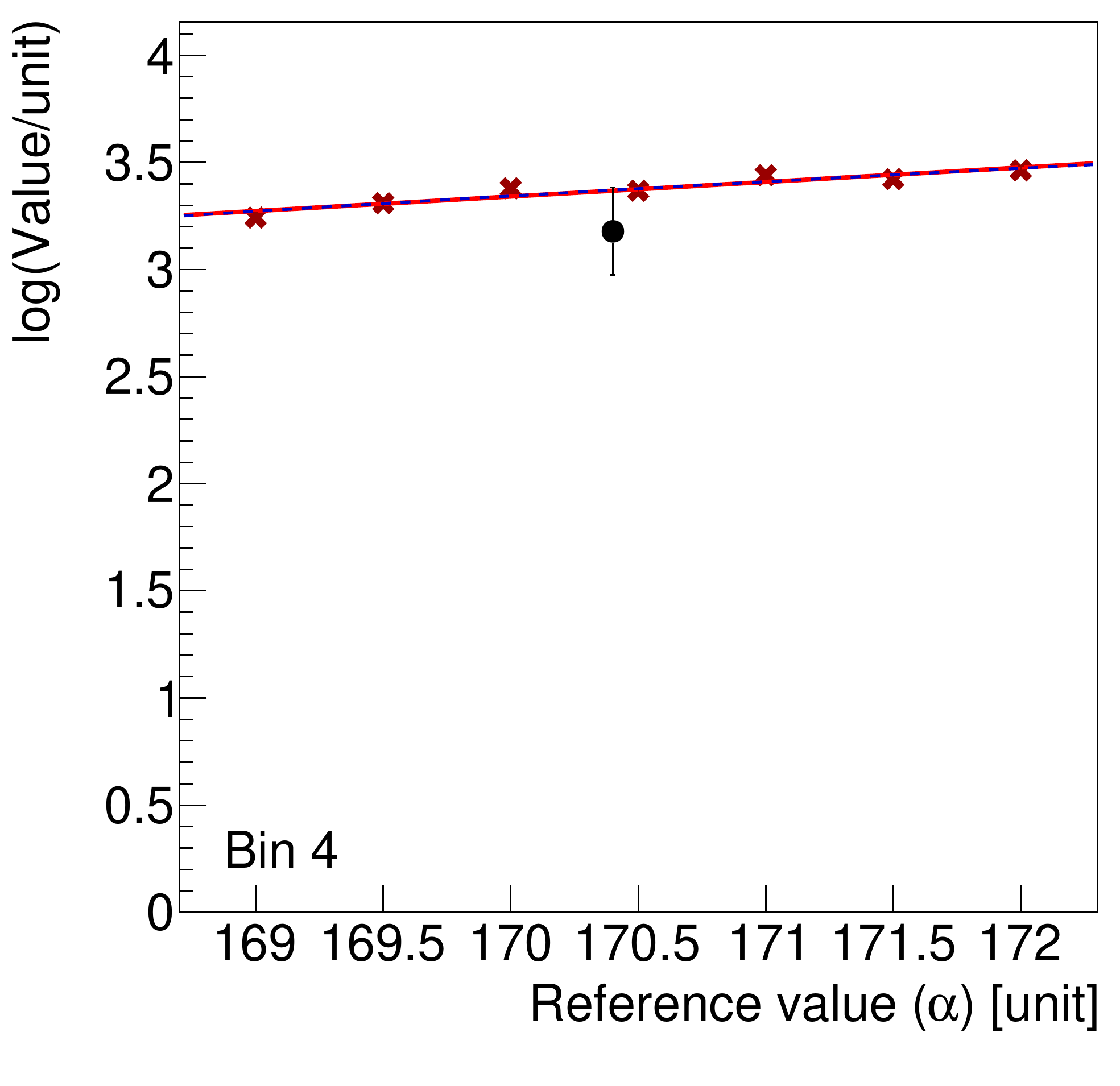}
       \includegraphics[width=0.24\textwidth,trim={0 0 10 0 },clip]{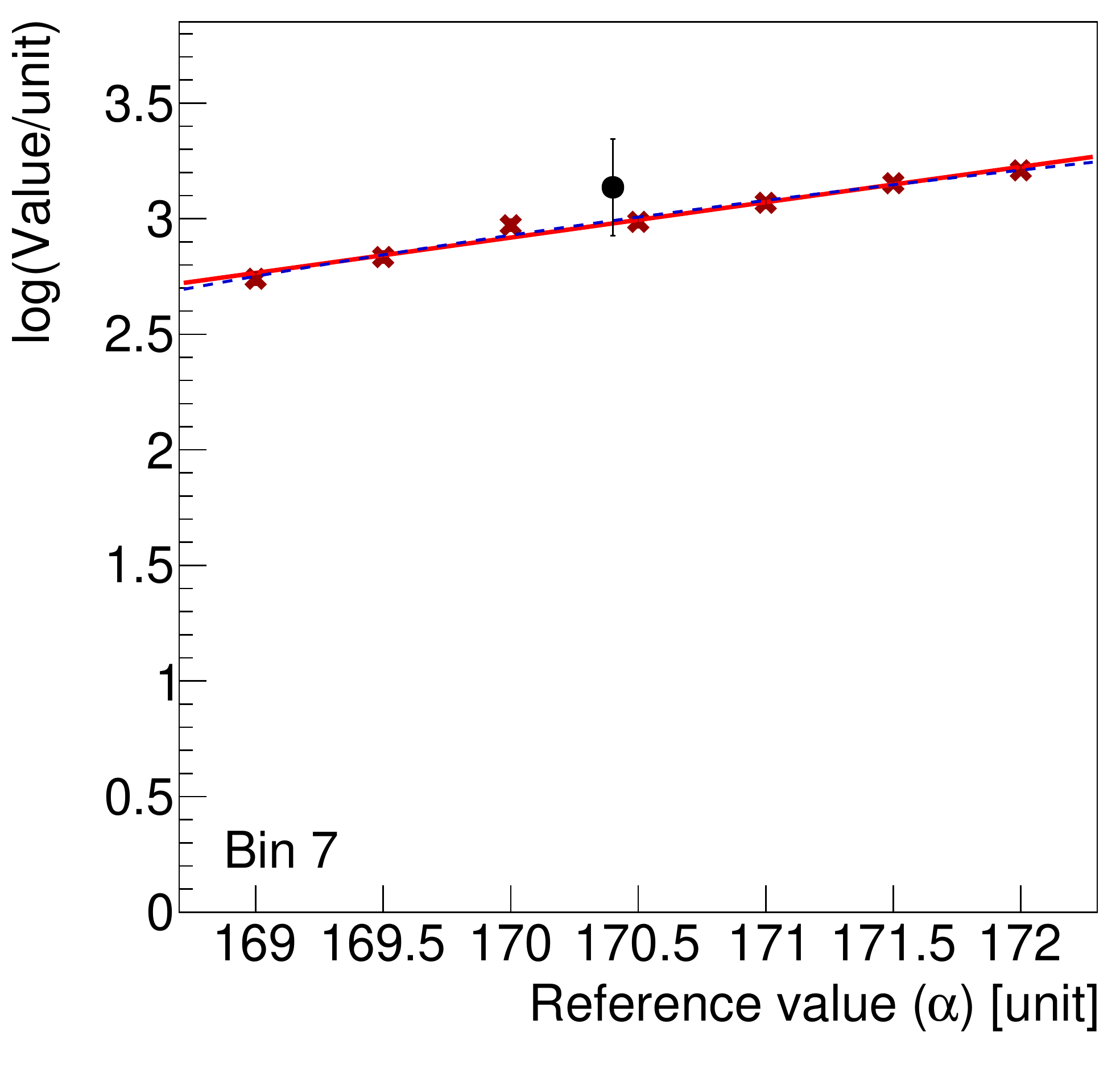}
       \includegraphics[width=0.24\textwidth,trim={0 0 10 0 },clip]{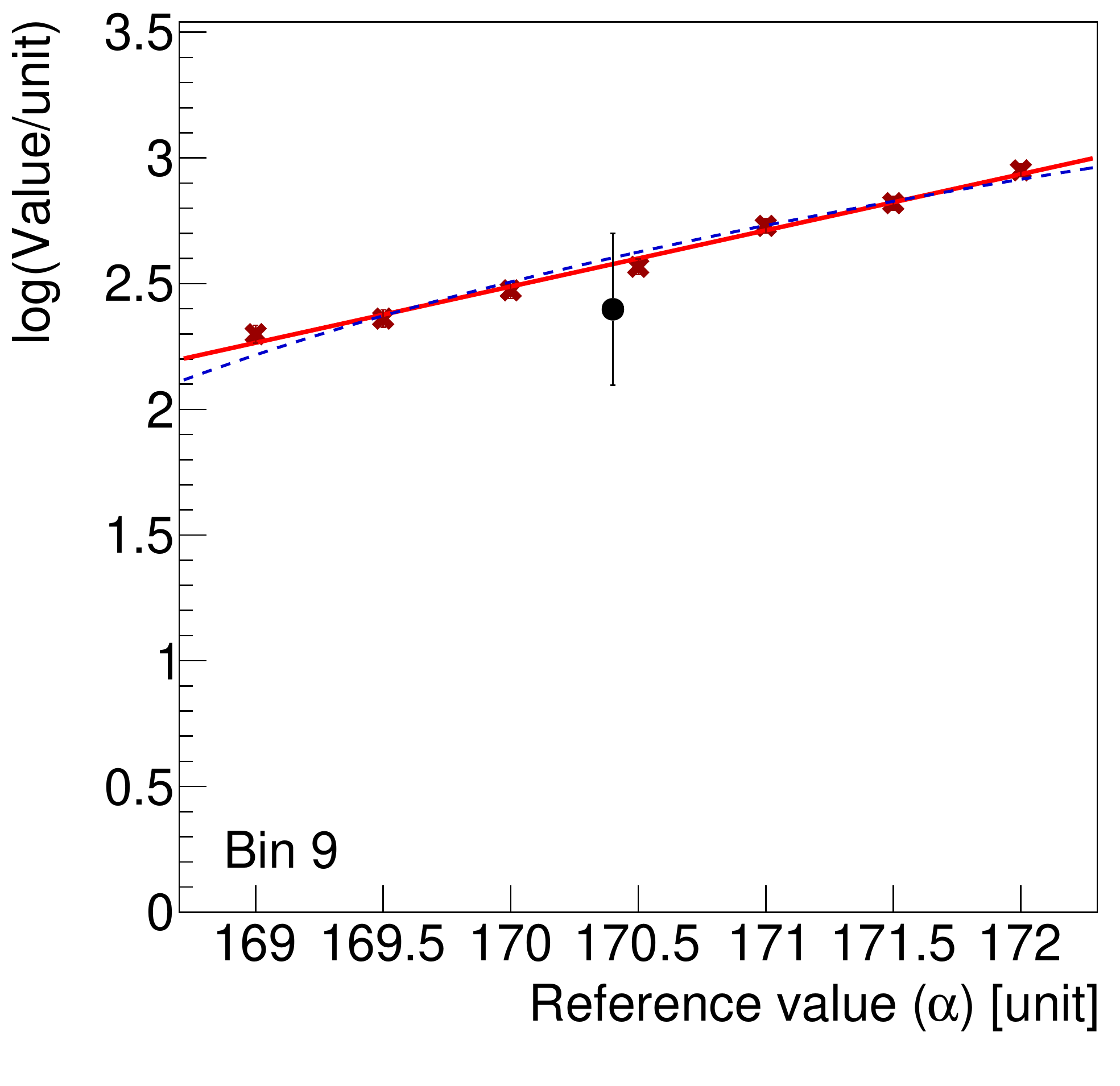}
\end{center}
\caption{
  The best estimate of the parameter dependence of the model,
  eq.~\eqref{eq:ymtlog} for four selected bins of Example 1, when
  performing a Linear Template Fit with log-normal distributed estimators. 
  Note that the $y$ axis indicates the logarithmized values of the
  templates.
  The pseudo-data are displayed for comparison at the best estimator
  $\hat\alpha$ of the Linear Template Fit. 
  For comparison, the linearized model from eq.~\eqref{eq:ymt} is also
  displayed, which is otherwise not used when log-normal-distributed 
  estimators are considered.
}
\label{fig:logmodel}
\end{figure}
As an illustration, the linearized model from eq.~\eqref{eq:ymtlog} is
displayed in figure~\ref{fig:logmodel} for some
selected bins of Example 1 (section~\ref{sec:example1}) and compared
with the approximations of eq.~\eqref{eq:ymt2}.
The two are reasonably similar in that example, since the templates are
present in the vicinity of the best estimator and in a range where a
linear approximation is valid. 
Although the parameter dependence in Example 1 is in fact truly linear,
from the generated statistical precision of the templates, the
templates cannot discriminate between these two best model estimators.

When using log-normal distributed variables, which corresponds to relative
uncertainties, the objective function \chisq\  of the Linear Template Fit becomes
\begin{equation}
  \chisq =
  \left( \bm{\log d} -{\textstyle\sum}_l\epsilon_l \mathbf{s}_{(l)} - \bm{\hat{\mu}}(\bm\alpha) \right)^{\text{T}}
  \mathbb{W}
  \left( \bm{\log d} -{\textstyle\sum}_l\epsilon_l \mathbf{s}_{(l)} - \bm{\hat{\mu}}(\bm\alpha) \right) + {\textstyle\sum}_l\epsilon_l^2\,.
  \label{eq:chi2log}
\end{equation}
The covariance matrix $\mathbb{V}=\mathbb{W}^{-1}$
comprises normal-distributed \emph{relative} uncertainties,
and also the uncertainties with bin-correlations $\mathbf{s}$ enter the
calculation through their relative values.
The notation $\bm{\log d}$ denotes a coefficient-wise application
of the logarithm.
A common relation between relative and absolute uncertainties would
be
\begin{equation}
  \mathrm{s}_{(l),i} \simeq \frac{s_{(l),i}}{d_{i}}\,,
  \label{eq:srel}
\end{equation}
and although this equation does not strictly translate between normal and
log-normal uncertainties, it is often used in practical applications.
Note that we use a Roman font s for relative uncertainties,  and 
italic $s$ for absolute uncertainties.

This \chisq\ function is again a linear least-squares problem, and the best
estimators for $\bm\alpha$ and $\bm\epsilon$ in the Linear Template Fit are 
\begin{equation}
  \ahut :=
  \left(\begin{matrix}\bm{\hat\alpha}\\\bm{\hat{\epsilon}}\\\end{matrix}\right)
    =
    \mathcal{F} \big( \bm{\log d} - \log(Y)\mb\big)\,,
    \label{eq:full2log}
\end{equation}
where the matrix $\mathcal{F}$  is defined in a similar way as
eq.~\eqref{eq:full0}, but from relative uncertainties in $\mathbb{V}$
and $\mathrm{s}$ ($S$) and employing the substitution $\Y\to\log(Y)\tilde{M}$.
Due to the treatment of systematic uncertainties as relative
uncertainties, the Linear Template Fit with the log-normal probability distribution functions may
 be preferred in practical applications over the Linear Template
Fit with absolute uncertainties.

\noindent
\fbox{
  \parbox[c]{0.97\textwidth}{
  \small
When repeating Example 1 (section~\ref{sec:example1}) with
relative uncertainties, the best estimator for the mean value is found
according to eq.~\eqref{eq:full2log} to be
$\hat\alpha=170.40\pm0.46$, which is in good agreement to the
generated value of 170.2. Some details are further displayed in
figure~\ref{fig:logmodel}.

\vspace{0.4em}
When considering Example 2 (section~\ref{sec:example2}) under the
assumption of log-normal distributed pseudo-data, the best estimators are
found to be $170.18\pm 0.47$ for the mean and
$6.33\pm0.39$ for the standard deviation, with a correlation coefficient
of $-0.3$.
When considering further a normalisation uncertainty of the size of
10\,\%, so a relative uncertainty, the
best estimators for the mean and variance become
$\alpha_0=170.20\pm0.76$ and $\alpha_1=6.33\pm 0.49$, respectively.
Their correlation coefficient is $-0.6$, and the nuisance parameter of
the normalisation uncertainty is $\hat\epsilon=-0.02\pm0.5$.
}
}

\clearpage
\section{Errors and error propagation}
\label{sec:Errors}
In many applications and fields, like particle physics, it
is of great importance to provide detailed uncertainties together with the fit parameters
and to propagate uncertainties from all relevant input quantities.
Due to its analytic and closed form, the Linear Template Fit provides unique opportunities
for detailed error propagation and analysis, that is otherwise not
easily possible in other inference algorithms.
For example, every single uncertainty component can be propagated
separately and analytically to the fit results.
In the following, the equations for error propagation are discussed, based
on the equation of the Linear Template Fit in eq.~\eqref{eq:full2}.

\subsection{Statistical uncertainties and uncertainties without bin-to-bin correlations}
\label{sec:errpropDatastat}
\vspace{-0.5em}
Statistical or epistemic uncertainties in the data are represented by
the covariance matrix $\mathbb{V}$.
The matrix $\mathbb{V}$ may include further uncertainty components
with or without bin-to-bin correlations, like systematic or aleatoric uncertainties.
%
%
From linear error propagation~\cite{Cowan:PDG:2020}, the covariance matrix is propagated to
the best estimators $\ahut$ as
\begin{equation}
  \hat{\mathbb{V}}_{\ahut}  = \mathcal{F} \mathbb{V} \mathcal{F}^{\text{T}}\,,
  \label{eq:errpropV}
\end{equation}
using $\mathcal{F}$ as defined in eq.~\eqref{eq:full0}.
Due to the Gau\ss--Markov theorem the estimators
$\ahut$ for the given objective function $\chisq$,
eqs.~\eqref{eq:chisqNuisance} and~\eqref{eq:chi2log},
have the least variance among all possible estimators.  

If $\mathbb{V}$ is a sum of
multiple individual matrices, like
$\mathbb{V} = \mathbb{V}_\textrm{stat.} +\sum\mathbb{V}_\textrm{uncorr.} + \sum\mathbb{V}_\textrm{corr.}+\ldots$,
each covariance matrix can be propagated separately using
equation~\eqref{eq:errpropV}. 
By doing so, the contribution from each uncertainty component to the full
uncertainty is determined separately.

\subsection{Systematic uncertainties or uncertainties with bin-to-bin correlations}
\label{sec:errpropDatassys}
\vspace{-0.5em}
Systematic uncertainties associated with the input data are often represented as
uncertainties with full bin-to-bin correlations and are represented either as systematic
shifts, $\bm{s}_{(l)}$ (eq.~\eqref{eq:S}), or as covariance matrices,
$\mathbb{V}_l^\textrm{corr} = \bm{s}_{(l)}^{ }\bm{s}_{(l)}^{\text{T}}$.
Using linear error propagation, these are propagated to the best estimators $\bm{\ahut}$ as
\begin{equation}
  \bm\sigma_{\ahut}^{(\bm{s}_{(l)})}=
  \mathcal{F}\bm{s}_{(l)}\,.
  \label{eq:errpropS}
\end{equation}
Alternatively, eq.~\eqref{eq:errpropV} can be employed.
For the Linear Template Fit with relative uncertainties, one literally
has the same equation but using relative values $\mathbf{s}_{(l)}$ (compare
eq.~\eqref{eq:srel}).

\noindent
\fbox{
  \parbox[c]{0.97\textwidth}{
  \small
In Example 2, where a normalisation uncertainty of 10\,\% was
considered 
the uncertainty breakdown for the two
parameters would be
$\sigma_{\hat\alpha_0}=\pm0.56_\text{(stat.)}\pm 0.51_\text{(norm.)}$
and
$\sigma_{\hat\alpha_1}=\pm0.42_\text{(stat.)}\pm -0.25_\text{(norm.)}$.
  }
}

When systematic uncertainties are included in the Linear Template Fit, their
corresponding nuisance parameters $\epsilon_l$ are constrained by data, and the
obtained reduction of the systematic uncertainty may be used for further
analyses.
In applications where the fitted parameters are dominated by
systematic uncertainties in the input data, it is good practice to
constrain the associated systematic shifts by adding additional
measurements to the fitting
problem~\cite{Abulencia:2005aj,D0:2007rwe,ATLAS:2012aj,CMS:2012sas}. This practice provides 
additional constraints on the related parameters, $\epsilon_l$,
and smaller experimental uncertainties can be obtained.
The uncertainty in the nuisance parameter becomes smaller than unity.

\subsection{External uncertainties}
The generalized transfer matrix $\mathcal{F}$ provides the straightforward opportunity
for linear error propagation; see eqs.~\eqref{eq:errpropV} 
or~\eqref{eq:errpropS}.
While these equations are predominantly employed to propagate the
uncertainties that are included in the Linear Template Fit,
equivalent equations are also applicable to propagate further
uncertainties that are not used in the fit, but which would still need to be
propagated to the fit result.
For example, these may be uncertainties in the underlying theory
prediction which cannot be constrained by the data.
An actual case for such an uncertainty is given below for
the non-perturbative correction factors (cf.\ Fig.~\ref{fig:asCMSErrors}).

%

\subsection{Unconstrained uncertainties with full bin-to-bin correlations}
\label{sec:unconstrained}
\vspace{-0.5em}
A special case of an uncertainty with full bin-to-bin correlations may be considered through an
\emph{unconstrained systematic uncertainty}.
These are defined by including the respective
systematic shift $s_{(l)}$ into the Linear Template Fit, but in contrast
to common \emph{constrained} systematic uncertainties, these have a
zero value (instead of unity) in the respective diagonal element of the
$l\times l$ matrix ${1}_l$ in eq.~\eqref{eq:full0}.
As a consequence, the Linear Template Fit can exploit the degree of freedom of that
systematic shift, but the data does not provide any constraint on that uncertainty source.
A possible application would be fits to data, where the normalization is 
unknown and is a free, unconstrained, systematic uncertainty.

\subsection{Uncertainties in the templates}
\label{sec:errpropMCstat}
\vspace{-0.5em}
The template distributions $Y$ may be associated with uncertainties from
various sources. These can be statistical uncertainties in the
elements $Y_{ij}$ or systematic uncertainties with
correlations among all entries of $Y$.
One may think about $Y$ being determined from two distinct models, or
templates that were generated with different parameters, and the resulting
difference in $Y$ may be  considered as a systematic uncertainty. 

The uncertainties in $\bm{\hat a}$ due to uncertainties in $Y$ are obtained from linear error
propagation.
The uncertainty in a single element of $Y$ is denoted as
$\sigma_{Y_{ij}}$, and using the shorthand notation
$\mathcal{D}$ from eq.~\eqref{eq:full0},
the partial derivative of eq.~\eqref{eq:full2} is
\begin{align}
  \frac{\partial\ahut}{\partial Y_{ij}}
  =
  \mathcal{D}^{-1}  \Bigg[
  &
      \begin{pmatrix}\mathbb{1}_{(ij)}\tilde{M}  & {0}_{il} \end{pmatrix}^{\text{T}} \W ( \dt  - Y \mb )
      -
      \begin{pmatrix}Y\tilde{M}  & S \end{pmatrix}^{\text{T}} \W \mathbb{1}_{(ij)} \mb\nonumber\\
      &-
      \left[ \begin{pmatrix} Y \tilde{M}  & S \end{pmatrix}^{\text{T}} \W \begin{pmatrix}\mathbb{1}_{(ij)}\tilde{M}  & {0}_{il}\end{pmatrix} +  \begin{pmatrix}\mathbb{1}_{(ij)}\tilde{M}  & {0}_{il} \end{pmatrix}^{\text{T}} \W \begin{pmatrix} Y \tilde{M}  & S \end{pmatrix} \right] \ahut
      \Bigg]\,,
\end{align}

where $\mathbb{1}_{(ij)}$ denotes an $i \times j$ zero matrix with only the
element $(i,j)$ being unity, 
and ${0}_{il}$ denotes an $i\times l$ zero-matrix.
It should be noted, that the usage of the matrices $\mathbb{1}_{(ij)}$ and ${0}_{il}$ in
numerical calculations is discouraged since a naive matrix
multiplication becomes numerically inefficient, but that it
remains instructive to write the equations in that extensive format.
The resulting size of the uncertainty of the best estimators due to
$\sigma_{Y_{ij}}$ becomes
\begin{equation}
  \bm\sigma^{(Y_{ij})}_{\ahut} =
  \frac{\partial\ahut}{\partial Y_{ij}}
  \sigma_{Y_{ij}}\,.
  \label{eq:errpropYij}
\end{equation}

Hence, uncertainties without entry-to-entry correlations, such as
statistical uncertainties, are propagated to the estimated parameters as
\begin{equation}
  \bm\sigma^{(Y_\text{uncorr.})}_{\ahut} =
  \sqrt{ \sum_{i,j} \bigg( \frac{\partial\ahut}{\partial Y_{ij}} \sigma_{Y_{ij}} \bigg)^2}\,,
  \label{eq:errpropYuncorr}
\end{equation}
where the square root and the exponentiation are performed
coefficient-wise,
while uncertainties that have full entry-to-entry correlations are propagated as
\begin{equation}
  \bm\sigma^{(Y_\text{corr.})}_{\ahut} =
  \sum_{i,j}   \frac{\partial\ahut}{\partial Y_{ij}}\sigma_{Y_{ij}} \,.
  \label{eq:errpropYcorr}
\end{equation}

\noindent
\fbox{
  \parbox[c]{0.97\textwidth}{
  \small
In Example 1 and 2 (sections~\ref{sec:example1}
and~\ref{sec:example2}), the templates are each generated with 40\,000
\emph{events} and independently from each other.
Hence, all their entries have statistical uncertainties.
Using eq.~\eqref{eq:errpropYuncorr}, the statistical uncertainties
from $Y$ under gaussian approximation are $\pm0.025$ in Example 1, and
$\pm0.042$ and $\pm0.060$ for the two parameters in Example 2.
  }
}

The equations for error propagation of the Linear Template Fit with relative
uncertainties (section~\ref{sec:LogN}) are altogether similar to
  the presented equations and do not need to be repeated.
However, since relative uncertainties are considered, the partial
derivatives for error propagation in eqs.~\eqref{eq:errpropYuncorr}
and~\eqref{eq:errpropYcorr} become
\begin{equation}
  \frac{
    \tiny\partial\ahut}{\partial Y_{tj}}
  \to
  \frac{
    \tiny\partial\ahut}{\partial \log{Y_{tj}}}\,.
\end{equation}

\noindent
\fbox{
  \parbox[c]{0.97\textwidth}{
  \small
Example 2 with log-normal uncertainties would report statistical
uncertainties from $Y$ of $\pm0.052$
and $\pm0.066$ for the two fit parameters.
  }
}

Note that the uncertainties in $Y$ are not considered within the fit,
but propagated separately, since it is assumed that these are
negligibly small, and independent of the model parameter
$\bm{\alpha}$, and cannot be constrained by the data.

\subsection{Full uncertainty}
\vspace{-0.5em}
The covariance matrix for the best estimators, including all
uncertainty components that are considered in the Linear Template Fit,
are calculated from eqs.~\eqref{eq:errpropV} 
and~\eqref{eq:errpropS}:
\begin{equation}
  V_{\ahut}  =
  \mathbb{V}_{\ahut}
  + {\textstyle\sum}_l \bm\sigma_{\ahut}^{(\bm{s}_{(l)})}
  (\bm\sigma_{\ahut}^{(\bm{s}_{(l)})})^{\text{T}}
  \label{eq:errorfull}
\end{equation}
Further uncertainties, which are not explicitly considered in the fit,
can be propagated to the result by using eqs.~\eqref{eq:errpropV}
and~\eqref{eq:errpropS} or also
eqs.~\eqref{eq:errpropYuncorr} and~\eqref{eq:errpropYcorr}.

\subsection{Error (re-)scaling}
\label{sec:errorrescaling}
\vspace{-0.5em}
When multiple data points are considered in the fit and they are
correlated through some (systematic) uncertainties, the best-fit estimators may become
biased~\cite{Lyons:1989gh,Lincoln:1993yi,DAgostini:1993arp,Takeuchi:1995xe,D'Agostini:642515,Ball:2009qv}.
This is mainly because uncertainties with bin-to-bin correlations are commonly valid 
as relative uncertainties, or so-called multiplicative uncertainties,
such as normalization uncertainties, but those are included
in the fit with absolute values 
(i.e., as variances). Since the data are subject to random noise from
statistical fluctuations, or simply because of the different
size of the values, for example when considering different data sets,
the result may become biased since smaller values are effectively
preferred by the fit.
While such a bias is largely reduced or even absent from first principles when
working with relative uncertainties, it  may still be useful 
to discuss possible solutions to this problem and several solutions
are discussed in the literature (see e.g.\ Ref.~\cite{Ball:2009qv}).

A common solution is to consider (multiplicative)
uncertainties with bin-to-bin correlations through their relative values in 
the \chisq\ computation, and their relative values are
multiplied coefficient-wise with the prediction(s) rather than the
data values~\cite{D'Agostini:642515}.
However, since the minimization of \chisq\ is then no longer
linear in the fit parameters, there is no closed analytic solution
for the best estimators.
However, an approximately equivalent result is obtained when
calculating the covariance matrices and systematic shifts
(cf.\ eq.~\eqref{eq:Vsum}) from 
theoretical predictions~\cite{DAgostini:1993arp}, and
the coefficients of the shifts are recalculated according to
\begin{equation}
  s_{(l),i} \to \frac{s_{(l),i} y_i}{d_i}\,= \mathrm{s}_{(l),i} y_i,
\end{equation}
and analogously for covariance matrices.
Various options for the predictions $\bm{y}$ for this \emph{error
  rescaling} may be considered, and the prediction $\bm{y}$ could be chosen
to be
\begin{compactitem}
\item One of the template distributions, $\bm{y}_{(j)}$;
\item The prediction of the linearized model with ad-hoc selected
  parameters $\bm{\alpha}$, $\mathbf{\hat{y}}(\bm{\alpha})$;
\item The best estimator of the linear model, $\mathbf{\hat{y}}(\ahut)$; or
\item The Linear Template Fit may be iteratively repeated with the
  best estimator, $\mathbf{\hat{y}}(\ahut)$.
\end{compactitem}
The rather compact equations of the Linear Template Fit are well
suited for an iterative algorithm.
Such procedures are equivalent to an alternative formulation of the
\chisq\ quantity,
but the convergence of such iterative algorithms should be critically assessed.

\section{Linear Template Fit with a detector response matrix}
\label{sec:A}
Many measurements need to be corrected for detector effects like
resolution or acceptance in order to compare the measurement
with theoretical predictions. This is referred to as
\emph{unfolding}  (for reviews see
e.g.~\cite{Blobel:1984ku,Behnke:1517556,Cowan:2002in,Blobel:2011fih,Spano:2013nca,Schmitt:2016orm,Brenner:2019lmf})
and is commonly performed by first simulating the detector response and
representing it in terms of a \emph{response matrix} $A$,
\begin{equation}
  \bm{y} = A\bm{x}\,,
\end{equation}
where $\bm{y}$ denotes the measured detector-level distribution
and $\bm{x}$ the ``true'' underlying distribution.
However, it is not straightforward to apply the inverse response
matrix to the distribution $\bm{y}$, since the unfolding problem represents an ill-posed
inverse problem, or the matrix $A$ needs to be a square matrix, and thus
 more sophisticated unfolding algorithms need to be employed to
 determine the ``true'' distribution
 $\bm{x}$~\cite{Blobel:run,DAgostini:1994fjx,Hocker:1995kb,Malaescu:2009dm,Schmitt:2012kp,Andreassen:2019cjw}. 
When performing a parameter determination, however, it is
equivalent whether using the unfolded
distribution $x$ together with templates of the predictions, or alternatively using the
detector-level measurement $y$ and applying the response matrix
to the templates instead.
These two procedures are supposed to be equivalent if the unfolding
problem and algorithm are unbiased, and are referred to
as \emph{folding}, \emph{up-folding}, or \emph{forward folding}.
Because the simulation of the detector effects and the determination of
the response matrix $A$ may be computationally expensive, as is
 the case for complex particle physics experiments, it may
be reasonable to determine the matrix $A$ only once and with
high precision, and then apply it to the templates.

The inclusion of the detector effects in the Linear Template Fit is
straightforward and is realized by the simple substitution
\begin{align}
  Y &\rightarrow  AX
  \label{eq:AX}
\end{align}
in eqs.~\eqref{eq:full2} and~\eqref{eq:full0}, 
where  $X$ represents 
the template matrix without detector effects at the ``truth'' or
``particle'' level.
When using relative uncertainties, the substitution becomes
\begin{equation}
  \log(Y) \to \log(AX)\,,
\end{equation}
and the full expression of the Linear Template Fit becomes
\begin{equation}
  \begin{small}
  \begin{pmatrix}\bm{\hat\alpha}\\\bm{\hat\epsilon}\end{pmatrix}
  \end{small}
  = \mathcal{F}
  \big( \bm{\log d} - \log(AX)\mb\big)
  ~~~\text{using} \nonumber\\
\end{equation}
\begin{equation}
\mathcal{F}=
  \left(
  \begin{pmatrix}\log(AX)\M & ~S\end{pmatrix}^{\text{T}}
    \W
    \begin{pmatrix}\log(AX)\M & ~S\end{pmatrix}
      + \begin{pmatrix} {0}_k & 0 \\ 0 & {1}_l\end{pmatrix}
        \right)^{-1}
        \begin{pmatrix}\log(AX)\M & ~S\end{pmatrix}^{\text{T}} \W\,.
\end{equation}

Since $Y$ and $AX$ are $i\times j$ matrices, the index
$t$ for the number of entries (``bins'') on the truth level was introduced.
Thus, $A$ is an $i\times t$ matrix and $X$ has dimension $t\times j$, and
in order to obtain a higher resolution, it may be reasonable to
consider a non-square response matrix $A$, i.e., $i\neq t$.
It is important that the detector response $A$ is required to be
independent from the physics parameter of interest.  

The response matrix $A$ is typically associated with uncertainties, and
these may be entry-to-entry correlations (like systematic or ``model''
uncertainties) or without such correlations (like statistical
uncertainties). Note, that the response matrix is commonly normalized
and therefore the statistical uncertainties are without entry-to-entry
correlations only approximately.

Uncertainties of the best estimators that arise from uncertainties
in $A$ can be obtained from linear error propagation.
After applying eq.~\eqref{eq:AX}, the partial derivative of 
eq.~\eqref{eq:full2} is 
\begin{align}
  \frac{\partial\ahut}{\partial A_{it}}
  =
  \mathcal{D}^{-1}  \Bigg[
  &
      \begin{pmatrix}\mathbb{1}_{(it)}X\tilde{M}  & {0}_{il} \end{pmatrix}^{\text{T}} \W ( \dt  - AX \mb )
      -
      \begin{pmatrix}AX\tilde{M}  & S \end{pmatrix}^{\text{T}} \W  \mathbb{1}_{(it)} X \mb \nonumber\\
      &-
      \left[
        \begin{pmatrix} AX \tilde{M}  & S \end{pmatrix}^{\text{T}} \W \begin{pmatrix}\mathbb{1}_{(it)}X\tilde{M}  & {0}_{il} \end{pmatrix}
        +
        \begin{pmatrix}\mathbb{1}_{(it)}X\tilde{M}  & {0}_{il} \end{pmatrix}^{\text{T}} \W \begin{pmatrix} AX \tilde{M} & S \end{pmatrix}
        \right]
      \ahut
      \Bigg]\,,
\end{align}
Consequently, uncertainties in the
response matrix $A$ without or with entry-to-entry
correlations are propagated to the fitted results as
\begin{equation}
  \bm\sigma^{(A_\text{uncorr.})}_{\ahut} =
  \sqrt{ \sum_{i,t} \left( \frac{\partial\ahut}{\partial A_{it}} \sigma_{A_{it}} \right)^2}
  ~~~\text{and}~~~
  \bm\sigma^{(A_\text{corr.})}_{\ahut} =
  \sum_{i,t} \frac{\partial\ahut}{\partial A_{it}} \sigma_{A_{it}} \,,
  \label{eq:errpropA}
\end{equation}
respectively,
where $\sigma_{A_{it}}$ denotes the size of the uncertainty in the element $A_{it}$,
and the power and square-root operations are applied coefficient-wise.
When using a response matrix, the uncertainties in the template matrix $X$
(cf.\ section~\ref{sec:errpropMCstat})
are propagated using
\begin{align}
  \frac{\tiny\partial\ahut  }{\partial X_{tj}}
  =
  \mathcal{D}^{-1}  \Bigg[
    &
      \begin{pmatrix}A\mathbb{1}_{(tj)}\tilde{M}  & {0}_{il} \end{pmatrix}^{\text{T}} \W \left(\dt  - AX \mb \right)
      -
      \begin{pmatrix}AX\tilde{M}  & S \end{pmatrix}^{\text{T}} \W A\mathbb{1}_{(tj)} \mb \nonumber\\
      &-
      \left[
        \begin{pmatrix} AX \tilde{M}  & S \end{pmatrix}^{\text{T}} \W \begin{pmatrix}A\mathbb{1}_{(tj)}\tilde{M}  & {0}_{il} \end{pmatrix} 
        +
        \begin{pmatrix}A\mathbb{1}_{(tj)}\tilde{M}  & {0}_{il} \end{pmatrix}^{\text{T}} \W \begin{pmatrix} AX \tilde{M}  & S \end{pmatrix}
        \right]
      \ahut
      \Bigg]\,.
  \label{eq:partialX}
\end{align}

Note that for the Linear Template Fit with relative uncertainties, it is
reasonable to consider uncertainties in $A_{it}$ as 
absolute uncertainties, rather than relative,
because the elements $A_{it}$ represent already relative quantities
and may be interpreted as probabilities.
Therefore, one employs the partial derivative
$\partial\ahut/\partial A_{ij}$ for error propagation, instead of
$\partial\ahut/\partial \log{A_{ij}}$, although elsewhere relative
uncertainties are considered.
This partial derivative is straightforward to derive.
The propagation of uncertainties in the templates $X_{tj}$, however, becomes
somewhat more complicated when using a response matrix $A$, since one
has to calculate the partial derivative 
$\partial\ahut/\partial\log{X_{tj}}$ for error propagation of the
relative uncertainties $\sigma_{X_{tj}}$.\footnote{In brief, the
  partial derivative $\partial\ahut/\partial\log{X_{tj}}$ is similar
  to eq.~\eqref{eq:partialX}, but each element of the $i\times j$ matrix
$(A\mathbb{1}_{(tj)})$ becomes $(A\mathbb{1}_{(tj)})_{ij}=(A\mathbb{1}_{(tj)}X_{tj})_{ij}/(AX)_{ij}$.}

\section{Considerations, validations and cross checks}
\label{sec:considerations}
As in any optimization problem, the result of the Linear Template
Fit needs to be validated and cross checked.
Due to its close similarity to the iterative Gau\ss--Newton
minimization algorithm of nonlinear least squares~\cite{Blobel:1984cy,Gill:99800,madsen,Fletcher:1568710,Nocedal:1638144}, the Linear Template Fit is expected to
result in an unbiased and optimal result, in particular if
the templates provide a sufficiently accurate approximation of the
model and its Jacobi matrix at the best estimator.
In fact, when assuming that the model  $\bm{\lambda}(\bm{\alpha})$
could be used with continuous values of $\bm\alpha$ for inference, 
the best estimator from the Linear Template Fit becomes equivalent to the best
estimator that would be obtained when using
$\bm{\lambda}(\bm{\alpha})$ directly, if two approximations are exactly
fulfilled,\footnote{One would have to further require that there is no second
  minima when using $\bm{\lambda}(\bm{\alpha})$ within the range covered by the reference values.}
\begin{align}
  \bm{\lambda}(\bm{\hat\alpha})
  &\approx
  \mathbf{\hat{y}}(\bm{\hat\alpha}) = Y\mathbf{\bar{m}} + Y\tilde{M}\bm{\hat\alpha}
  ~~~~~\text{and}~ \\
  \left.\tfrac{\partial\bm{\lambda}(\bm{\alpha})}{\partial\bm{\alpha}}\right|_{\bm{\alpha}=\bm{\hat\alpha}}
  &\approx
  \left.\tfrac{\partial\mathbf{\hat{y}}(\bm{\alpha})}{\partial\bm{\alpha}}\right|_{\bm{\alpha}=\bm{\hat\alpha}}
  =Y\tilde{M}\,.
\end{align}
The first equation requires that the model is correctly represented by the
interpolated templates at the best estimator, which is needed for the value of the best estimator,
and the second equation demands that the first derivatives are correctly
approximated, which is a requirement for the error propagation and minimization.
These equations taken together represent just a linear approximation
of $\bm{\lambda}(\bm{\alpha})$ at $\bm{\hat\alpha}$, but using the
templates only.
However, since for the motivation of the  Linear Template Fit it was assumed that
$\bm{\lambda}(\bm{\alpha})$ cannot be used for inference itself,
these two approximations also cannot be validated directly.
In the following, some possible studies for the validation of the results and the
practical application of the Linear Template Fit are discussed.

\subsection{Template range}
\label{sec:templrange}
\vspace{-0.5em}
If the model is nonlinear in the parameters $\bm{\alpha}$, an unbiased result may
only be obtained if the best estimator $\bm{\hat{\alpha}}$ lies within the
interval of the reference values,
\begin{equation}
  \bm{\hat{\alpha}} \in [\bm{\dot\alpha}_{(1)},\bm{\dot\alpha}_{(j_\text{max})}]\,.
  \label{eq:validation1}
\end{equation}
This is because the templates provide the approximation of the model
and its first derivatives, but an extrapolation beyond the template
range is inappropriate for a non-linear model.
If eq.~\eqref{eq:validation1} does not hold,
additional templates at further reference points should be generated
and the Linear Template Fit should be repeated with an improved
selection of reference values.
For similar reasons, the templates should be in the close vicinity of the
best estimator, such that the linear approximations are justified.
As a rule of thumb, the distance between the reference points
should not greatly exceed the size of the variance of the best estimator
\begin{equation}
  \bm{\dot\alpha}_{(j)} - \bm{\dot\alpha}_{(j+1)} \lesssim \bm{\sigma}_{\bm{\hat\alpha}}\,.
  \label{eq:validation2}
\end{equation}
For similar reasons, the best-estimator should ideally be found in the
center of the template range, rather than at its boundary,
\begin{equation}
  \bm{\ahut}-\bm{\dot\alpha}_{(0)} \approx \bm{\dot\alpha}_{(j_\text{max})}-\bm{\ahut}\,.
  \label{eq:validation4}
\end{equation}
However, for rather linear parameter dependencies these requirements
may be more relaxed. 

The choice of the reference values may be related to the size of the
uncertainties of the best estimator.
If one considers the uncertainties of the best estimator to be
normal distributed,\footnote{A reasonable argument for a
  normal-distributed parameter can also be obtained from previous
  analyses, e.g.\ if these published normal-distributed
  uncertainties, or if the results are input to some combination
  procedure which considers uncertainties to be normal
  distributed~\cite{aitken_1936,Lyons:1988rp,Cowan:1998ji,Valassi:2003mu,Nisius:2014wua}.}
as a consequence, the model also needs to exhibit a sufficiently
linear behavior at least within a few $\sigma$.
As a rule of thumb, one could
demand that the reference values are approximately within $3$ standard deviations:
\begin{equation}
  \bm{\dot\alpha}_{(j)} \in [\bm{\hat\alpha}-3\bm{\sigma}_{\bm{\hat\alpha}},
    \bm{\hat\alpha}+3\bm{\sigma}_{\bm{\hat\alpha}}]\,.
  \label{eq:validation3}
\end{equation}

\subsection{\boldmath The value of \hchisq, its uncertainty and the partial \chisq}
\vspace{-0.5em}
The value of \chisq\ at the minimum, \hchisq, is a meaningful quantity to assess the
agreement between the data and the model.
Since the calculation of \chisq\ is not explicit in the Linear Template
Fit, its value needs to be calculated separately, and \hchisq\ is calculated
from data and the templates at $\bm{\hat{a}}$ using the somewhat simplified equation
\begin{equation}
  \hchisq
  = \left( \dt - Y\mb - \Y\ahut \right)^{\text{T}} W \left(\dt-Y\mb\right)\,.
  \label{eq:chihat}
\end{equation}
Since the data and the templates are associated with uncertainties,
the value of \hchisq\ consequently may also be considered as uncertain.
The uncertainty in \hchisq\ is obtained from linear error propagation
of the input uncertainties and calculated as
\begin{equation}
  \sigma_{\hchisq} = \sqrt{\bm\xi^{\text{T}} \mathbb{V}\bm\xi + \sum_l(\bm\xi \bm{s}_{(l)}^{\text{T}})^2}\,,
\end{equation}
where the $i$-vector $\bm\xi$ is introduced and whose coefficients represent the partial
derivatives $\tfrac{\partial\hchisq}{\partial d_i}$, which are calculated
as $\xi_i = 2W_i (\dt - \mathbf{\hat{y}}(\ahut))$, and $W_i$ denotes
the $i$th row of the matrix $W$.
Also, the uncertainties in the templates $Y$ may be propagated to
\hchisq\ with similar equations.
The interpretation of $\sigma_{\hchisq}$ is not trivial, but it
may well be that this quantity has interesting features and may serve
as an additional quality parameter in the future.

The value of \chisq\ is inherently obtained from a sum of various
sources of uncertainties.
It is instructive to calculate the
contribution from any uncertainty source $\ell$ to \hchisq\ separately using~\cite{Johnson:2017ttl}
\begin{equation}
  \hchisq_\ell = (\dt-\mathbf{\hat{y}}(\ahut))^{\text{T}} W V_\ell W (\dt-\mathbf{\hat{y}}(\ahut))\,,
\end{equation}
where $V_\ell$ is any of the covariance matrices,
see eq.~\eqref{eq:Vsum}.
It is obvious that $\hchisq=\sum_\ell\hchisq_\ell$.
This may become particularly useful when multiple measurements are considered in the fit,
and thus their individual contributions to \hchisq, i.e.\ the partial-\chisq's,
may be calculated by summing the respective $\hchisq_\ell$ values.

\subsection{\boldmath Interpretation of the $\chisq_j$ values as a
  cross check}
\label{sec:parabfit}
\vspace{-0.5em}
An obviously simple validation of the result from the Linear Template Fit is
performed by comparing $\hat\chi^2$ 
with the \chisq\ values calculated from the individual templates,
defined as
\begin{equation}
  \chisq_j = (\dt - \bm{y}_{(j)})^{\text{T}} W (\dt -\bm{y}_{(j)})~~~\text{or}~~~
  \chisq_j = (\bm{\log\dt} - \bm{\log{\bm{y}}}_{(j)})^{\text{T}} W (\bm{\log{\dt}} -\bm{\log{\bm{y}}}_{(j)})\,,
  \label{eq:chi2j}
\end{equation}
whether
absolute or relative normal distributed uncertainties are considered.

\begin{figure}[tb!]
\centering
   \includegraphics[width=0.44\textwidth,trim={0 0 10 0 },clip]{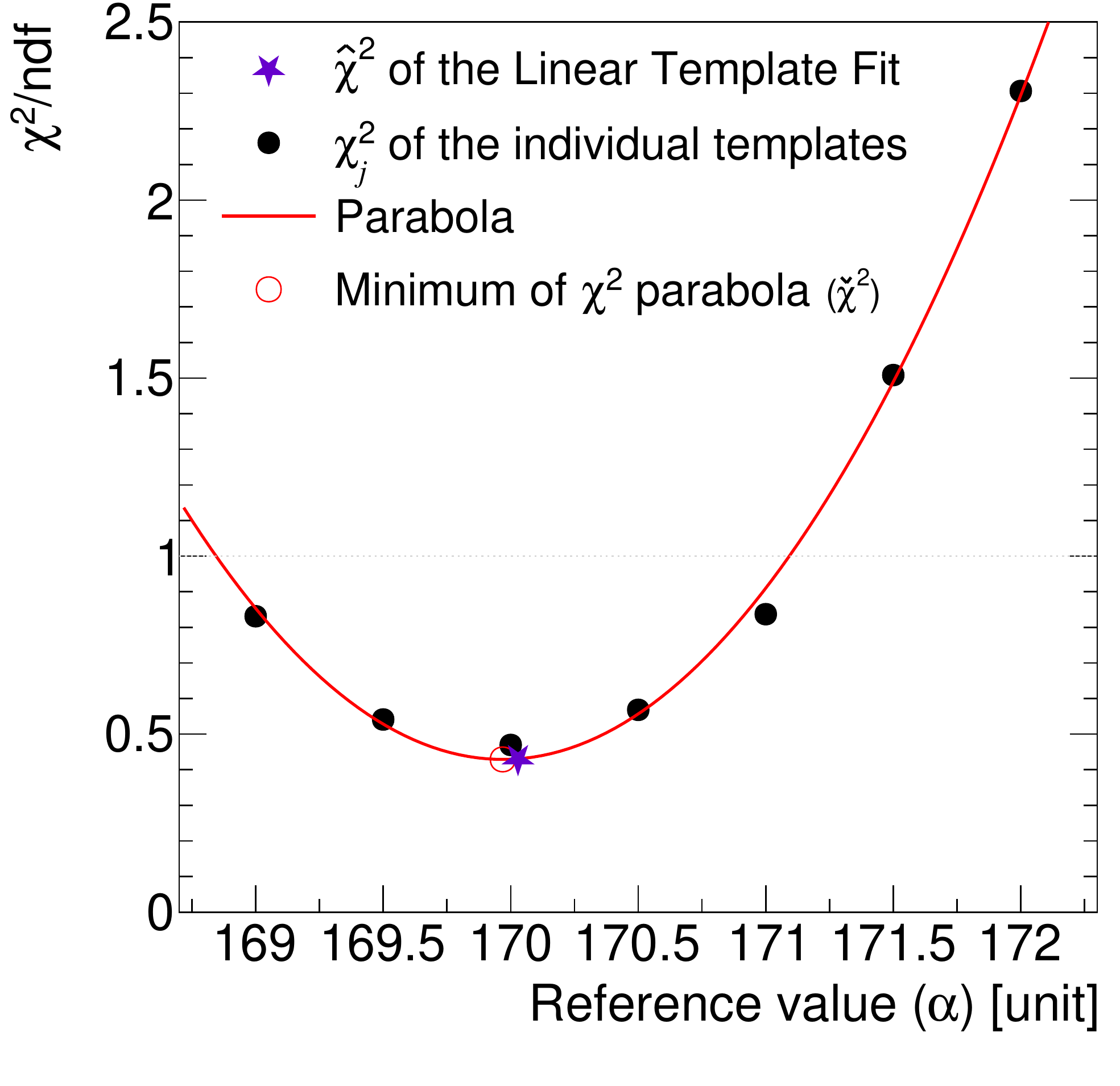}
\caption{
  Illustration of various $\chisq/n_\text{df}$ values in the scope of the Linear
  Template Fit for Example 1 (section~\ref{sec:example1}).
  The red star indicates the best-fit \chisq\ of the Linear Template
  Fit, \hchisq.
  The full circles display the \chisq\ values of the individual
  templates and the red line displays the \chisq-parabola calculated
  from them, and its minimum is inicated with an open circle
  ($\check\chi^2$, see text).
}
\label{fig:chisq}
\end{figure}

An alternative best estimator, denoted in the following as
$\bm{\check\alpha}$, is obtained from the interpretation of
all of the $\chisq_j$ values, and serves as a validation of the 
best estimator from the Linear Template Fit ($\bm{\hat\alpha}$).
Based on the assumption that the $\chisq$ distribution
$\chisq(\bm\alpha)$ follows a second-degree polynomial, the so-called
\chisq-parabola, its minimum is interpreted as best estimator.
For a univariate problem, when introducing
\begin{equation}
  \bm{\chisq} = \begin{pmatrix} \chisq_1 \\ \vdots \\ \chisq_j \end{pmatrix}
  \,,~~~
  C_\textrm{(pol-2)}=
  \begin{pmatrix}
    1 & \dot{\alpha}_1 & \dot{\alpha}_1^2 \\
    \vdots & \vdots& \vdots \\
    1 & \dot{\alpha}_j & \dot{\alpha}_j^2 \\
  \end{pmatrix}
  ~~~\text{and}~~~
  \bm{\vartheta} = \begin{pmatrix} \vartheta_0 \\ \vartheta_1 \\ \vartheta_2 \end{pmatrix}\,,
  \label{eq:pol2fit}
\end{equation}
with $\chisq_j$ from eq.~\eqref{eq:chi2j},
the best estimator of the model parameter is given by the stationary
point of the polynomial
\begin{equation}
  \check{\alpha} = \frac{\hat\vartheta_1}{-2\hat\vartheta_2}\,,
  \label{eq:ahut}
\end{equation}
where the best estimators of the (three) polynomial parameters are
$\bm{\hat\vartheta} = (C^{\text{T}}C)^{-1}C\bm{\chisq}$.
An illustration for Example 1 is shown in figure~\ref{fig:chisq}.
The uncertainty in $\check\alpha$ is given by the criterion $\chisq_\text{min}+1$~\cite{Behnke:1517556} and hence
\begin{equation}
  \sigma_{\check\alpha} = \frac{\sqrt{\hat\vartheta_1^2 - 4\hat\vartheta_2(\hat\vartheta_0 - \check\chi^2 - 1)}}{-2\hat\vartheta_2}
  ~~~\text{using}~~~
  \check\chi^2 = (\bm\chisq - C\bm{\hat\vartheta})^{\text{T}}(\bm\chisq- C\bm{\hat\vartheta})\,.
\end{equation}
The equations for the multivariate case are straightforward and not
repeated here.
This alternative best estimator, $\check\alpha$, as well as its 
variance $\sigma_{\check\alpha}$ and the minimum \chisq, should be sufficiently consistent
with the results from the Linear Template Fit:
\begin{equation}
  \check\alpha\sim\hat\alpha\,,
  ~~~
  \sigma_{\check\alpha}\sim\sigma_{\hat\alpha}
  ~~~\text{and}~~~
  \check\chi^2\sim\hat\chi^2\,,
  \label{eq:validationchi2}
\end{equation}
and may serve as a valuable cross check.
As a quick and practical visual test, if the templates line up on the
parabolic fit (as seen in figure~\ref{fig:chisq}),
the problem is essentially linear and the Linear Template Fit will
provide an unbiased and optimal best estimator.
Note, that for the calculation of eq.~\eqref{eq:ahut} at least three templates are
required, whereas the minimum number of templates for the univariate Linear Template Fit is
only two.

\subsection{Power law factor for the linear regression}
\label{sec:gamma}
\vspace{-0.5em}
In the Linear Template Fit the functions $\mathbf{\hat{y}}(\bm\alpha)$
are (multi-dimensional) linear functions, see eqs.~\eqref{eq:ymt},
\eqref{eq:ymt2}, and~\eqref{eq:ymtlog}.
Although the use of higher-degree polynomials or multivariate polynomials with
interference terms could in principle be considered (see next
section~\ref{sec:NonLinearModel}), in such cases the optimization
problem cannot be solved analytically and no closed matrix expression for the
best estimators is found. 
However, an analytic solution is still obtained
when only power law factors are introduced, and
in a univariate Linear Template Fit
the model approximation becomes
\begin{equation}
  \mathrm{y}_i(\alpha;\theta_0,\theta_1) = \theta^{(i)}_0 + \theta^{(i)}_1 \alpha^\gamma
  \label{eq:gammafactor}
\end{equation}
with $\gamma>0$ being a real number.
The matrix $M$, eq.~\eqref{eq:M}, then has elements $M_{j,0}=1$ and
$M_{j,1}=\dot\alpha_j^{\gamma}$.
In the multivariate case, different power law factors
for  each fit parameter can be introduced, and 
one writes $\gamma_k$ and has $M_{j,(k+1)}=\dot\alpha_{(j),k}^{\gamma_k}$.
The best estimators of the multivariate Linear Template Fit, 
eq.~\eqref{eq:full2}, then become
\begin{equation}
  \hat\alpha_k= \left( \mathcal{F}( \dt-Y\mb) \right)_k^{\frac{1}{\gamma_k}}\,.
  \label{eq:gammaresult}
\end{equation}
This solution can immediately be seen by a variable substitution of
$\alpha^\gamma$ in eqs.~\eqref{eq:full2} and~\eqref{eq:gammafactor}.
The application of the power law factor(s) $\gamma$ when working
with relative uncertainties (see section~\ref{sec:LogN}) is straightforward,  
since all equations remain linear in $\alpha_k$.
Similarly, the equations for the error propagation (see
section~\ref{sec:Errors}) can easily be adopted with that variable
substitution.
For instance, the elements of the systematic shifts (see
eq~\eqref{eq:errpropS}) become
\begin{equation}
  (\bm\sigma_{\ahut}^{(\bm{s}_{(l)})})_k = \frac{\left(\mathcal{F}\bm{s}_{(l)}\right)_k }{\gamma_k\hat\alpha_k^{\gamma_k-1}  }\,,
  \label{eq:errpropSgamma}
\end{equation}
and similarly for other uncertainty components or covariance matrices.
The power law factor $\gamma$ may be of practical use in order to
validate the assumptions of the linearity of the regression analysis,
or to improve the fit, such as when it is known that the model is
proportional to $\alpha^2$.
This would be the case in determinations of the
strong coupling constant from $n$-jet production cross sections
at the LHC, which are proportional to $\alpha^n_s(m_Z)$ at the leading
order in perturbative Quantum Chromodynamics (QCD).

\subsection{Finite differences}
\vspace{-0.5em}
In a use case where a larger number of templates are available, various choices of
subsets of these templates can become input to the Linear Template
Fit, and
some of such potential choices have already been suggested in the
previous paragraphs. 
In the case where only two (or three) templates are
provided as input to the Linear Template Fit, the selected templates
may have a special meaning, because the templates can also be considered
as a numerical first derivative through finite differences.
Such numerically calculated derivatives are the essence of many
gradient descent optimization
methods~\cite{Gill:99800,Fletcher:1568710,Nocedal:1638144,madsen,James:1019859},
and it is instructive to study the relation.

As an example, let us consider a univariate problem where the
objective function \chisq\ is given by eq.~\eqref{eq:chisq1}.
Close to the minimum, \chisq\ is nearly quadratic, and the Newton
step, $\Delta\alpha=-(\chisq)^\prime/(\chisq)^{\prime\prime}$, exhibits
a good
convergence~\cite{Blobel:1984cy,Gill:99800,Fletcher:1568710,Nocedal:1638144}.
The first and second derivatives at an expansion
point $\alpha_0$ are 
\begin{equation}
  (\chisq)^\prime 
  = -2\bm\lambda^{\prime \text{T}}(\alpha_0) W (\bm{d} - \bm\lambda(\alpha_0))
  ~~~\text{and}~~~
  (\chisq)^{\prime\prime} = 2 \bm\lambda^{\prime \text{T}}(\alpha_0) W \bm\lambda^\prime(\alpha_0)\,, 
  \label{eq:finitediff1}
\end{equation}
where the second derivatives $\bm\lambda^{\prime\prime}$ are zero or
neglected in the latter equation.
Hence, one obtains the estimator from the Newton step as
\begin{equation}
  \hat\alpha_\text{(Newton)}=
  \alpha_0 + \left[ \bm\lambda^{\prime \text{T}}(\alpha_0) W \bm\lambda^\prime(\alpha_0)  \right]^{-1}
  \bm\lambda^{\prime \text{T}}(\alpha_0) W (\bm{d}-\bm\lambda(\alpha_0))\,,
  \label{eq:finitediff2}
\end{equation}
and the  calculation of the first derivative could be
done numerically using a finite difference with nonzero step size $h$
\begin{equation}
  \bm\lambda^\prime(\alpha_0) =
  \frac{\bm\lambda(\alpha_0+h) - \bm\lambda(\alpha_0)}{h}\,.
  \label{eq:finitediff0}
\end{equation}

The analogous Linear Template Fit makes use of the templates at
$\alpha_0$ and $\alpha_0+h$  and defines the following matrices
\begin{equation}
  Y=\begin{pmatrix}
  \bm\lambda(\alpha_0) &  ~\bm\lambda(\alpha_0+h)
  \end{pmatrix}
  ~~\text{and}~~
  M=\begin{pmatrix}
  1 & \alpha_0 \\
  1 & ~\alpha_0+h
  \end{pmatrix}\,,
  ~\text{hence}~
  M^+=\begin{pmatrix}
  \tfrac{\alpha_0+h}{h} & -\tfrac{\alpha_0}{h}\\
  -\tfrac{1}{h} & \tfrac{1}{h}
  \end{pmatrix}
  =
  \begin{pmatrix}
    \bm{\bar{m}}^{\text{T}} \\
    \bm{\tilde{m}}^{\text{T}}
  \end{pmatrix}\,.
  \label{eq:finitediffLTF1}
\end{equation}
From the Linear Template Fit, eq.~\eqref{eq:mtDetLev}, when resolving $\bm{\bar{m}}$, one
obtains the best estimator
\begin{equation}
  \hat\alpha=
  \alpha_0 + \left[(Y\bm{\tilde{m}})^{\text{T}} W (Y\bm{\tilde{m}})\right]^{-1}
  (Y\bm{\tilde{m}})^{\text{T}} W(\bm{d} - \bm\lambda(\alpha_0))
  \label{eq:finitediffLTF2}
\end{equation}
and since $Y\bm{\tilde{m}}=\bm\lambda^{\prime}(\alpha_0)$, it can be 
directly seen that it is equivalent to the estimator
$\hat\alpha_\text{(Newton)}$ from the Newton step using a numerical finite
difference.


\section{Nonlinear model approximation}
\label{sec:NonLinearModel}
The Linear Template Fit can be regarded such that linear functions
are fitted to every row $i$ of the template matrix $Y$.\footnote{ Conceptionally,
  (nonlinear) functions can also
be fitted to the elements of each column $j$, and so interpreting
the shape of the templates and the data distributions.
Subsequently, the resulting functions, when normalized, can be
interpreted as probabilities and are often used as input to binned or
un-binned maximum-likelihood optimizations~\cite{Abulencia:2005aj,ATLAS:2012aj,ATLAS:2018fwq}.
Similarly, the parameters of such a transformation can also
become again input to a (Linear) Template Fit.
An example of this approach is discussed in Appendix~\ref{sec:prefit}.} 
In the following, nonlinear functions are considered for the model
$\lambda_i(\bm{\alpha})$ in each ``bin'' $i$.  
%
For a simplified discussion, the formuluae for only a
univariate model are sometimes discussed, while the
multivariate case is commonly straightforward.
Also, the concepts of the $\gamma$ factor, relative
uncertainties, or ``systematic shifts'' $\bm{s}_l$ can equally be
considered in a multivariate formulation.

The model can be approximated from the templates in every bin $i$ with an
$n$th degree polynomial function
\begin{equation}
  \lambda_i(\alpha) \approx \mathrm{y}_i(\alpha;\bm{\hat\theta}_{(i)}) =: \hat{\mathrm{y}}_i(\alpha)
  ~~~\text{using}~~~
  \mathrm{y}_i(\alpha;\bm{\theta}_{(i)})= \theta_{0}^{(i)} + \theta_{1}^{(i)}\alpha + \hdots + \theta_{n}^{(i)}\alpha^n\,.
  \label{eq:nonliny}
\end{equation}
The best estimators for the polynomial parameters are obtained
from regression analysis, 
\begin{equation} 
  {\bm{\hat\theta}_{(i)}}
  = \left(\begin{smallmatrix}\hat\theta_{0}\\\hat\theta_{1}\\\vdots\\\hat\theta_{n}\end{smallmatrix}\right)_{(i)} =
   \mathcal{M}^+\begin{pmatrix} y_{(1),i} \\ \vdots \\ y_{(j),i} \end{pmatrix}\,,
    \label{eq:multtheta}
\end{equation}
where $y_{(j),i}$ is the $i$th value of the $j$th template, and the
$g$-inverse is
\begin{equation}
  \mathcal{M}^+ = \left(\mathcal{M}^{\text{T}} \mathcal{M}\right)^{-1}  \mathcal{M}^{\text{T}}
  ~~~~\text{using here}~~~~
  \mathcal{M} =
   \begin{pmatrix}
    1 & \dot{\alpha}_1 &\hdots & \dot{\alpha}_1^n \\
    \vdots & \vdots& \vdots& \vdots \\
    1 & \dot{\alpha}_j & \hdots& \dot{\alpha}_j^n \\
   \end{pmatrix}\,.
   \label{eq:nonlinM}
\end{equation}
For reasons outlined in section~\ref{sec:basicmodel} an unweighted
regression is considered.
Using this model approximation in the objective function of the
fitting problem,
eq.~\eqref{eq:chisqDetLev}, the \chisq\ expression for a univariate model becomes of order
$\mathcal{O}(2n)$ in $\alpha$, 
\begin{align}
  \chi^2_{\ord{(2n)}}
  :&=
  \left( \bm{d} - \bm{\hat\mathrm{y}}(\alpha)\right)^{\text{T}}
  W
  \left( \bm{d} - \bm{\hat\mathrm{y}}(\alpha)\right) \\
  &=
  \left( \bm{d} - Y\bm{m_0} - Y\bm{m_1}\alpha - \hdots - Y\bm{m_n}\alpha^n\right)^{\text{T}}
  W
  \left( \bm{d} - Y\bm{m_0} - Y\bm{m_1}\alpha - \hdots - Y\bm{m_n}\alpha^n\right)\,,
  \label{eq:chi2ord2n}
\end{align}
where the vectors $\bm{m}$ were used as the representation of
$\mathcal{M}^+$ according to
\begin{equation}
  \mathcal{M}^+
  =:
  \left(\begin{smallmatrix}
    \bm{m_0}^{\text{T}} \\
    \bm{m_1}^{\text{T}} \\
    \vdots \\
    \bm{m_n}^{\text{T}} \\
  \end{smallmatrix}\right)  \,.
\end{equation}
It is obvious that for $n=1$, this is the Linear Template Fit, and the
generalizations to the multivariate fit
(eq.~\eqref{eq:chisqNuisance}), and with relative uncertainties 
(eq.~\eqref{eq:chi2log}), are  straightforward.
Note that for different degrees $n$, the vectors $\bm{m}$ differ unless
the equations are not rewritten for orthogonal polynomials~\cite{Blobel:1984cw}.

Since the first derivative of $\chi^2_{\ord{(2n)}}$ is nonlinear in
$\alpha$, no analytic solution of the optimization problem is found.
Therefore, in the following, two iterative algorithms to find the minimum of the objective
function $\chi^2_{\ord{(2n)}}$ are discussed, which refer to the
Gau\ss--Newton and the Newton algorithm 
(see also Refs.~\cite{Blobel:1984cy,Gill:99800,Fletcher:1568710,Nocedal:1638144}). 
In both cases, if the starting
value is identified with the best estimator from the Linear Template Fit,
$\alpha=\hat\alpha$, the step size of the next iterative step can be
interpreted as convergence criterion or uncertainty of the Linear
Template Fit. 
Reasonable values for $n$ are $n=2$, 3, or at most 4 (since
$\mathcal{M}$ becomes quickly ill-conditioned for large $n$,
e.g.~Refs.~\cite{Gautschi:10.1007/BF01437212,pan2016bad} and therein), while
for $n=1$, consistent results with the Linear Template Fit are found.

\subsection{\boldmath{Linearization of the nonlinear approxmation $\hat{\mathrm{y}}(\alpha)$}}
\vspace{-0.5em}
An iterative minimization algorithm based on first derivatives is
defined through linearly approximating the (nonlinear) function
$\bm{\hat{\mathrm{y}}}(\alpha)$, eq.~\eqref{eq:nonliny}.
For the purpose of validating the result from the Linear Template Fit,
we consider $\hat\alpha$ as the expansion point, although this could
be any other value as well.
Hence, the linear approximation of $\bm{\hat{\mathrm{y}}}(\alpha)$ in
every bin $i$ is 
\begin{align}
  \tilde{y}_i(\alpha)
  &\approx \hat{\mathrm{y}}_i(\hat\alpha) +
  \left.\frac{\partial\hat{\mathrm{y}}_i(\alpha)}{\partial\alpha}\right|_{\alpha=\hat\alpha} (\alpha-\hat\alpha) \nonumber\\
  &= Y(\textstyle{\sum}_0^n (1-n)\bm{m}_{(n)}\hat\alpha^n) + Y\left(
  \textstyle{\sum}_0^n n\bm{m}_{(n)}\hat\alpha^{n-1} \right)\alpha \nonumber \\
  &=: Y\bm{\tilde{\bar{m}}}
  + Y\bm{\tilde{\tilde{m}}}\alpha\,.
  \label{eq:approxm}
\end{align}
The last term defines the shorthand notations $\bm{\tilde{\bar{m}}}$ and $\bm{\tilde{\tilde{m}}}$.
From comparison with eq.~\eqref{eq:ymt}, it is seen that
an alternative best estimator of the template fit
(see eq.~\eqref{eq:master}) is obtained as
\begin{equation}
  \tilde{\hat\alpha} =   \frac{(Y\bm{\tilde{\tilde{m}}})^{\text{T}}W}{(Y\bm{\tilde{\tilde{m}}})^{\text{T}}WY\bm{\tilde{\tilde{m}}} } ( \dt - {Y\bm{\tilde{\bar{m}}}})\,.
  \label{eq:nonlinahat}
\end{equation}
The difference $\hat\alpha-\tilde{\hat\alpha}$ is supposed to become
small for a valid application of the Linear Template Fit, and one
should find
\begin{equation}
  |\hat\alpha - \tilde{\hat\alpha}| \ll \sigma_{\hat\alpha}\,.
  \label{eq:nonl1}
\end{equation}
If eq.~\eqref{eq:nonl1} holds for a particular application, then
nonlinear effects are negligible, and  
the best estimator of the Linear Template Fit is insensitive to
nonlinear effects in the model $\lambda$.
If eq.~\eqref{eq:nonl1} does not hold, then it should be studied if
the nonlinearity is present in the vicinity of $\hat\alpha$ or if it
is rather introduced by a long-range nonlinear behavior of $\lambda(\alpha)$. In
the latter case, the range of the templates should be reduced such that
one has about
\begin{equation}
  \dot\alpha_\text{max}-\dot\alpha_\text{min} \lesssim 2\sigma_{\hat\alpha}\,
  \label{eq:range}
\end{equation}
and the Linear Template Fit should be repeated.
Note that if the templates are subject to statistical fluctuations and if only
a few templates are provided, then a nonlinear approximation is more
sensitive to those fluctuations than linear approximations, and
$|\hat\alpha - \tilde{\hat\alpha}|$ may become larger. In such an
application $\sigma_{\hat\alpha}^Y$ should also be considered as a
relevant source of uncertainty
(see eqs.~\eqref{eq:errpropYuncorr} and~\eqref{eq:errpropYcorr}).

\noindent
\fbox{
  \parbox[c]{0.97\textwidth}{
  \small
In Example 1, when using a second-degree polynomial, a value 
$|\hat\alpha - \tilde{\hat\alpha}| = 0.02$  is found, which is much
smaller than $\sigma_{\hat\alpha}=\pm0.41$ and of similar size
than the statistical uncertainties from $Y$ (see
section~\ref{sec:errpropMCstat}).
When working with log-normal uncertainties, a value
$|\hat\alpha - \tilde{\hat\alpha}| = 0.05$ is found. Although the
model is highly nonlinear in $\log(y)$, the assumptions of the linear
template fit are well justified since the templates are in the close
vicinity of the best estimator.
In Example 2, for a second degree approximation values of
$|\hat\alpha_0 - \tilde{\hat\alpha}_0| = 0.03$ 
and $|\hat\alpha_1 - \tilde{\hat\alpha}_1| = 0.14$ are found.
When further considering interference terms, see Appendix~\ref{sec:nonlinndim},
the values are
$0.01$ and $0.11$, respectively
All these values are smaller than the respective statistical
uncertainties in the data, although the values suggest that for
$\alpha_1$ (the variance of the gaussian model) a finer grid for the
reference values may be preferred, for a more accurate linear
approximation.
  }
  }

Furthermore, the notation of eq.~\eqref{eq:approxm} suggests an alternative
interpretation of the non-linear behaviour of the model.
The vectors $\bm{\tilde{\bar{m}}}$ and $\bm{\tilde{\tilde{m}}}$ may
be interpreted in terms of uncertainties of $\bm{\bar{{m}}}$
and $\bm{\tilde{{m}}}$ when writing
\begin{equation}
  \bm{\sigma}_{\bm{\bar{{m}}}} := \bm{\tilde{\bar{m}}} - \bm{\bar{{m}}}
  ~~~~\text{and}~~~~
  \bm{\sigma}_{\bm{\tilde{{m}}}} := \bm{\tilde{\tilde{m}}} - \bm{\tilde{{m}}}\,.
\end{equation}
Through linear error propagation, one can define an uncertainty in
$\hat\alpha$ as
\begin{align}
  \sigma_{\hat\alpha}^{(m)} &=
  \sum_j \bigg( -FY\mathbb{1}_j\sigma_{\bar{m}_j}  
  +
  D^{-1} \left[ (Y\mathbb{1}_j)^{\text{T}}W ( d-Y\bar{m})  -
    \left[ (Y\mathbb{1}_j)^{\text{T}}W Y\tilde{m} +
      (Y\tilde{m})^{\text{T}}WY\mathbb{1}_j\right] \hat{a} \right]
    \sigma_{\tilde{m}_j} \bigg) \nonumber \\
    &=
   \sum_i \bigg( -F\mathbb{1}_i\sigma_{\overline{Ym}_i}  
   + D^{-1} \big[ \mathbb{1}_i^{\text{T}}W ( d-Y\bar{m})  -
    \left[ \mathbb{1}_i^{\text{T}}W Y\tilde{m} +
      (Y\tilde{m})^{\text{T}}W \mathbb{1}_i\right] \hat{a} \big]
   \sigma_{\widetilde{Ym}_i} \bigg)\,. 
   \label{eq:deltaalpham}
\end{align}
Note that the sum runs over $j$ in the first and over $i$ in the second
expression.
The second expression makes use of
\begin{equation}
  \bm{\sigma}_{\bm{\overline{Ym}}} := Y\bm{\sigma}_{\bm{\bar{m}}}
  ~~~~\text{and}~~~~
  \bm{\sigma}_{\bm{\widetilde{Ym}}} := Y\bm{\sigma}_{\bm{\tilde{{m}}}} 
\end{equation}
and it has the advantage that the summands represent a useful
quantifier for every bin $i$.
For a sufficiently linear problem, both the linear sum
(eq.~\eqref{eq:deltaalpham}) and the quadratic sum should
be sufficiently small. In particular, one should find
$\sigma_{\hat\alpha}^{(m)}\approx\tilde{\hat\alpha}-\hat\alpha$ for
reasonable $n$th degree polynomials, i.e.\ $n$=2, and also, according to
eq.~\eqref{eq:nonl1}, 
\begin{equation}
  |\sigma_{\hat\alpha}^{(m)}| \ll |\sigma_{\hat\alpha}|\,.
  \label{eq:sigmam}
\end{equation}

\subsection{\boldmath{Minimization of $\chi^2_{\ord{(2n)}}$}}
\vspace{-0.5em}
When $\hat{\alpha}$ is obtained from the Linear Template Fit,
the objective function $\chi^2_{\ord{(2n)}}$,
eq.~\eqref{eq:chi2ord2n}, is expected to exhibit a 
minimum in the close vicinity of $\hat{\alpha}$. 
This  provides good motivation for the application of the
Newton optimization at $\hat{\alpha}$, which is
based on second derivatives of 
$\chi^2_{\ord{(2n)}}$, so a parabolic approximation.
The Newton distance to the minimum is given by
\begin{equation}
  \Delta\alpha = -\frac{(\chisq_{\ord{(2n)}})^\prime}{(\chisq_{\ord{(2n)}})^{\prime\prime}} =
  -\left(\left.\frac{\partial^2\chisq_{\ord{(2n)}}}{\partial^2\alpha}\right|_{\alpha=\hat\alpha}\right)^{-1}
  \left.\frac{\partial\chisq_{\ord{(2n)}}}{\partial\alpha}\right|_{\alpha=\hat\alpha} 
  \label{eq:newton1}
\end{equation}
and in the multivariate case by~\cite{Blobel:437773,Gill:99800}   
\begin{equation}
  \bm{\Delta\alpha} = \left.-H^{-1}\bm{g}\right|_{\bm{\alpha}=\bm{\hat\alpha}}\,.
  \label{eq:newton2}
\end{equation}
The gradient vector $\bm{g}$ and the Hesse matrix $H$ are defined by
the first and second derivatives of $\chi^2_{\ord{(2n)}}$, respectively.
Both quantities can be calculated directly at an  
expansion point $\bm{\alpha_0}$, which was identified in the above equations
with $\bm{\hat\alpha}$, since $\chisq_{\ord{(2n)}}$ is just a known
second-degree polynomial and dependent on $M$, $Y$, $\bm{d}$, $S$, and $W$.
The value of $\bm{\Delta\alpha}$ can be interpreted as the expected distance
to the minimum (EDM) based on the nonlinear model
approximation.\footnote{When the widely used approximation of the Hesse
  matrix with the Jacobi matrix of first derivatives is used,
  $H\approx J^{\text{T}}J$ (\emph{truncated} Hesse), then the Newton
  step becomes equivalent to the Gau\ss--Newton step, which is
  equivalent to linearization of the model, eq.~\eqref{eq:nonlinahat}, and 
  one obtains
  $\bm{\alpha} + \bm{\Delta\alpha}= \bm{\tilde{\hat\alpha}}$, when
  using $J^{\text{T}}J$ as $H$. }

For practical purposes and to validate the
result of the Linear Template Fit one should require that its value
be smaller than the uncertainty from the fit
\begin{equation}
  |\Delta\alpha| \ll \sigma_{\hat\alpha}\,.
\end{equation}
The most reasonable degree for this purpose is $n=2$  (which is
exploited in the following in greater detail), and one expects that $\Delta\alpha$ will be negligibly small.
This observation can also be interpreted such that the result from the
Linear Template Fit and an alternative result from a
minimization of $\chi^2_{\ord{(2n)}}$ are consistent.
If the reference points are in the vicinity of $\hat\alpha$, this is
to be expected because of the underlying statistical model of
(log-)normal distributed estimators.
In contrast, if $\Delta\alpha$ becomes large, it may well be
that in such a case not even the reported uncertainties in the best estimator, $\sigma_{\hat\alpha}$,
can be considered to be normal distributed.

\section{The Quadratic Template Fit}
\label{sec:algo}
Despite all the considerations and cross-checks as discussed in the
previous sections, 
in some problems the linearized model
 may simply be an inaccurate approximation of the true model
 (cf.\ eq.~\eqref{eq:ymt2}). 
Reasons could be that the templates cannot be generated in a
sufficiently close vicinity around the best estimator, perhaps for
 technical reasons, or in multivariate problems some parameters are highly
nonlinear, a dimension is poorly constrained and the reference points span
a large range, or interference terms between the parameters are
non-negligible.
In such cases, the quantities
$\bm{\tilde{\hat\alpha}}-\bm{\hat\alpha}$,
$\bm{\sigma_{\hat\alpha}^{(m)}}$,
or
$\bm{\Delta\alpha}$
(eqs.~\eqref{eq:nonl1},
\eqref{eq:deltaalpham} and~\eqref{eq:newton2}) 
  become non-negligible when 
they are calculated for second-degree polynomials $n=2$.

An improved approximation of the model is obtained when using a second order
approximation, which may also include interference terms.
The equations are summarized in Appendix~\ref{sec:nonlinndim} 
(eqs.~\eqref{eq:nonlinyndim}--\eqref{eq:nonlinMndim}).
Using such a second-degree polynomial model $\bm{\hat{\mathrm{y}}}(\bm{\alpha})$
that includes terms $\mathcal{O}(\alpha_k)$, $\mathcal{O}(\alpha^2_k)$,
and $\mathcal{O}(\alpha_{k_1}\alpha_{k_2})$ in the
optimization problem results in a nonlinear least-squares problem.
Therefore, an iterative algorithm is required to obtain the
best estimator.

A robust algorithm, denoted as \emph{Quadratic Template
  Fit}, to obtain the best estimator and to perform a full
error propagation is defined as
\begin{enumerate}
\item The Linear Template Fit is performed to obtain a first
  estimator $\bm{\hat\alpha}_{(0)}$, for example using eq.~\eqref{eq:full2};
\item The Newton
  algorithm~\cite{Blobel:1984cy,Gill:99800,Fletcher:1568710,Nocedal:1638144}, eq.~\eqref{eq:newton2},
  is employed with a few $m$ iterations in order to obtain improved best estimators
  $\bm{\hat\alpha}_{(m)} = \bm{\hat\alpha}_{(m-1)} + \bm{\Delta\alpha}_{(m)}$
  and for which the Hesse matrix $H$ is directly analytically calculable;
\item The best estimator and the error calculation are obtained using
  the linearized approximations for $\bm{\tilde{\bar{m}}}$ and $\tilde{\tilde{M}}$
  (see eqs.~\eqref{eq:approxm} or~\eqref{eq:approxmnDim}) in the
  equations of the Linear Template Fit. 
\end{enumerate}
The first step provides the starting point for the Newton
algorithm, and since it is obtained with the Linear Template Fit it is
already close to the minimum.
The second step provides a fast and stable convergence since the
starting point is already in the vicinity of the minimum, the Hesse
matrix is commonly positive definite and is analytically exactly
calculable,\footnote{The analytic calculation of the Hesse matrix results
in an exact application of the
Newton minimization, which is in contrast to many other iterative
minimization algorithms, where the Hesse matrix is only approximated,
iteratively improved, or numerically calculated. This results in a
quick convergence with only a few steps, and no adaptions to the Newton
steps are required.}, as well as the gradient vector, and the Newton method has
good convergence for nearly quadratic
functions~\cite{Blobel:1984cy,Gill:99800,Nocedal:1638144}, which is
the case for $\chi^2_{\ord{(4)}}$.
Although the algorithm is expected to converge quickly after a very few
iterations, several stopping criteria are thinkable, for example when
EDM$\ll|\bm{\Delta\alpha}|$ or
$\bm{\Delta\alpha}_{(m)}\approx\bm{\Delta\alpha}_{(m-1)}$.
The last step provides an even more improved best estimator, but mainly
gives access to the equations of the error propagation.
Note that in the first step, the true model $\bm{\lambda}(\bm{\alpha})$ is
linearly approximated, and in the latter steps, by quadratic functions.

\noindent
\fbox{
  \parbox[c]{0.97\textwidth}{
  \small
When applying the Quadratic Template Fit to Example 2, and using $m=2$
Newton-iterations, the best estimator for the mean is found to be
$169.90\pm0.44$ and for the variance it is $6.23\pm0.32$. 
Both values are very close to the best estimators from the
Linear Template Fit, see section~\ref{sec:example2}.
The EDM $\bm{\Delta\alpha}$ and
$\bm{\tilde{\hat\alpha}}-\bm{\hat\alpha}$ are found to be smaller than
$10^{-4}$ for both parameters, and $\bm{\sigma}_{\bm{\hat{\alpha}}}^{(m)}$ is
equally small.
  }
  }

\section{Example 3: the Quadratic Template Fit}
\label{sec:example3}
As an application of the Quadratic Template Fit, an example is constructed where
the model is nonlinear and the templates purposefully span a (too)
large range. Therefore, the criteria for a valid
application of the Linear Template Fit as discussed in
section~\ref{sec:considerations} are mainly not fulfilled.

\begin{figure}[tb!h]
  \centering
     \begin{minipage}[c]{0.48\textwidth}
       \centering
       \includegraphics[width=0.98\textwidth,trim={0 0 05 0 },clip]{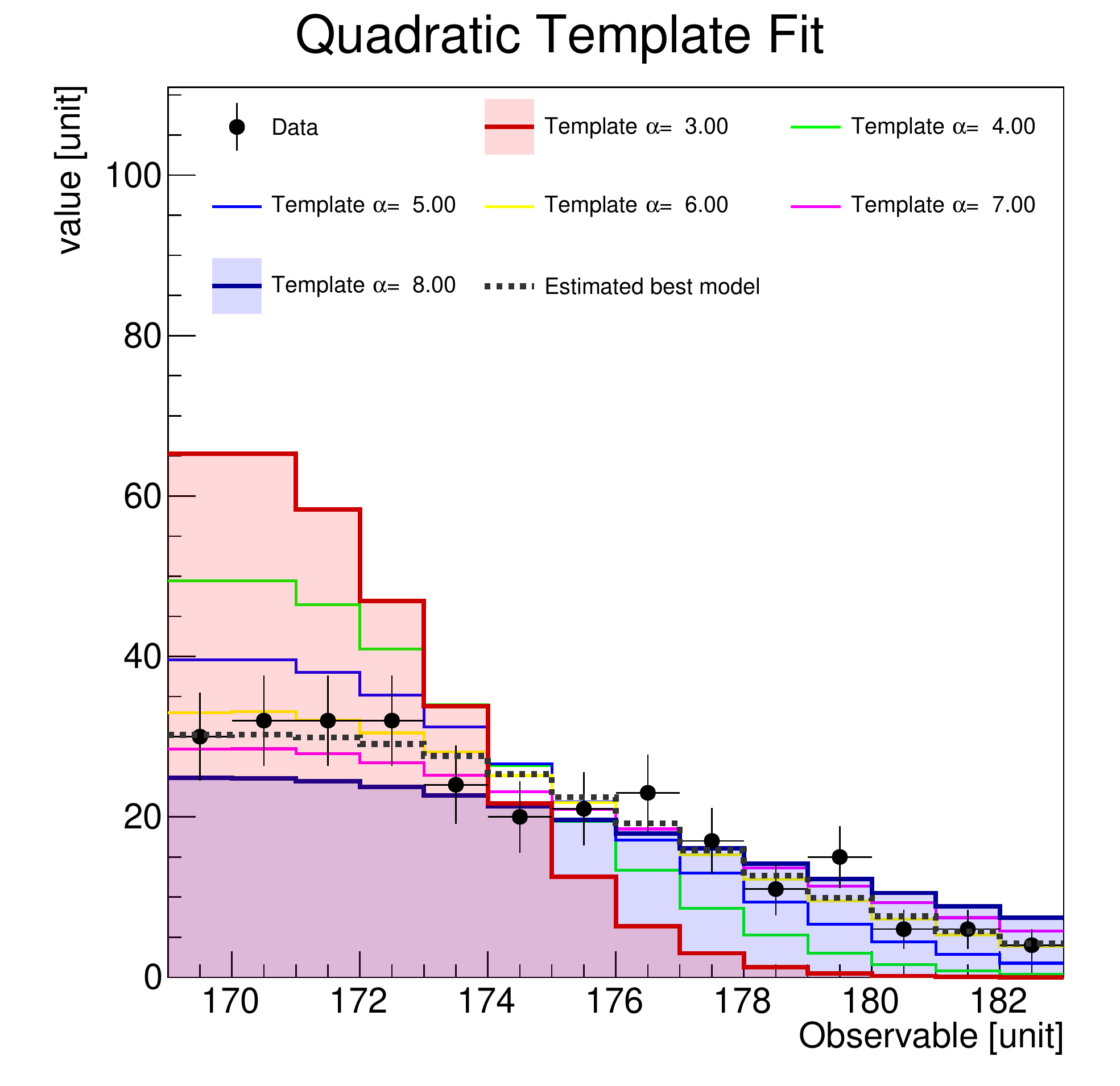}
     \end{minipage}
     \begin{minipage}[c]{0.48\textwidth}
       \centering
    \includegraphics[width=0.44\textwidth,trim={0 0 05 0 },clip]{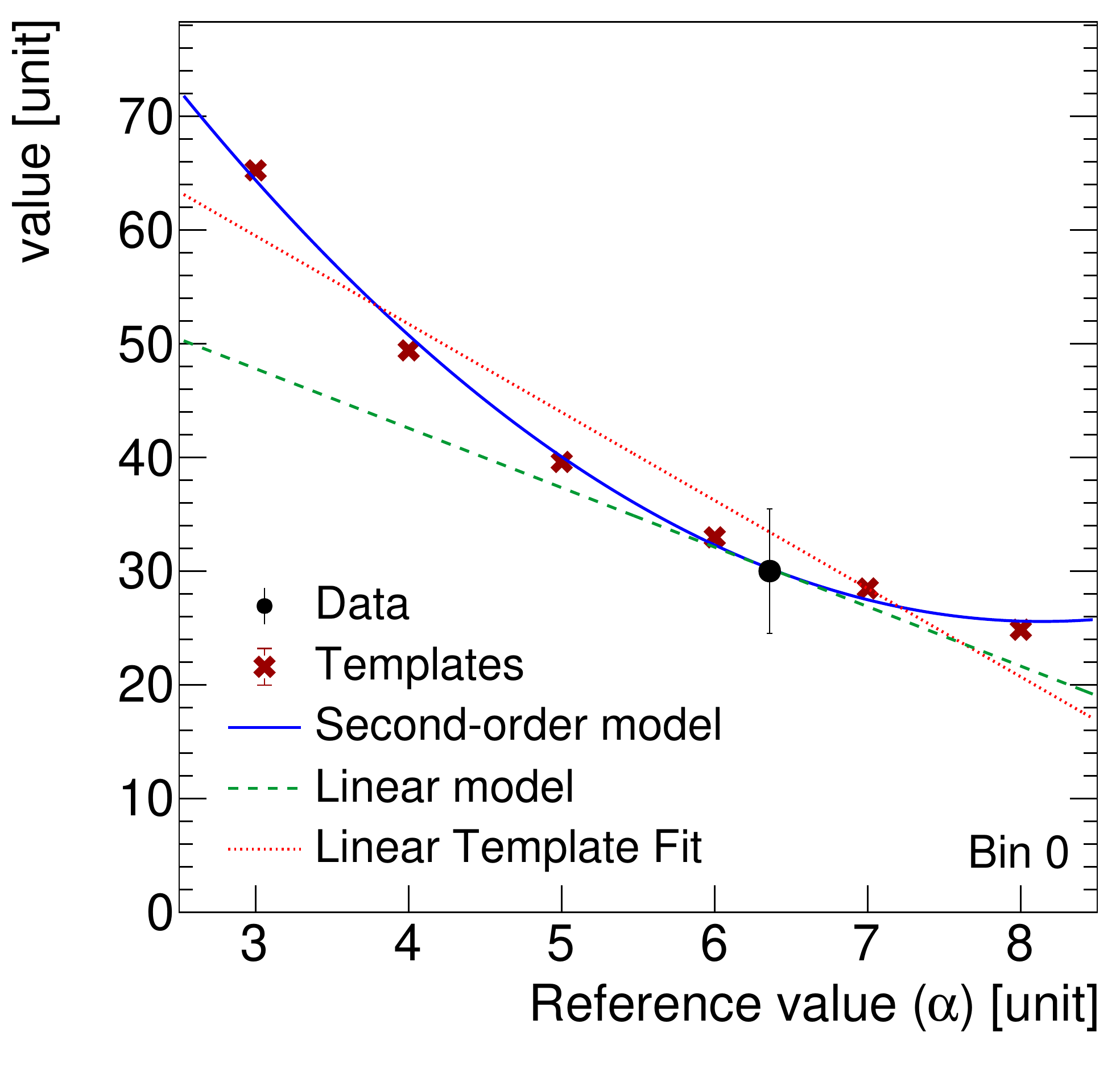}
    \includegraphics[width=0.44\textwidth,trim={0 0 05 0 },clip]{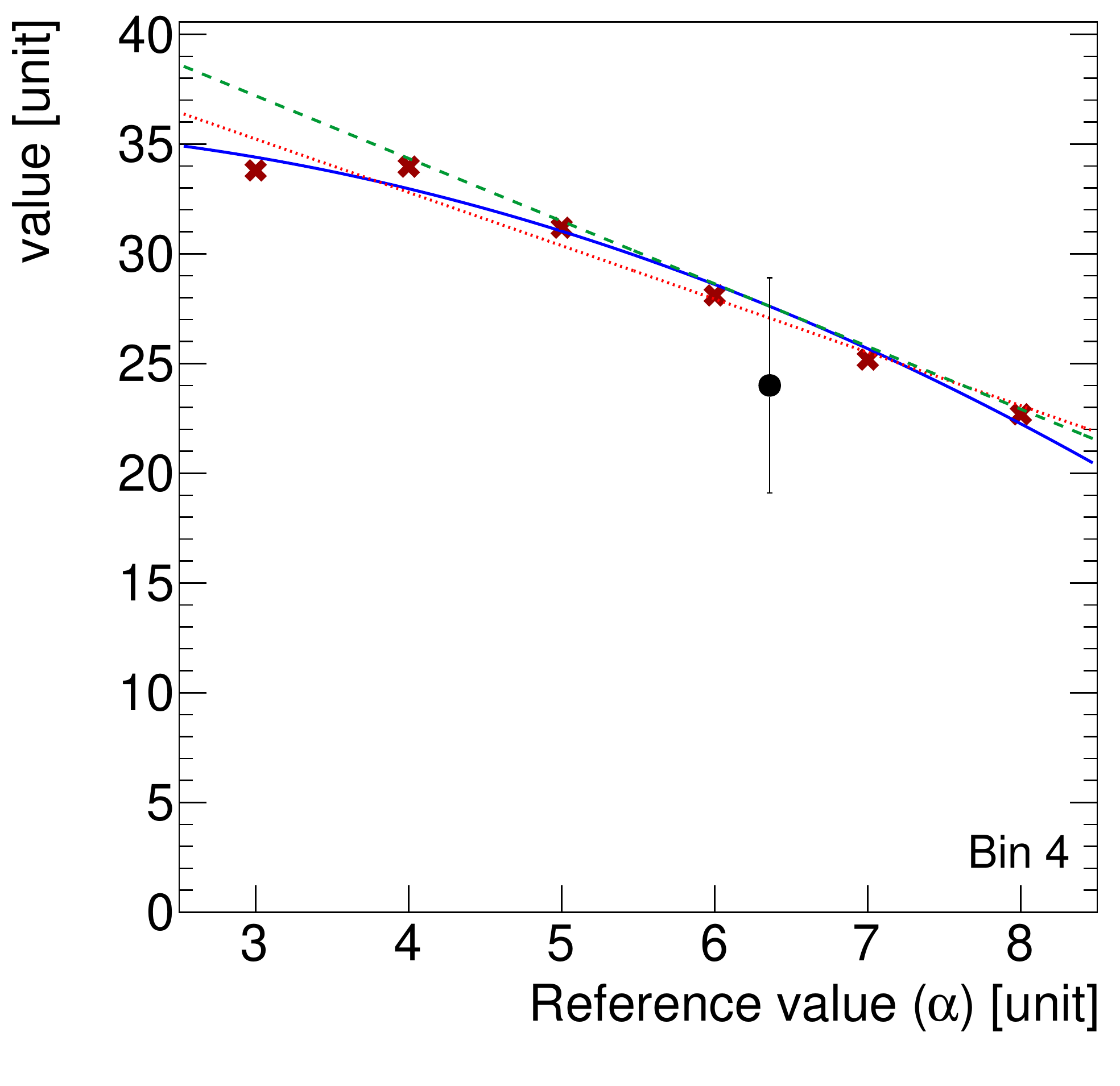}\\
    \includegraphics[width=0.44\textwidth,trim={0 0 05 0 },clip]{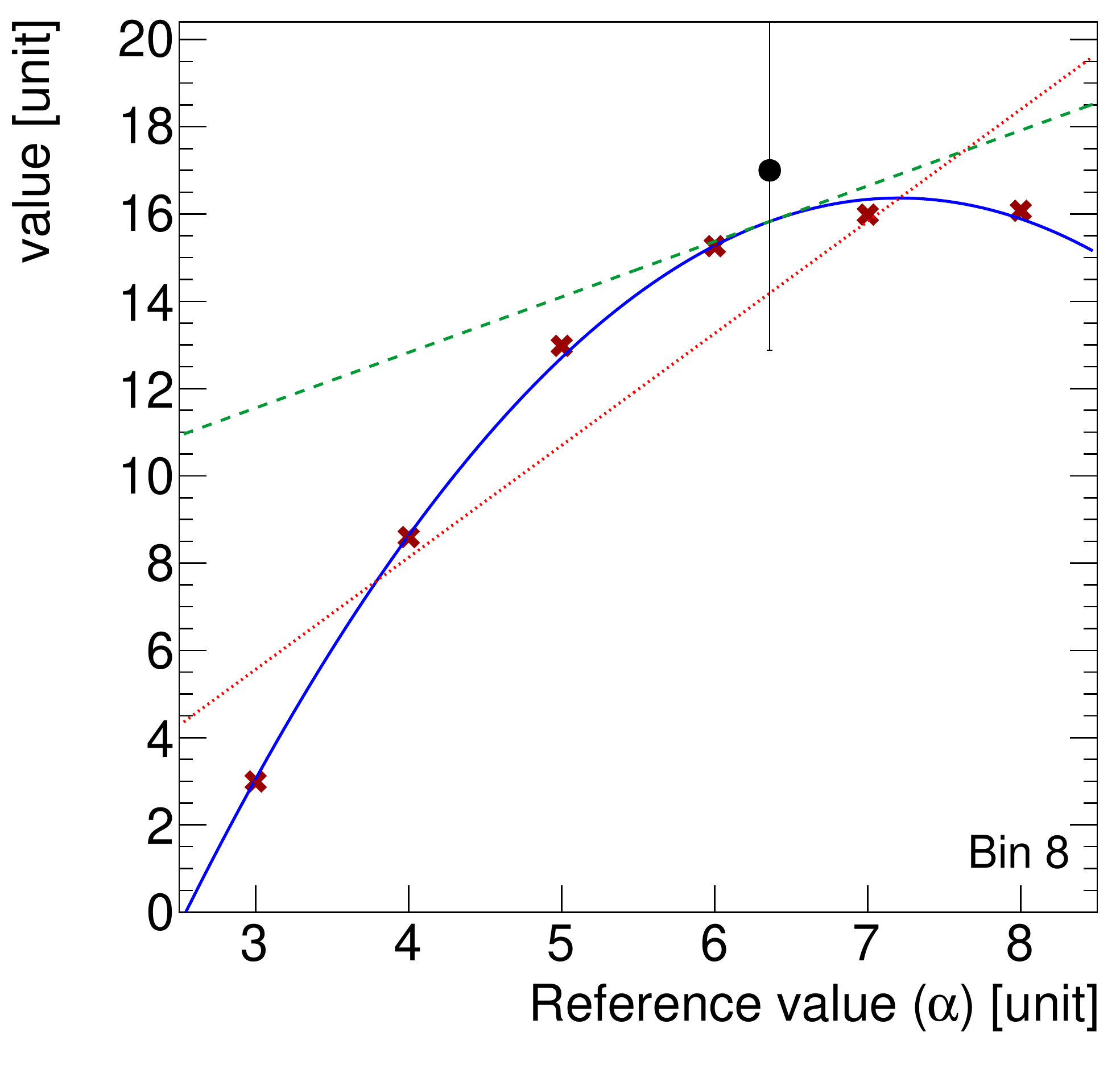}
    \includegraphics[width=0.44\textwidth,trim={0 0 05 0 },clip]{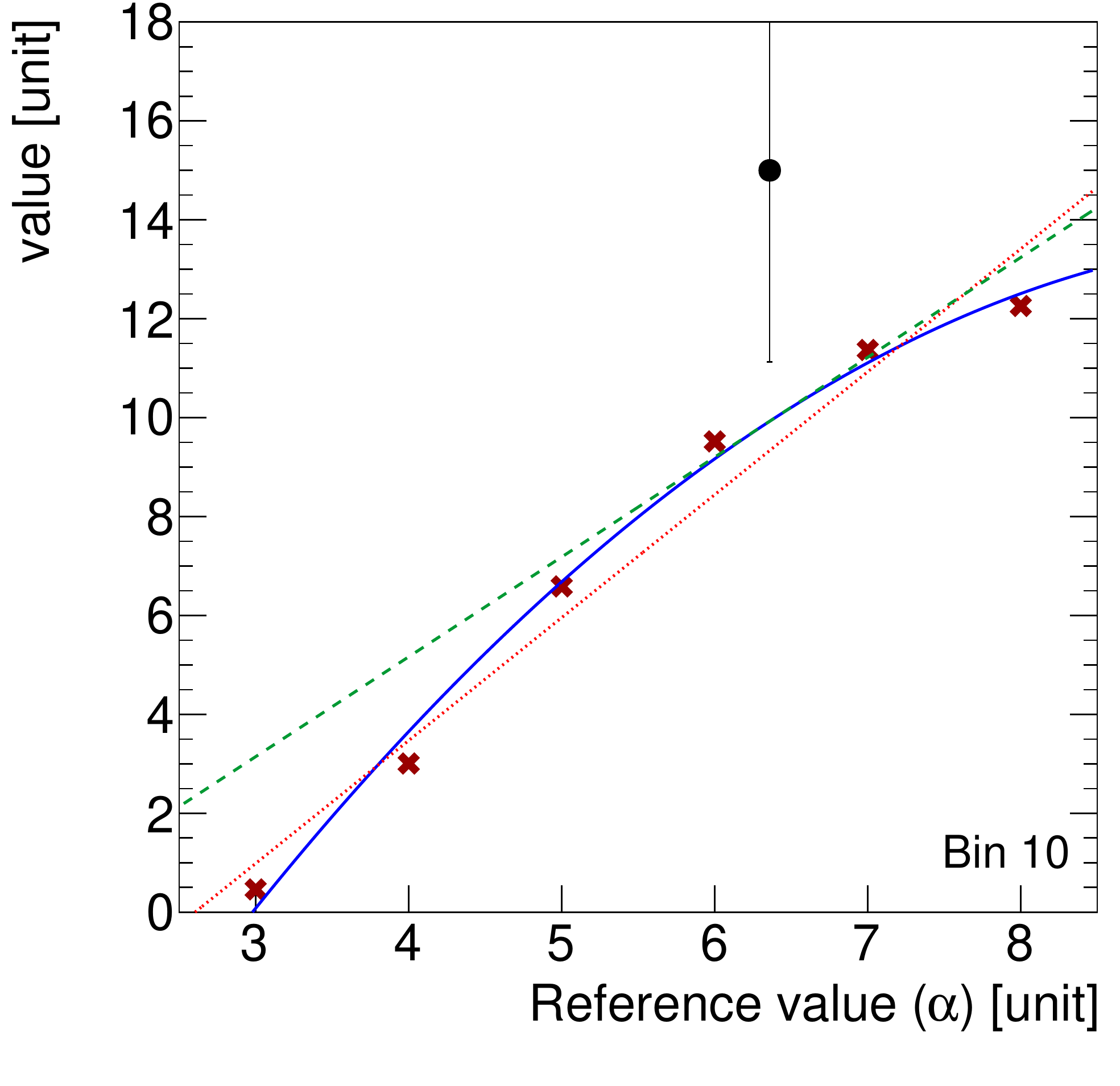}
     \end{minipage}
  \caption{
    Some illustrations of Example 3.
    Left: the template distributions $Y$ as a function of the
    observable, and more details are given in the caption of
    figure~\ref{fig:example1}.
    Right: the templates $Y$ for some selected bins.
    The blue line indicates a second-degree polynomial regression of
    the templates, as is used by the Quadratic Template Fit.
    The green dashed line indicates the linearization at the best
    estimator, and the red dotted line indicates the linear regression
    of the templates as it would  be used by the plain Linear Template Fit.
  }
  \label{fig:example3a}
\end{figure}

As in Example 1 and 2 (sections~\ref{sec:example1} and~\ref{sec:example2}),
the model is considered to be a normal distribution, and
the same pseudo-data as in the previous examples are also used.
In contrast to Examples 1 and 2, only the variance of the
normal distribution shall be determined. 
Six templates are generated similarly to Examples 1 and 2, with variances
between 3 and 8, and each template is generated using $4\cdot10^6$ events.
The pseudo-data and the templates are displayed in
figure~\ref{fig:example3a}.
Recall that the pseudo-data were generated with a variance of 6.2 (due to
statistical fluctuations, the true variance is about 6.3). 

\begin{figure}[tb!h]
  \centering
     \includegraphics[width=0.44\textwidth,trim={0 0 05 0 },clip]{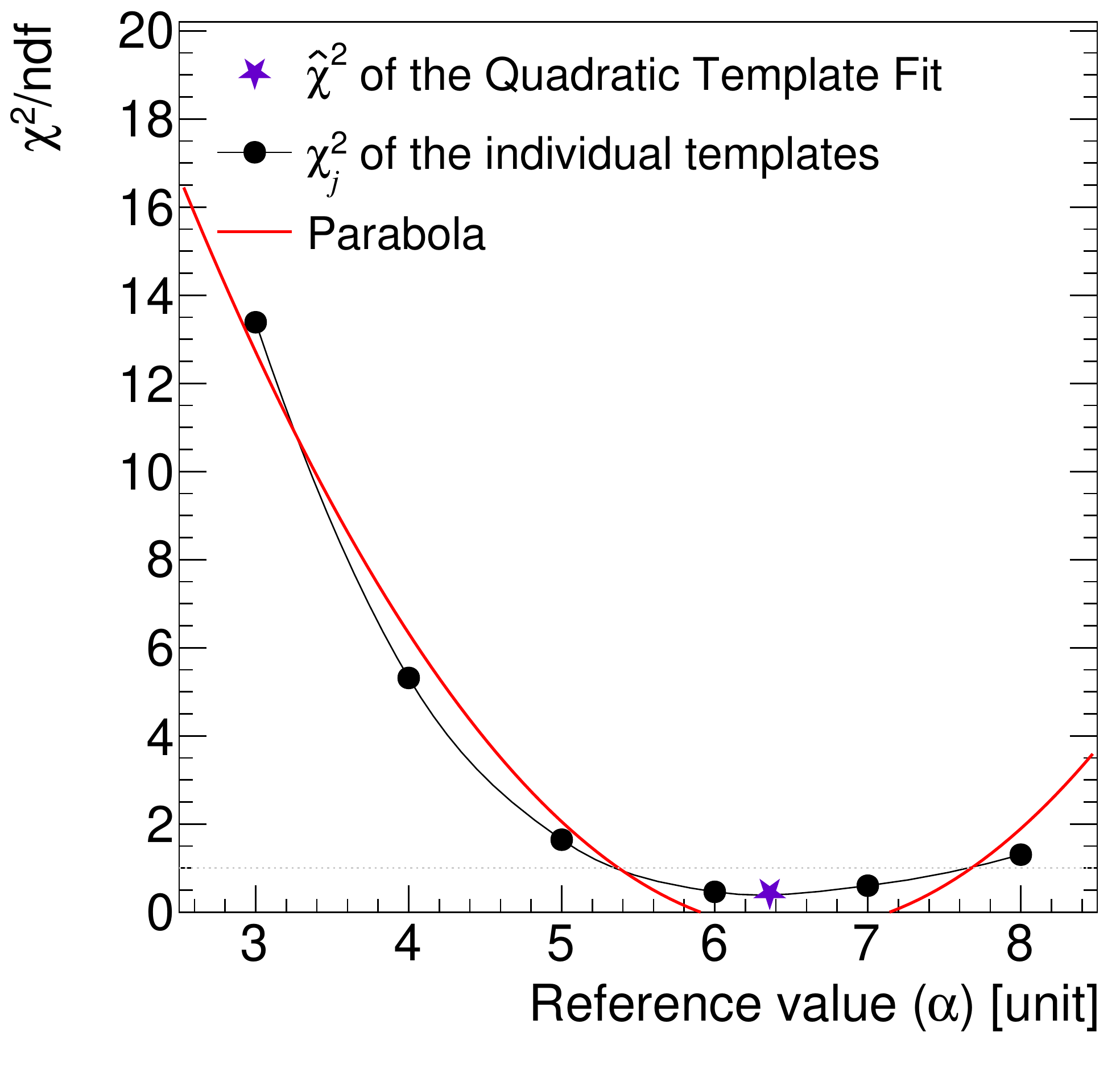}
  \caption{
    Visualization of the \chisq\ values of the templates in Example 3. A thin line
    connects these values.
    The red line displays a fitted parabola, which is not able to
    describe the \chisq\ values adequately, which is indicative of a
    nonlinear problem.
    The star indicates the best estimator from the Quadratic Template Fit.
  }
  \label{fig:example3b}
\end{figure}

In this example, the Linear Template Fit results in a best estimator
of $6.61\pm0.32$, and is thus not quite consistent with the expectation.
The EDM is comparably large, with a value of $-0.27$ , and
$\sigma_{\hat\alpha}^{(m)}=-0.27$, and the minimum of the
$\chisq$-parabola results in $6.52\pm0.27$.
Consequently,
eqs.~\eqref{eq:validation2},~\eqref{eq:validation4},~\eqref{eq:validationchi2},
and~\eqref{eq:sigmam} are not well fulfilled. 
In figure~\ref{fig:example3a} (right), the parameter dependence in some
selected bins is displayed, and it is clearly seen that the templates
are not described by a linear function.
It is also observed from the $\chisq$ values of the individual
templates that these cannot be described by a parabola as would be
the case for a linear problem, as displayed in figure~\ref{fig:example3b}.

The Quadratic Template Fit, in contrast, results in a best estimator of
$6.36\pm0.40$ and is thus in excellent agreement with the expectation.
This is because the algorithm employs a second-degree polynomial in
each bin to represent the model. This provides an adequate
representation of the parameter dependence even of a nonlinear model,
as also seen in figure~\ref{fig:example3a} (right).
A linearization at the value of the best estimator is used only for
linear error propagation, and appears to be appropriate given the size
of the final uncertainties, $\pm0.40$.

\section{
  Validation study:
  determination of the strong coupling
  constant from inclusive jet cross section data}
\label{sec:CMSalphas}
As an example application of the Linear Template Fit, the strong
coupling constant in Quantum Chromodynamics (QCD), \asmz, is determined from
a measurement of inclusive 
jet cross sections in proton--proton collisions at a center-of-mass
energy of $\sqrt{s}=7\,$TeV~\cite{CMS:2012ftr}.
The data were taken with the CMS detector at the LHC at CERN.
Inclusive jets were measured as a function of the transverse momentum
$p_\text{T}$ in five regions of absolute rapidity $|y|$.
Altogether 133 data points are available, and 24 different sources of
uncertainties are associated with the data.

The model is obtained from a calculation in next-to-leading order
perturbative QCD (NLO pQCD) using the program
\texttt{nlojet++}~\cite{Nagy:2001fj,Nagy:2003tz} that was  
interfaced to the tool
\texttt{fastNLO}~\cite{Kluge:2006xs,Britzger:2012bs,Britzger:2015ksm}. The
latter makes it possible to provide
templates for any value of \asmz\ and different parameterizations of
parton distribution functions of the proton (PDFs) without rerunning
the calculation of the pQCD matrix elements.
Multiplicative
non-perturbative (NP)~\cite{Skands:2010ak} and electroweak
corrections~\cite{Dittmaier:2012kx} are applied to the NLO
predictions.

The value of \asmz\ was already determined earlier from these data
and in NLO accuracy in two independent
analyses~\cite{CMS:2014qtp,Britzger:2017maj}.
The two analyses differ in the assumption
of the statistical model, the inference algorithm, and the
uncertainties considered in the fit.
This provides a comprehensive testing ground to compare
the results from the Linear Template Fit.

In Ref.~\cite{CMS:2014qtp}, the value of \asmz\ is determined  from
the minimum of a \chisq-parabola, where the \chisq\ is derived from
normal distributed probability distribution functions.
In order to avoid a bias in that
procedure~\cite{Ball:2009qv,D'Agostini:642515,Lyons:1989gh}, some
uncertainty 
components are treated as multiplicative~\cite{CMS:2014qtp}, which
means that the covariance matrix is calculated from relative
uncertainties that are multiplied with theory predictions. We
chose predictions obtained with
$\asmz=0.116$ for that purpose, and the total covariance matrix in \chisq\ is
calculated as described in Ref.~\cite{CMS:2014qtp}.

\begin{figure}[tb!]
  \centering
    \includegraphics[width=0.23\textwidth,trim={0 0 0 0 },clip]{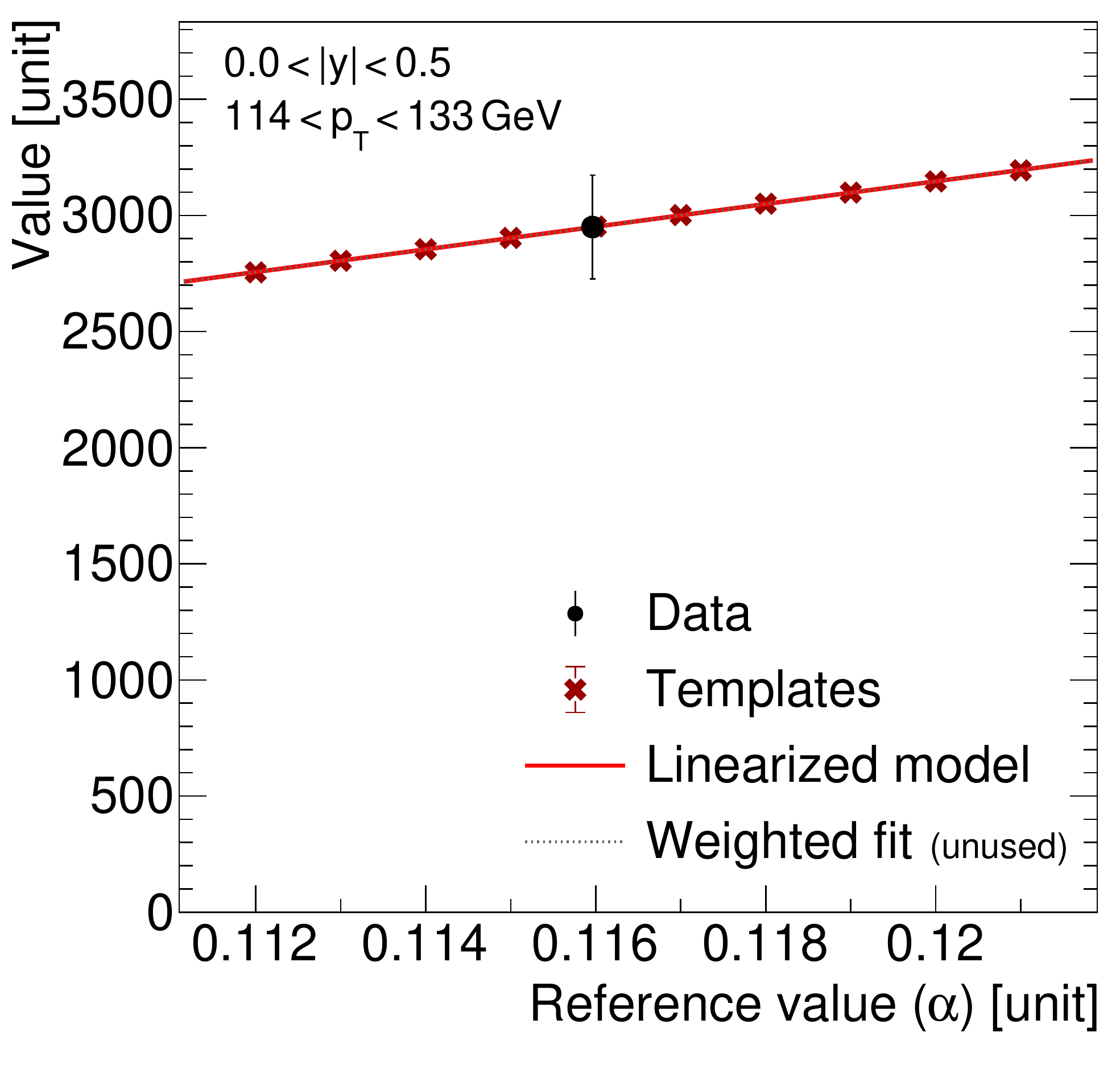}
    \includegraphics[width=0.23\textwidth,trim={0 0 0 0 },clip]{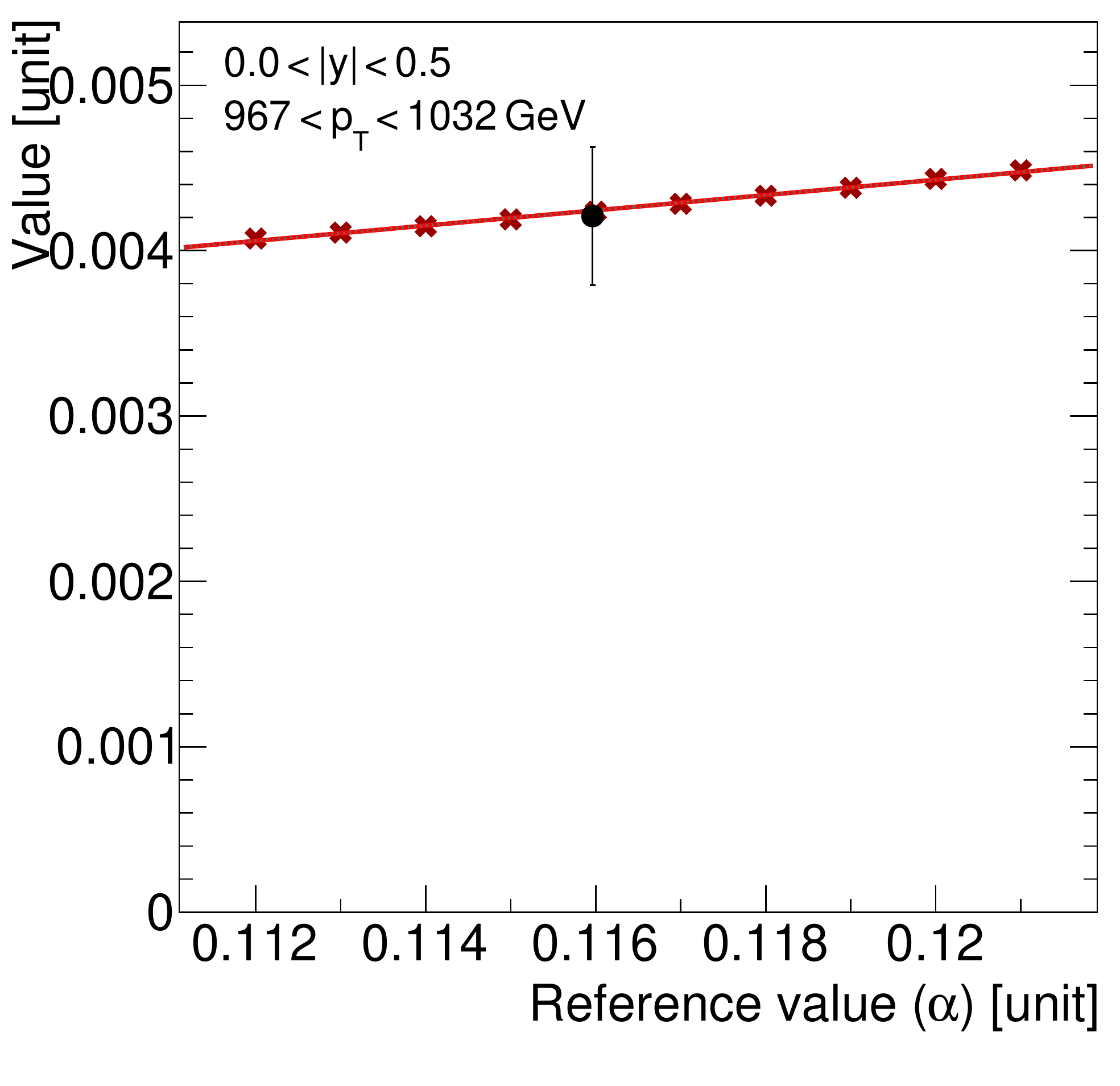}
    \includegraphics[width=0.23\textwidth,trim={0 0 0 0 },clip]{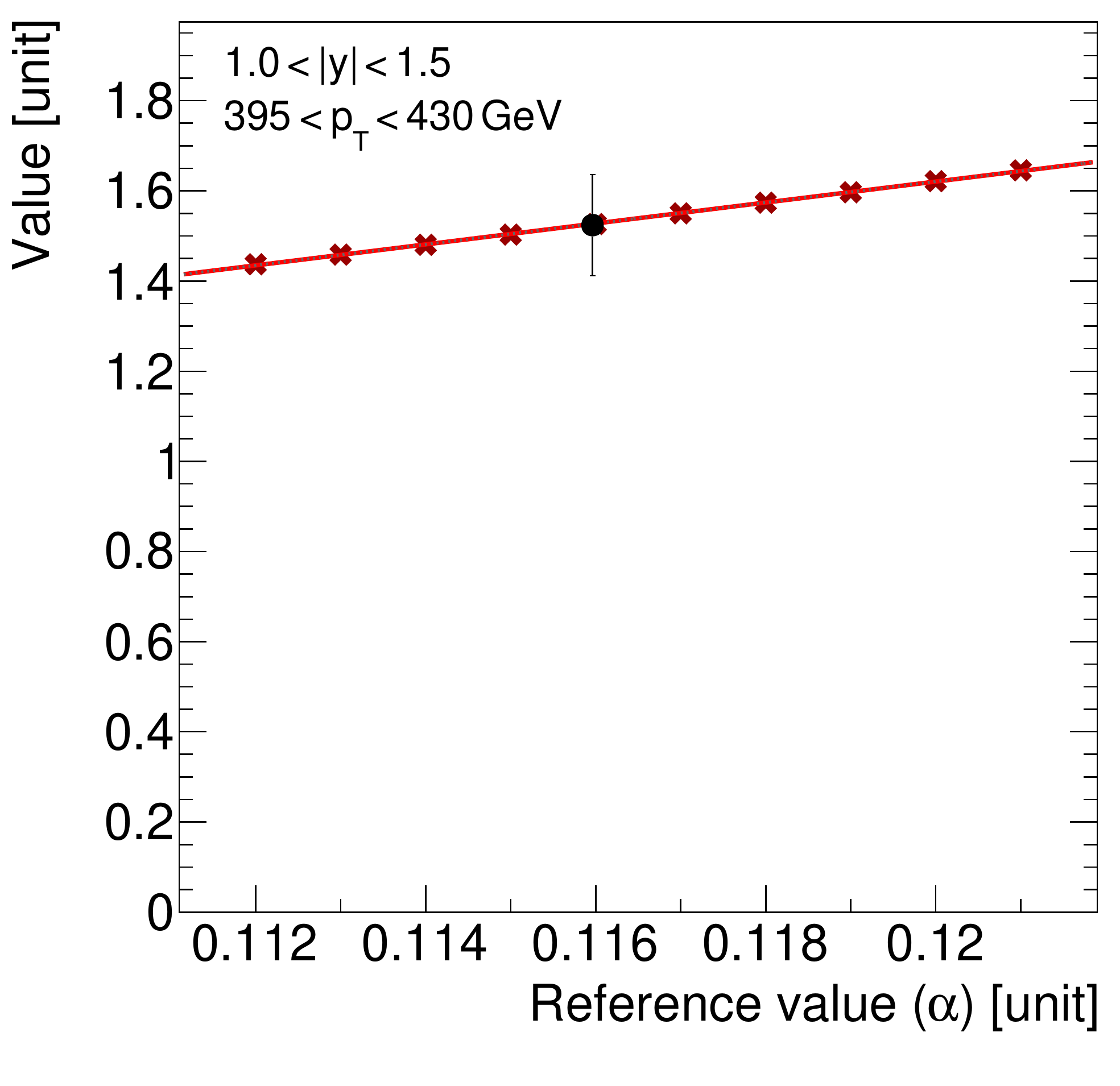}
    \includegraphics[width=0.23\textwidth,trim={0 0 0 0 },clip]{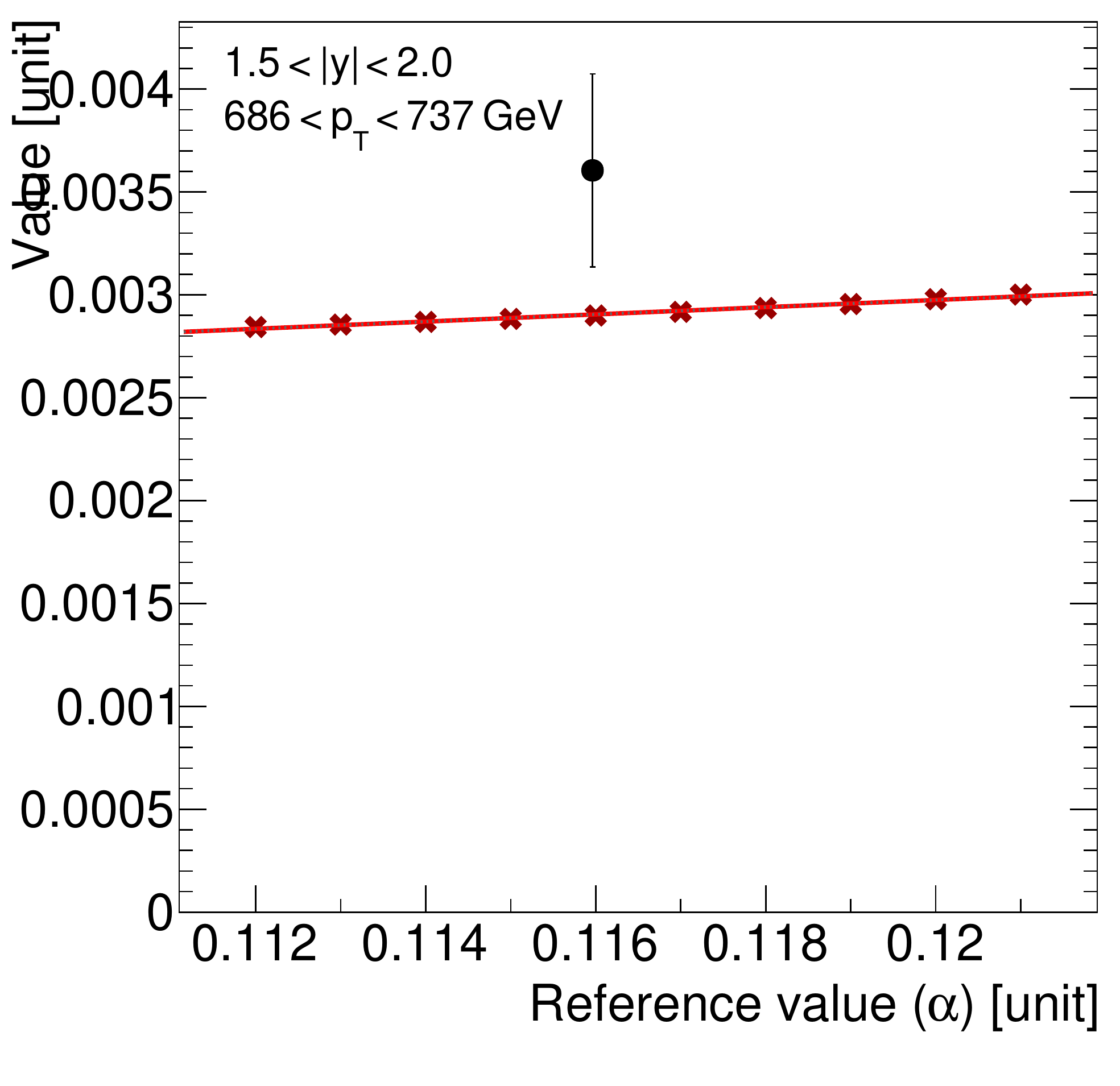}
  \\                
  \caption{
    Visualization of the templates, the data, and the implicit
    linearization of the model in some selected bins of the Linear
    Template Fit to the CMS inclusive jet cross sections.
    More details are given in the caption of figure~\ref{fig:example1}.
  }
  \label{fig:asCMSTmplts}
\end{figure}

The templates are generated in the range $0.112\leq\asmz\leq0.121$ in
steps of 0.001 using the MSTW PDF set~\cite{Martin:2009iq}.
In this methodology, the model predictions are not available for
continuous values of \asmz, since the PDFs als exhibit an
\as\ dependence, but are provided only in a
limited range and for discrete \asmz\ values~\cite{Martin:2009iq}.
Therefore, this kind of \asmz\ inference is a typical use case for the 
Linear Template Fit
The templates and the linear model are displayed for some selected
bins in  figure~\ref{fig:asCMSTmplts}.

\begin{table}[tbhp!]
  \small
  \centering
    \begin{tabular}{lll}
      \toprule
      Fit method & Best estimator \asmz  & \chisq/\ndf \\
      \midrule
      Linear Template Fit    & $0.1159\pm0.0014_\text{(exp)}\pm0.0011_\text{(pdf)}\pm0.0001_\text{(NP)}$ &  $107.4/132$ \\ 
      Quadratic Template Fit &$0.1160\pm0.0014_\text{(exp)}\pm0.0011_\text{(pdf)}\pm0.0001_\text{(NP)}$ & $107.2/132$ \\
      \chisq-parabola        & $0.1160\pm0.0018_\text{(exp,PDF)}$ & $107.2/132$ \\
      \midrule
      CMS~\cite{CMS:2014qtp} & $0.1159\pm0.0012_\text{(exp)}\pm0.0014_\text{(pdf)}\pm0.0001_\text{(NP)}$ & $107.2/132$ \\
      \bottomrule
    \end{tabular}
    \caption{
      The results from a Linear Template Fit of NLO pQCD predictions using the MSTW PDF
      set to CMS inclusive jet cross section data.
      The results are compared with the Quadratic Template Fit and to
      the analytic calculation of the minimum of the
      \chisq-parabola, as well
      as to the published results by the CMS collaboration
      in the last row.
      Shown are the best estimators for \asmz, the quadratic sum of all
      experimental uncertainties (exp), the propagated PDF
      uncertainties (PDF), the propagated NP uncertainties (NP), and
      the last column shows the \chisq/\ndf\ values.
      The difference between the Linear and the Quadratic Template Fit
      is only $0.000017$, but appears larger due to a rounding effect.
      Differences to the CMS result may also be due to numerical 
      limitations of the input data, or how the PDF uncertainties are
      symmetrized.
      The total uncertainties that are considered in the fit sum
      to $\pm0.0018$ in all cases.
    }
    \label{tab:asCMS}
\end{table}

\begin{figure}[tb!]
  \centering
  \includegraphics[width=0.41\textwidth,trim={0 0 10 0 },clip]{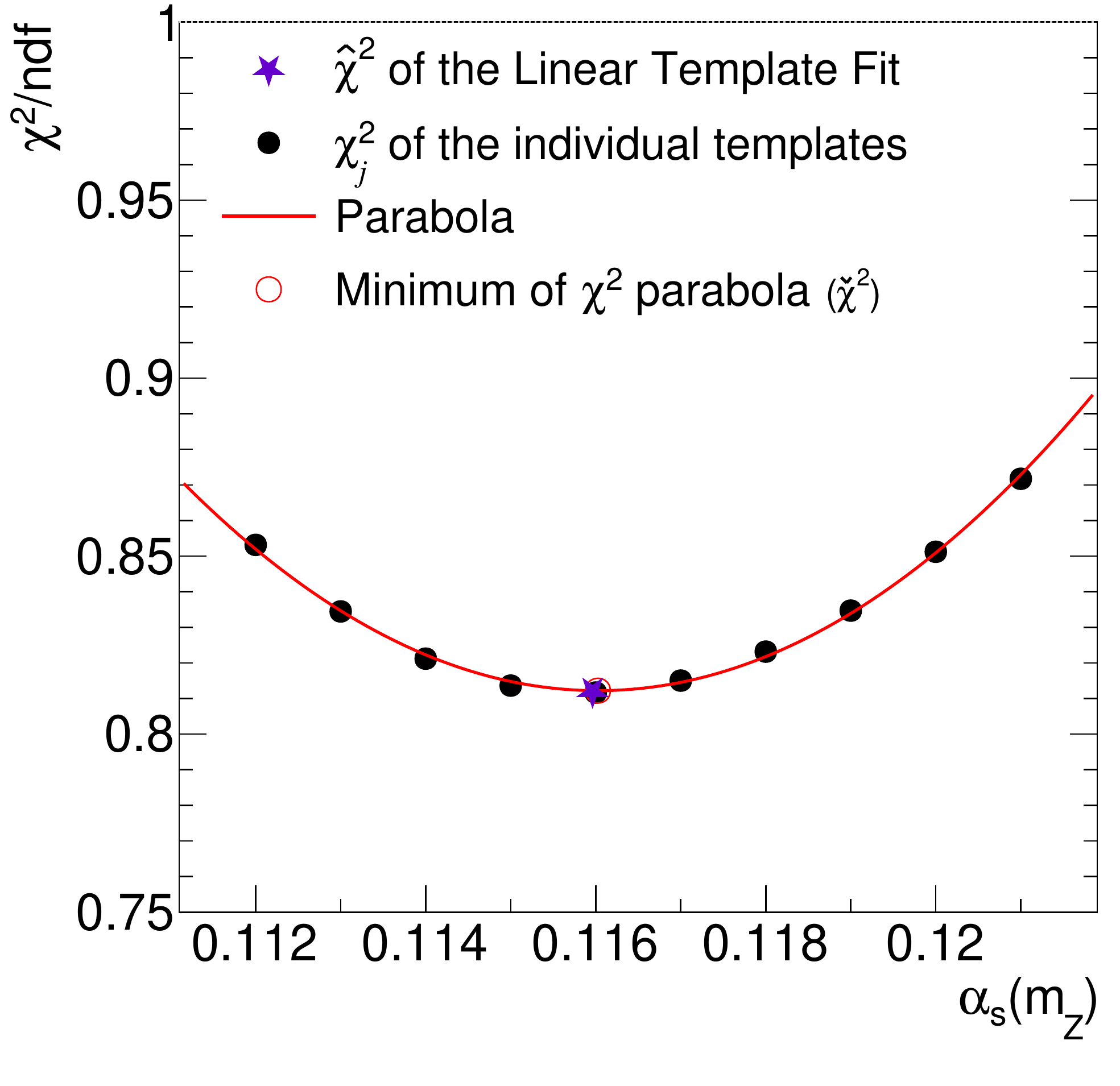}
  \hspace{0.04\textwidth}
  \includegraphics[width=0.41\textwidth,trim={0 0 10 0 },clip]{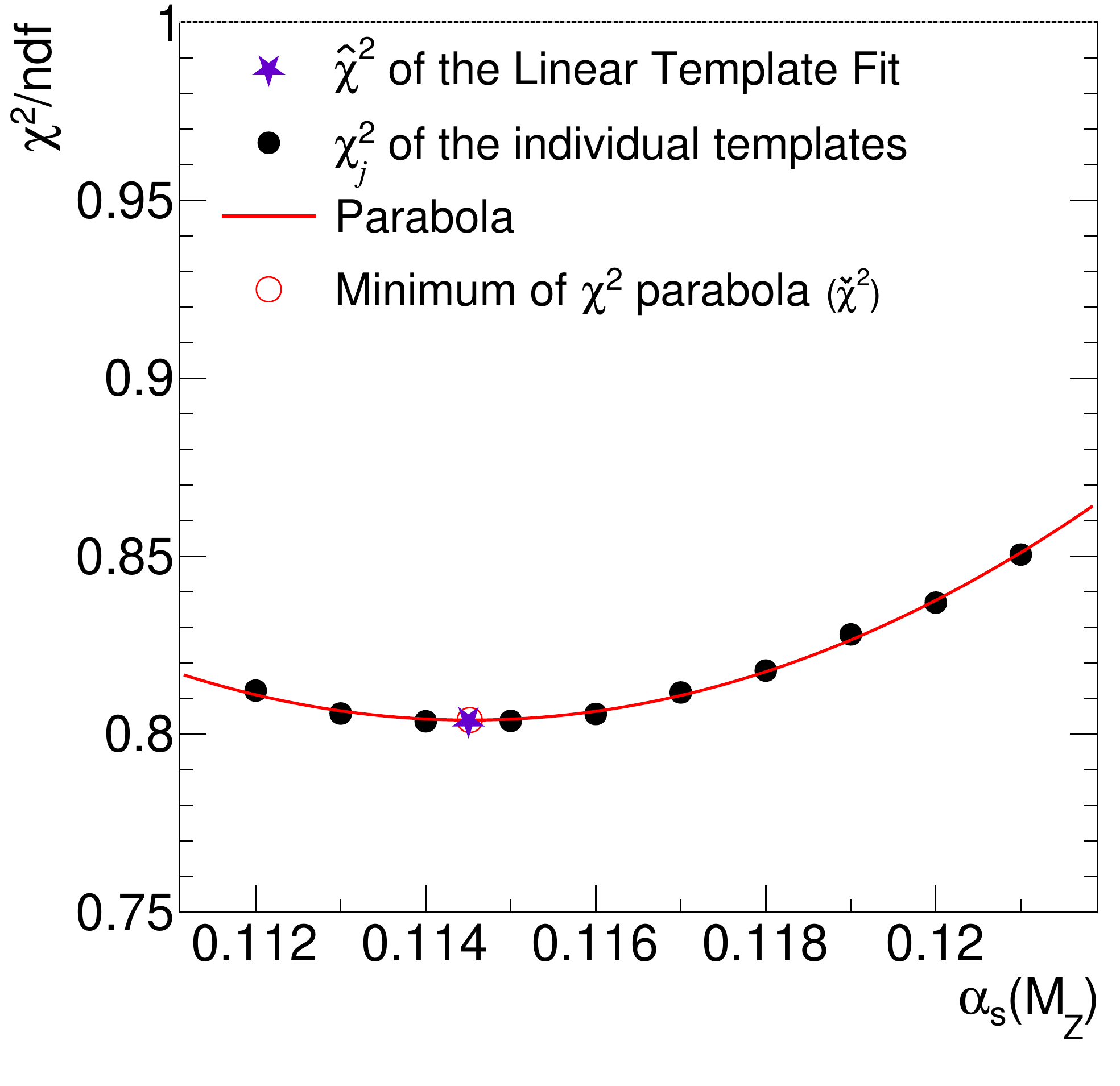}
  \caption{
    Illustration of \chisq/\ndf\ values in the fit of NLO pQCD
    predictions to CMS inclusive jet cross section data.
    Shown are the \chisq/\ndf\ value of the Linear Template Fit at its
    best estimator (star), the \chisq/\ndf\ values of the individual templates (full
    circle), and the calculated \chisq~parabola  (see sect.~\ref{sec:parabfit}) and
    its minimum (open circle).
    Left: \chisq/\ndf\ values of the Linear Template Fit with
    normal-distributed uncertainties.
    Right: \chisq/\ndf\ values of the linear template with log-normal
    distributed uncertainties.
    Note that the left figure uses the PDF set MSTW for the NLO
    predictions with a varying \asmz\ value, whereas the right one uses
    the PDF set NNPDF3.0.
  }
  \label{fig:asCMS}
\end{figure}

The results from the Linear Template Fit (eq.~\eqref{eq:full2}),
the Quadratic Template Fit (section~\ref{sec:algo}), and
the parabolic fit (section~\ref{sec:parabfit}) are compared with the
published results from CMS in table~\ref{tab:asCMS}.
An illustration of the \chisq\ values is displayed in figure~\ref{fig:asCMS}
(left).
\begin{figure}[tbh!]
  \centering
  \includegraphics[width=0.32\textwidth,trim={0 0 0 0 },clip]{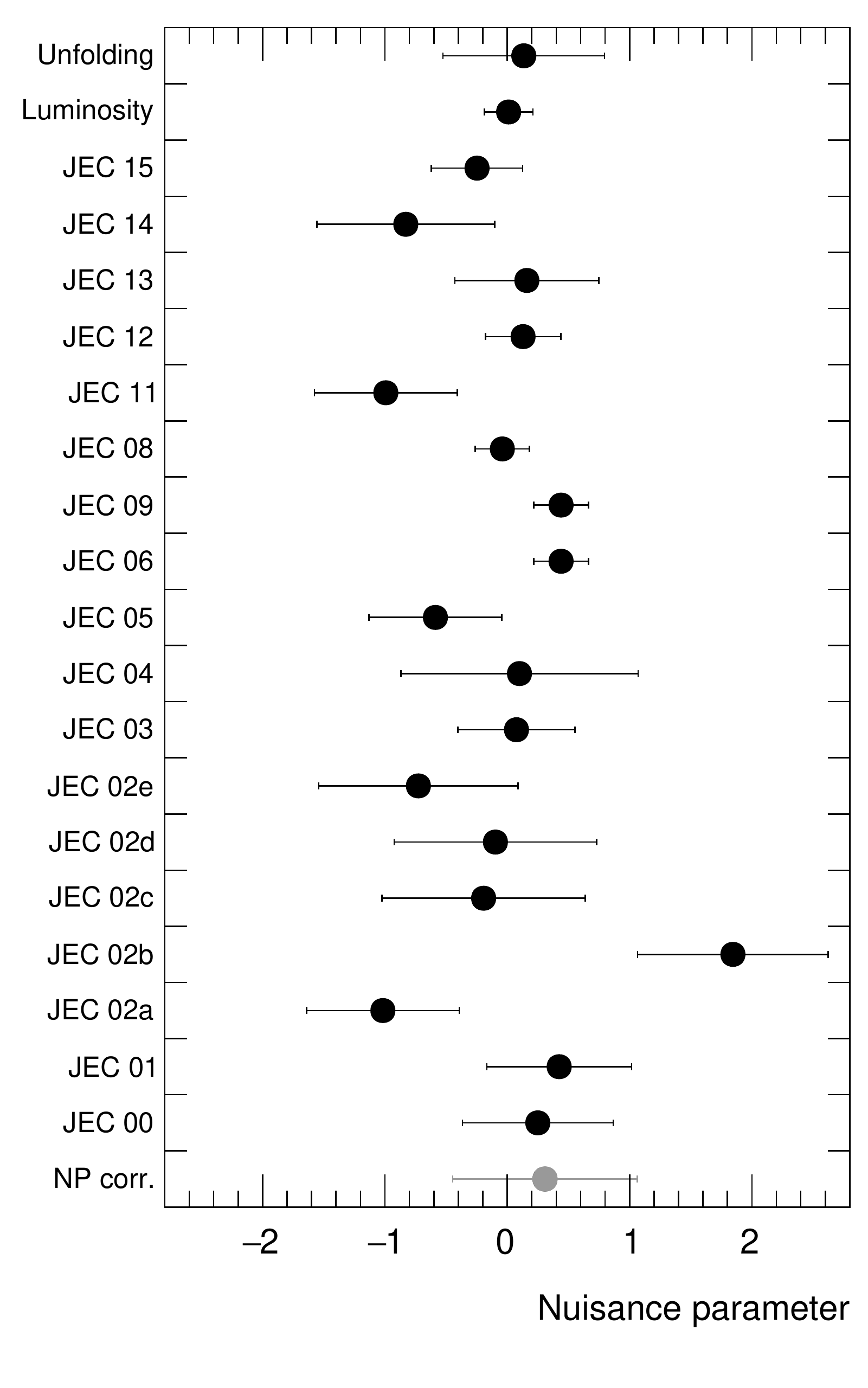}
  \includegraphics[width=0.32\textwidth,trim={0 0 0 0 },clip]{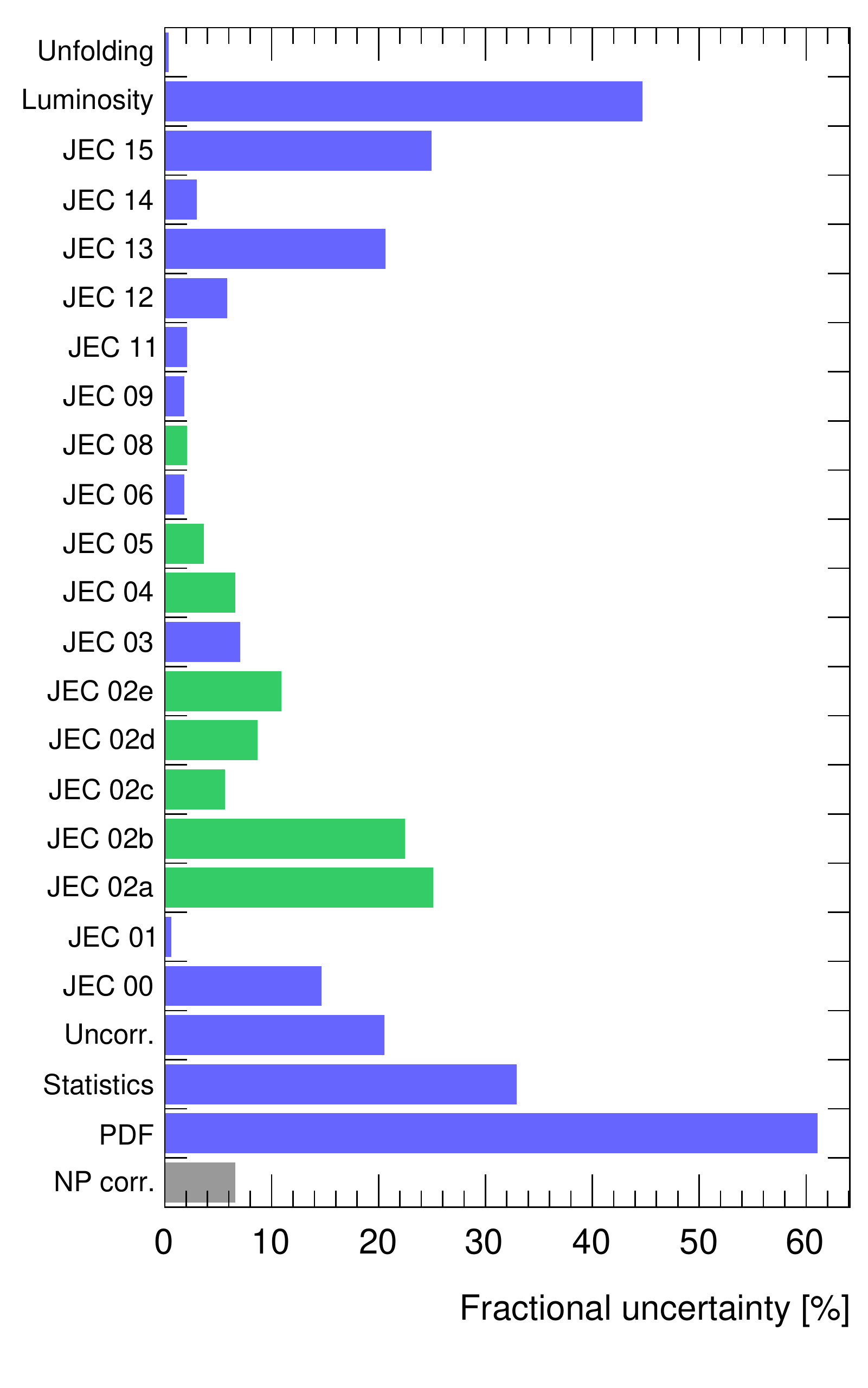}
  \includegraphics[width=0.32\textwidth,trim={0 0 0 0 },clip]{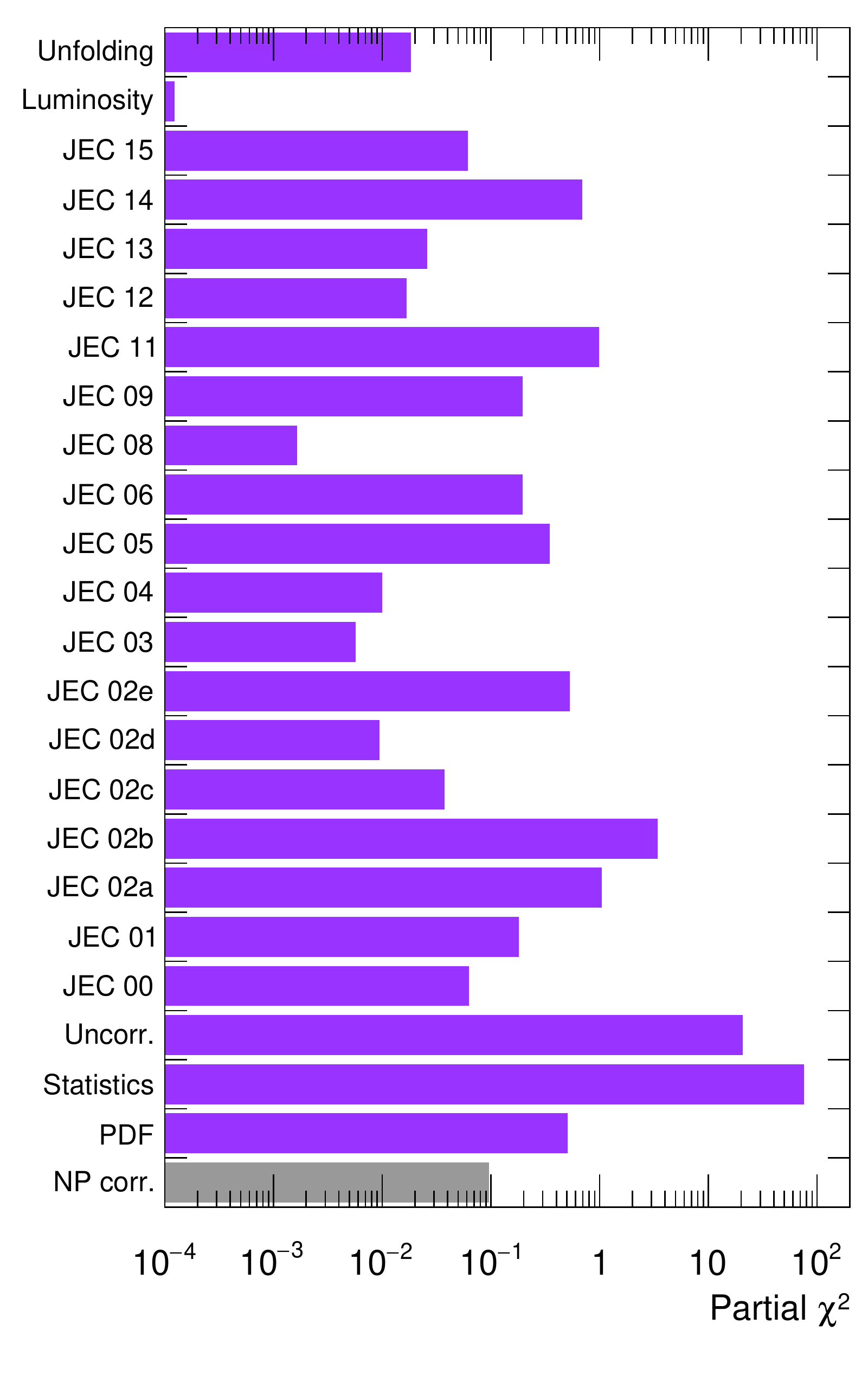}
  \caption{
    Left: nuisance parameters of the systematic uncertainties in the \asmz-fit to CMS inclusive jet
    cross section data. Details on these uncertainties are found
    in Refs.~\cite{CMS:2012ftr,CMS:2014qtp}.
    The gray value(s) are not included in the
    \chisq~computation but are propagated separately.
    Middle: size of the individual uncertainty components to the total
    uncertainty of the best estimator, \asmz, in the Linear Template Fit to CMS inclusive
    jet cross section data. The color indicates the sign of the
    uncertainty, which, however, is of relevance only for multivariate
    fits. 
    Right: partial \chisq\ from the individual uncertainty components.
  }
  \label{fig:asCMSErrors}
\end{figure}
The best estimator for \asmz, its uncertainties, and the value of \chisq\ are in
good agreement among each other, and with the published CMS
values~\cite{CMS:2014qtp}, and differences are only in the rounding digit.
It is also found that the total ``fit'' uncertainty is reproduced,
which comprises the experimental and the PDF uncertainties,
$\delta\asmz=\delta\as^\text{(exp)}\oplus\delta\as^\text{(PDF)}=\pm0.0018$.
The small differences in the individual
uncertainty components between Ref.~\cite{CMS:2014qtp} and the Linear Template
Fit are likely because the error breakdown in Ref.~\cite{CMS:2014qtp}
is only approximate, whereas the Linear Template Fit provides an analytic error propagation at the value of
the best estimator.
In fact, all 24 uncertainty components of the data,
the 20 symmetrized PDF uncertainties, and the NP uncertainties are
propagated separately (cf.\ section~\ref{sec:Errors}) and are
displayed in figure~\ref{fig:asCMSErrors}.
It is observed that the largest individual experimental uncertainty source is
the luminosity uncertainty and the statistical uncertainty.
The nuisance parameters are quite similar to those from (yet
another) fit in Ref.~\cite{CMS:2014qtp}.
The uncertainty component with the largest contribution to the total
uncertainty stems from the PDFs and the luminosity uncertainty.
The statistical and uncorrelated uncertainties have the largest contribution to \hchisq.
The NP uncertainties are not included in the nominal \chisq~computation
and are propagated separately.
The EDM of the Linear Template Fit is $7\cdot10^{-5}$ and thus smaller
than the rounding digit of the result.

\begin{figure}[tbh!]
  \centering
  \includegraphics[width=0.23\textwidth,trim={0 0 0 0 },clip]{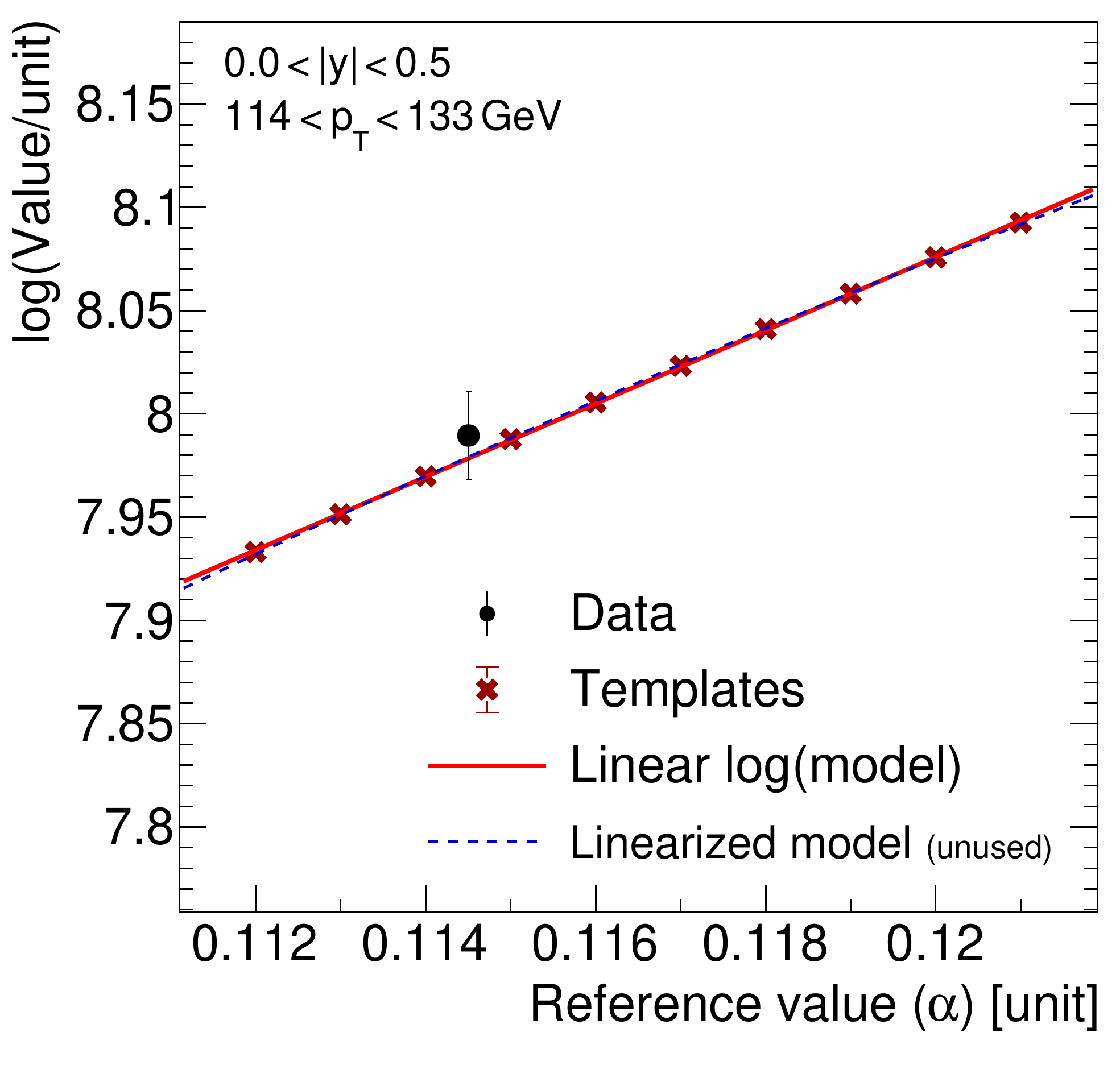}
  \includegraphics[width=0.23\textwidth,trim={0 0 0 0 },clip]{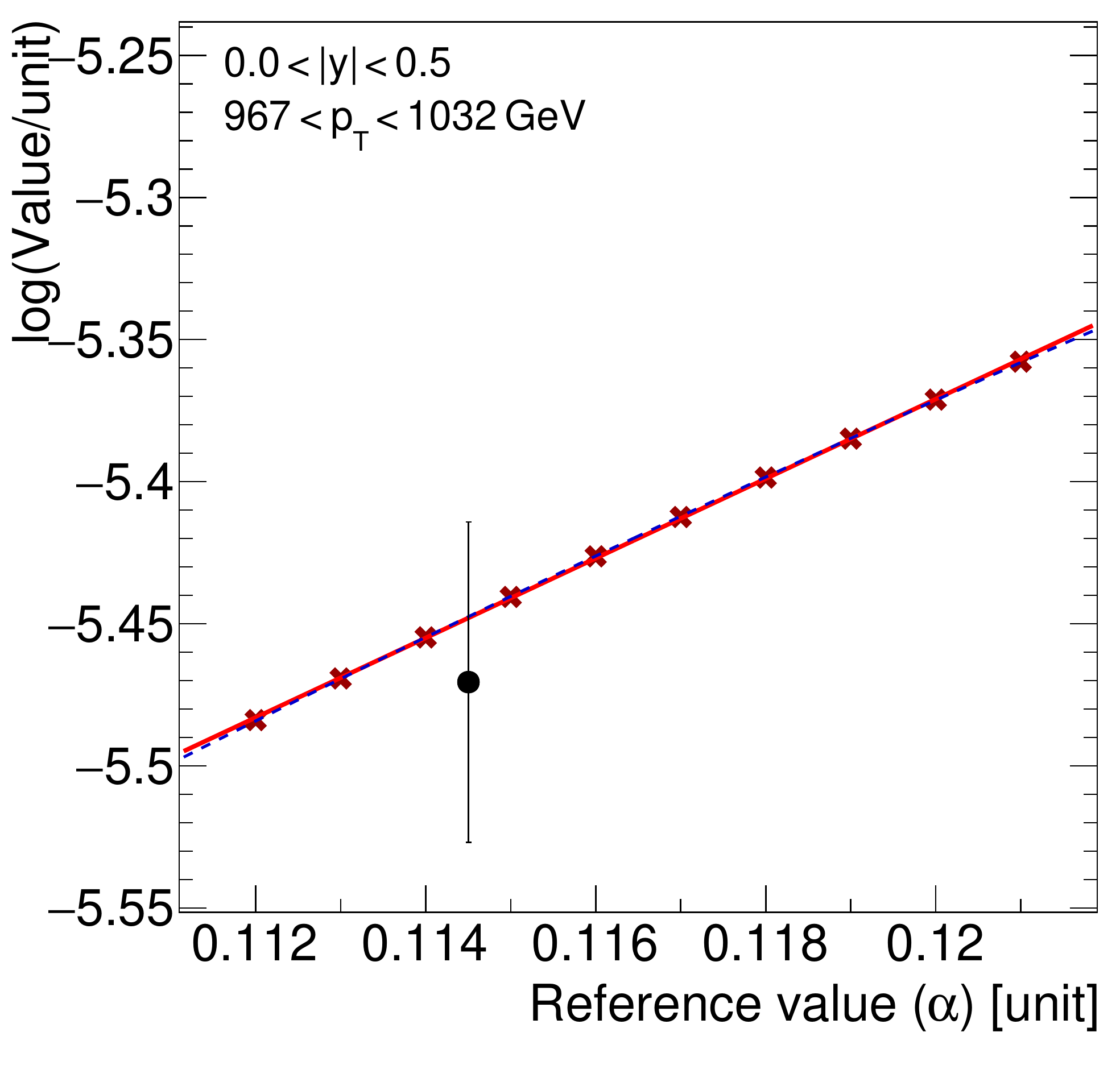}
  \includegraphics[width=0.23\textwidth,trim={0 0 0 0 },clip]{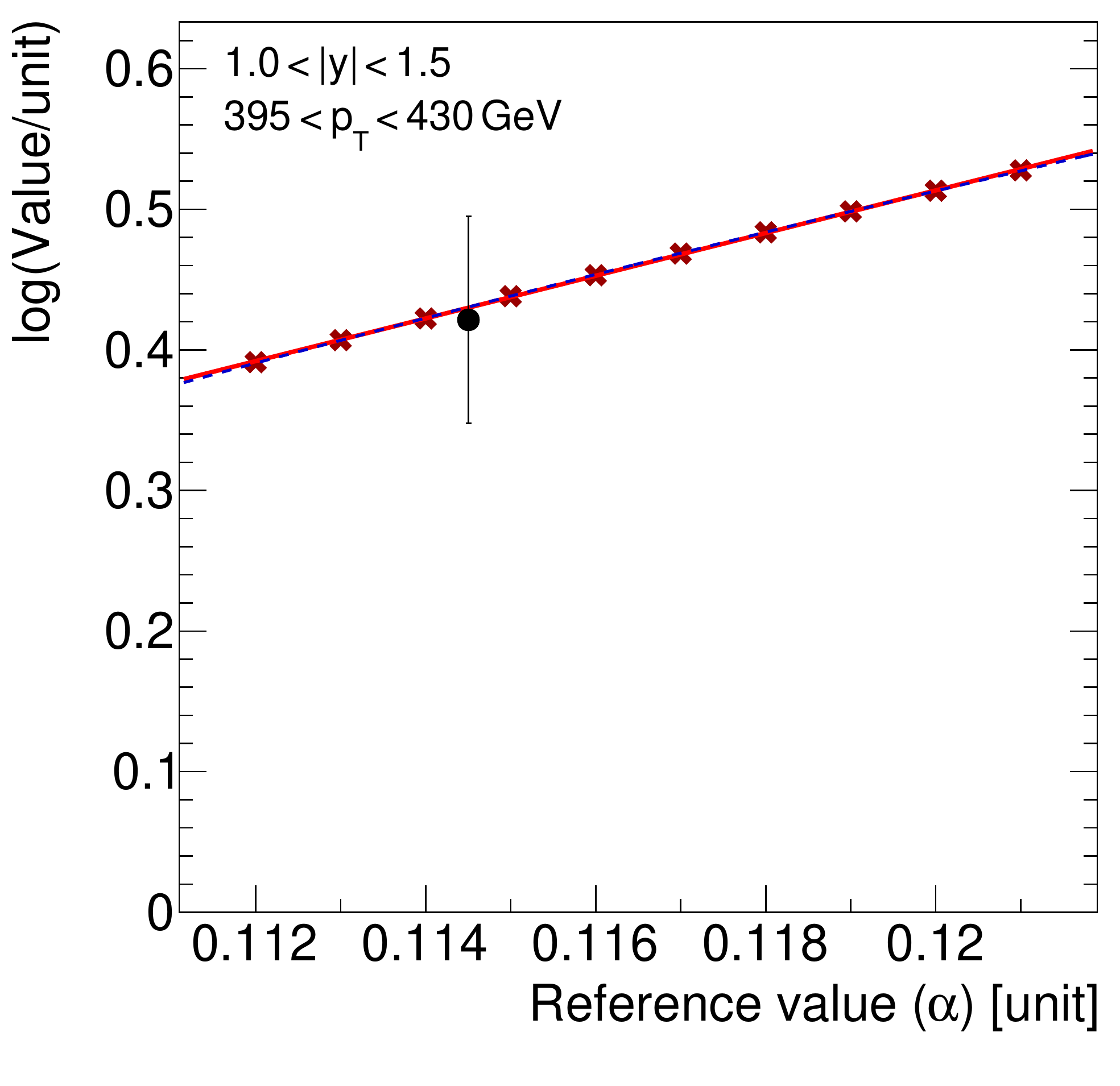}
  \includegraphics[width=0.23\textwidth,trim={0 0 0 0 },clip]{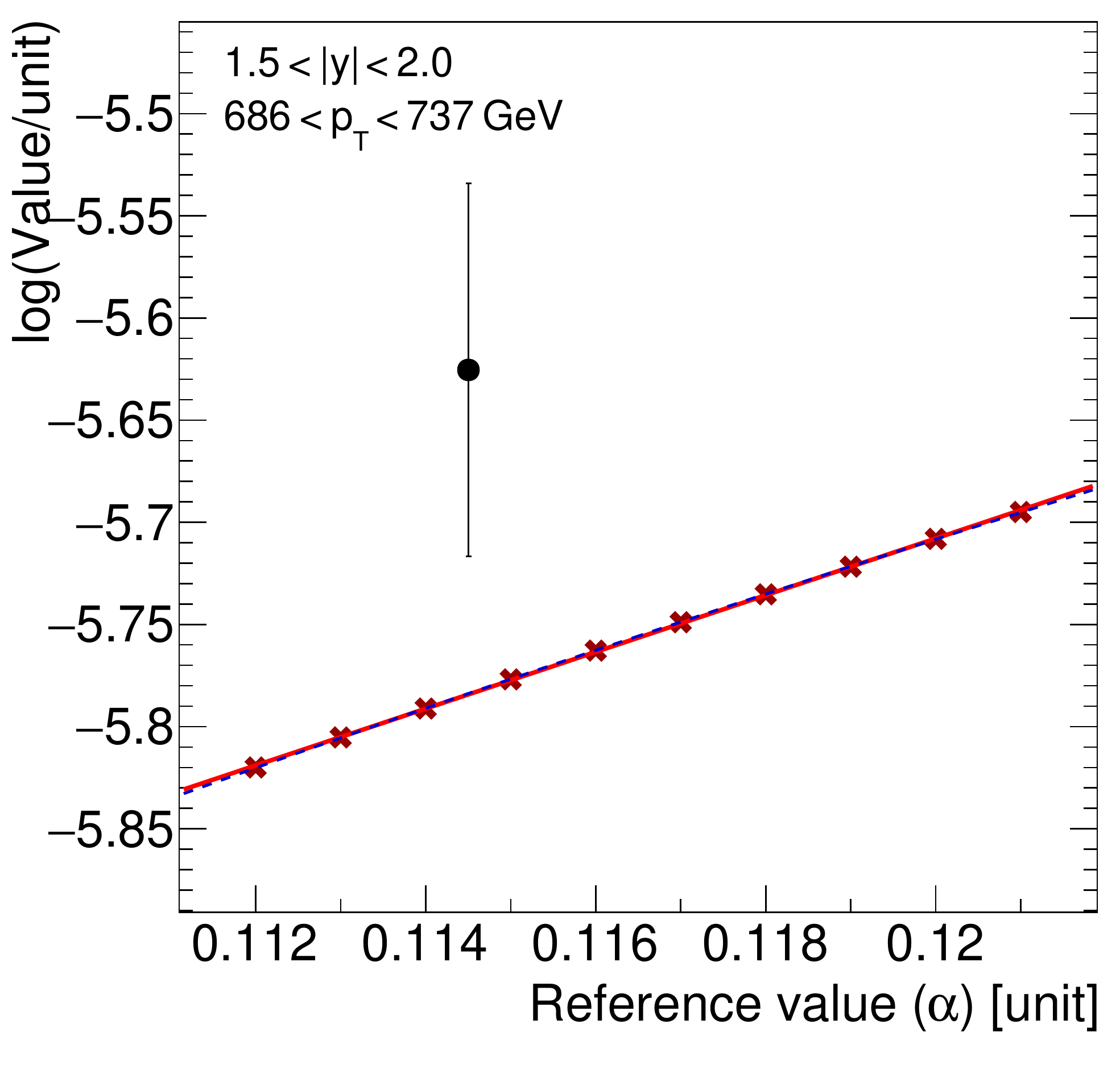}
  \caption{
    Similar visualization of the templates than in figure~\ref{fig:asCMSTmplts}, but
    using relative uncertainties (log-normal distributed probability distribution functions)
    in the Linear Template Fit.
  }
  \label{fig:asCMSTmpltsLog}
\end{figure}
In Ref.~\cite{Britzger:2017maj}, as an intermediate result, the value
of \asmz\ was determined from CMS inclusive jet cross sections as well.
In contrast to Ref.~\cite{CMS:2014qtp}, the estimators are assumed to
follow a log-normal distribution, and some theoretical uncertainties are further
considered in the \chisq\ expression, which was then minimized with
Minuit~\cite{James:1975dr}. The NNPDF3.0 PDF sets~\cite{NNPDF:2014otw} are used for the NLO predictions and provide
a covariance matrix with PDF uncertainties.
This inference provides a useful testing ground for the Linear
Template Fit with relative uncertainties (section~\ref{sec:LogN}).
Templates are generated in the range $0.112\leq\asmz\leq0.121$, and the
same uncertainty components as in Ref.~\cite{Britzger:2017maj} are
considered. Some selected bins are displayed in figure~\ref{fig:asCMSTmpltsLog}.
The results from the Linear Template Fit (eq.~\eqref{eq:full2log}),
the Quadratic Template Fit, and the parabolic fit are compared with the
published results in table~\ref{tab:asBRSSW},
and an illustration of \chisq\ values, which here are calculated from
normal-distributed relative uncertainties (eq.~\eqref{eq:chi2log}), is displayed in
figure~\ref{fig:asCMS} (right).
\begin{table}[tbhp]
  \small
  \centering
    \begin{tabular}{lll}
      \toprule
      Fit method & Best estimator \asmz  & \chisq/\ndf \\
      \midrule
      Linear Template Fit    & $0.1144\pm0.0024_\text{(exp)}\pm0.0011_\text{(pdf)}\pm<$$0.0001_\text{(NP)}$ &  $106.1/132$ \\ 
      Quadratic Template Fit & $0.1145\pm0.0024_\text{(exp)}\pm0.0011_\text{(pdf)}\pm<$$0.0001_\text{(NP)}$ & $106.1/132$ \\
      \chisq-parabola        & $0.1145\pm0.0026_\text{(exp,PDF)}$ & $106.1/132$ \\
      \midrule
      Ref.~\cite{Britzger:2017maj} & $0.1144\pm0.0022_\text{(exp)}\pm0.0014_\text{(pdf)}\pm0.0001_\text{(NP)}$ & 0.81 \\
      \bottomrule
    \end{tabular}
    \caption{
      Results from a Linear Template Fit with relative uncertainties
      where NLO pQCD predictions using the NNPDF3.0 PDF 
      set are fitted to CMS inclusive jet cross section data.
      The results are compared
      with the Quadratic Template Fit and
      to the analytic calculation of the minimum of the
      \chisq~parabola,
      and to previously published results in the last row.
      Shown are the best estimators for \asmz, the quadratic sum of all
      experimental uncertainties (exp), the propagated PDF
      uncertainties (PDF), the propagated NP uncertainties (NP), and
      the last column shows the \chisq/\ndf\ values.
    }
    \label{tab:asBRSSW}
\end{table}
The best estimator, the uncertainties, and the value of \chisq\ are in
good agreement among each other and with the result from
Ref.~\cite{Britzger:2017maj}.
The results differ only in the rounding digit, and different sizes of
the uncertainty breakdown (into (exp), (PDF), and (NP) uncertainties)
may be explained since the error 
breakdown in Ref.~\cite{Britzger:2017maj} is only approximate, but the total
uncertainty is consistently $\delta\as=\pm0.0026$.

It is worth noting that due to the application of the
factorization theorem~\cite{Collins:1989gx}, both the hard
coefficients and the PDFs exhibit an $\as$ sensitivity.
Consequently, in the two approaches discussed above,
not only is the value of \asmz\ determined, but the
PDFs are determined simultaneously as well~\cite{Paukkunen:2014zia}, which is realized
through the inclusion of the PDF uncertainties in the Linear 
Template Fit, and which is commonly referred to as \emph{PDF profiling}.
This technique avoids a possible bias that is present when the PDFs
are kept fixed~\cite{Forte:2020pyp}, and it is particularly valid since
the PDFs and the value of \asmz\ are only weakly correlated in PDF
determinations, and thus their correlations are negligible~\cite{Lai:2010nw}.
The different size of the uncertainties in tables~\ref{tab:asCMS} and~\ref{tab:asBRSSW} is then a consequence of how the PDFs are considered in the fit.
Since the PDFs themselves are determined from data as well, the ansatz from CMS also exploits, to some extent, the \as\ sensitivity of these further data by using the \as\ dependent PDF, whereas the other ansatz from Ref.~\cite{Britzger:2017maj} exploits only the jet data and their sensitivity to \asmz\ in the hard coefficients.

Consequently, these examples represent a comprehensive application of the
Linear Template Fit, where in addition to determining the value of \asmz,
a PDF profiling is also performed, as well as a simultaneous
constraint of the systematic uncertainties.
Such constraints are obtained when the uncertainties in the nuisance
parameters become smaller than unity.
A PDF determination from these data alone, instead of a PDF profiling
of an existent PDF, would be possible by considering the PDF uncertainties as \emph{unconstrained}
uncertainties in the Linear Template Fit, as described in section~\ref{sec:unconstrained}. Such a
fit exploits the PDF parameter space of the PDF set that is used to
derive the PDF uncertainties.
For the given example of CMS inclusive jet cross sections,
however, this is not possible, since these data do not have sufficient
constraints on the separate PDF flavors.

\section{Summary and conclusions}
\label{sec:summary}
In this article, the equations of the Linear Template Fit were presented.
The Linear Template Fit provides an analytic expression for
a maximum likelihood estimator in a uni- or multivariate parameter
estimation problem.
The underlying statistical model is constructed from normal or
log-normal probability distribution functions and refers to a maximum
likelihood or a minimum \chisq\ method.
For parameter estimation with the Linear Template Fit, the model
needs to be provided at a few \emph{reference} values in the parameter(s) of
interest: the templates.
The multivariate Linear Template Fit can be written as
\begin{equation}
  \bm{\hat\alpha}
    =
    \left((Y\M)^{\text{T}}WY\M\right)^{-1}(Y\M)^{\text{T}} W ( \dt-Y\mb)\,,
    \nonumber
\end{equation}
where $Y$ is the template matrix, $\bm{d}$ is the data vector (random
vector), $W$ is the inverse  
covariance matrix, and $\mb$ and $\M$ are calculated from the reference
points (eq.~\eqref{eq:Mbar}).

The equation of the Linear Template Fit is derived using two key arguments:
(i) in the vicinity of the best estimator the model is linear, and
(ii) the templates for different reference values are all generated in the same
manner.
The first precondition is commonly justified if the templates are provided
within a reasonably small range around the (to be expected) best estimator.
From the second, it follows that the bin-wise polynomial regression is
unweighted, and thus the identical regression matrix is applicable in
all bins.
Several quantities to validate the applicability of the Linear
Template Fit in a particular application were discussed.

If the linearity of the model is not justified, the Quadratic Template Fit
is a suitable alternative algorithm for parameter estimation.
It considers second-degree polynomials for the parameter dependence of
the model and employs the quickly converging exact Newton
minimization algorithm. 
For error propagation, the equations of the Linear Template Fit are applicable.

Both the Linear and the Quadratic Template Fit implicitly transform the
discrete representation of the model (the templates)  into analytic
continuous functions using polynomial regression.
These expressions themselves may also become useful in several other
applications where templates are available but continuous functions
would be needed.

The equations for error propagation of different sources of
uncertainties were presented.
The analytic nature of the Linear Template Fit allows the straightforward
propagation of each uncertainty component separately, which
provides additional insights into the parameter estimation analysis.

As an example application, previously published results on the
determination of the strong coupling constant \asmz\ from inclusive
jet cross section data taken at the LHC were repeated. The Linear and the Quadratic
Template Fits reproduce these previously published results within the
rounding digit.

In summary, key features of the Linear and the Quadratic Template Fit
are its profound statistical model based on normal- or log-normal
distributed probability distribution functions, its simple formulae, stable results, low computational
demands, full analytic error propagation, and its simple applicability.
It is believed that these template fits may become useful for
statistical inference in a large variety of problems, such as performance-critical
applications, and also in several
fields outside of high energy particle physics.

\section*{Acknowledgements}

I would like to thank Stefan Kluth and Roman Kogler for numerous discussions
on the subject and valuable comments to the manuscript. 
I am also thankful to Olaf Behnke, Richard Nisius, Klaus
Rabbertz, and Stefan Schmitt for useful discussions, several important
suggestions, and valuable comments on the manuscript.
Many thanks to Ludovic Scyboz for introducing me to the subject of
template fits.

\section*{Data Availability Statement}
An implementation of the Linear Template Fit in C++ using the linear algebra package
\texttt{Eigen}~\cite{eigenweb} can be found in
\url{https://github.com/britzger/LinearTemplateFit}.
This manuscript has no associated data.

\clearpage
\appendix

\section{Notation}
\label{sec:notation}
\begin{table}[tbhp]
  \scriptsize
  \centering
    \begin{tabular}{lll}
      \toprule
      Letter(s) & Description  & Example(s) or definition \\
      \midrule
      $d_i$, $\bm{d}$& Set of random variables $\{d_i\}$ (random
      vector; data distribution) &  E.g., histogram of an observable \\ 
      $V$            & Covariance matrix including all uncertainty sources&  $V=\mathbb{V}+\sum \bm{s}_{(l)}\bm{s}^{\text{T}}_{(l)}$\\
      $\mathbb{V}$   & Covariance matrix (not including uncertainties $s$)& \\
      $\bm{s}_{(l)}$, $\bm{\mathrm{s}}_{(l)}$  & Uncertainties with
      full bin-to-bin correlations (shifts) &\\
      $S$            & Matrix of uncertainties $s$, eq.~\eqref{eq:S}&  $S=\begin{pmatrix}\s_{(1)} \,\hdots\,  \s_{(l)} \end{pmatrix}$ or $\begin{pmatrix}\bm{\mathrm{s}}_{(1)} \, \hdots\,  \bm{\mathrm{s}}_{(l)} \end{pmatrix}$\\
      $W$, $\mathbb{W}$   & Inverse of $V$ or $\mathbb{V}$ & $W=V^{-1}$, $\mathbb{W}=\mathbb{V}^{-1}$ \\
      $A$            & A detector response matrix (migration matrix) [optional]  &  \\

      \cmidrule(lr){1-3}
      $\alpha$, $\alpha_k$, $\bm\alpha$     & Model parameter(s) of interest  &  \\
      $\bm{\lambda}(\bm{\alpha})$ & The (physics)model with free parameter(s) $\bm\alpha$\\ 
      $\bm{\hat\mathrm{y}}(\bm{\alpha})$ & Best estimator of the linearized model & $\bm{\lambda}(\bm{\alpha})\approx\bm{\hat\mathrm{y}}(\bm{\alpha}) \equiv \bm{\mathrm{y}}(\bm{\alpha};\bm{\hat\theta})$ \\
      $\epsilon_l$, $\bm{\epsilon}$ & Nuisance parameter(s) \\
      $\hat\alpha$, $\hat\alpha_k$, $\bm{\hat\alpha}$ & The best estimator(s) of the model parameter (an MLE) \\
      $\hat\epsilon_l$, $\bm{\hat\epsilon}$ & The best estimator(s) of the nuisance parameter(s)\\
      $\ahut$ & A $k$+$l$ vector for the best estimator and nuisance parameters  & $\ahut=\begin{pmatrix}\bm{\hat\alpha}\\\bm{\hat\epsilon}\end{pmatrix}$  \\
      \cmidrule(lr){1-3}
      $\dot\alpha_{j}$, $\dot\alpha_{j,k}$, $\bm{\dot\alpha}$     &      Reference values of the ($j$th) templates (for a multivariate \\
      &       Linear Template Fit the vector $\bm{\dot\alpha}$ becomes a $j\times k$ matrix )\\
      $\bm{y}_{(j)}$, $\bm{x}_{(j)}$  & A template distribution represented as an $i$ vector (or $t$ vector), & E.g., histograms of an observable\\
        & i.e.\ a model prediction for the reference value $\bm{\dot\alpha}$ &  $\bm{y}_{(j)} = \bm\lambda(\bm{\dot\alpha}_{(j)})$\\  
        &  value(s) to match the data distribution in the fit \\
      $Y$, $X$       & Template matrix, eq.~\eqref{eq:Y} & $Y=\begin{pmatrix}\bm{y}_{(1)} & \hdots&\bm{y}_{(j)} \end{pmatrix}$\\
      $M$          & Matrix of reference values (i.e., given
      $\dot\alpha_j$) and a unit vector & 
      Eqs.~\eqref{eq:M} or \eqref{eq:MnD} (also eq.\ \eqref{eq:nonlinM}) \\
      
      $M^+$          & The $g$-inverse of the linear regression & $M^+=(M^{\text{T}}M)^{-1}M^{\text{T}}$\\
      $\bm{\bar{m}}$ & First column of ${M^+}^{\text{T}}$, eqs.~\eqref{eq:mbar} and~\eqref{eq:Mbar}  & \\
      $\M$, $\bm{\tilde{m}}$ & A submatrix of ${M^+}^{\text{T}}$  (i.e., all but $\bm{\bar{m}}$), eqs.~\eqref{eq:mbar} and~\eqref{eq:Mbar}  & $\begin{pmatrix}\bm{\bar{m}} & {\M}\end{pmatrix}={M^+}^{\text{T}}$ \\

      \cmidrule(lr){1-3}
      $F$, $\mathcal{F}$   & Solution matrix of Linear Template Fit & \\

      \cmidrule(lr){1-3}
      $i$   &   \multicolumn{2}{l}{Index of an entry of the data vector; number of data points} \\
      $j$   &   \multicolumn{2}{l}{Index of the reference points; number of templates and reference points} \\
      $k$   &   \multicolumn{2}{l}{Index of the fit parameter for multivariate Linear Template Fit; number of fit parameters}  \\
      $l$   &   \multicolumn{2}{l}{Index for uncertainties with full bin-to-bin correlations; number of systematic uncertainties} \\
      $t$   &   \multicolumn{2}{l}{Number of entries (bins) on truth level, if a response matrix $A$ is used}  \\
      $n$   &   \multicolumn{2}{l}{Degree of a polynomial function, or $n$th order expansion} \\
      \bottomrule
    \end{tabular}
    \caption{
      Summary of the notation used.
      Capital letters denote a matrix, and small bold letters denote a column vector.
      The letters $i$, $j$, $k$, $l$, and $t$ are indices.
      Throughout this article,
      a vector notation is used and, for instance, the vector $\bm{d}$
      denotes a set of (random) variables, so $\bm{d}=\{d_i\}$.
      The same letters for indices and for the maximum value of an 
      index are used.
      A subscript index denotes the entry of a vector (or matrix),
      and a bracketed index denotes one vector out of several others.
      Some letters have multiple meanings, if e.g.\ normal- or log-normal-distributed
      random variables are considered, or if nonlinear effects are
      discussed, while the meaning always becomes clear from the context.
      Occasionally, a single
      entry of a vector will be denoted as a \emph{bin}, like it would be a
      histogram entry.
      The dot notation, ``$\dot\alpha$'', denotes the \emph{reference}
      values and looks similar to a ``point'' of a
      grid, but it should not be mistakenly understood as a derivative.
    }
    \label{tab:asresults}
\end{table}

\clearpage
\section{Pre-fitting the template  distributions}
\label{sec:prefit}
It is tempting to fit the distributions of the templates with some
well-motivated function, and then use the best estimators of the function parameters as
 input parameters to a (linear) template fit for the determination of
 a model parameter.
In this common practice, it is often believed that the fitted function
may better exploit the sensitivity of the shape of the template
distribution to the parameter of interest and thus enhance the quality
of the results.

Let us consider to fit the template distribution with a polynomial of
degree $n$. Unless there is no particular motivation for any
particular other $n$+1~parameter function, e.g.\ from the underlying
physics model, a polynomial is an equally
valid function as any other $n$+1~parameter function for that purpose.
In matrix notation, the polynomial is expressed using
\begin{equation}
  P\bm{\theta} = 
  \begin{pmatrix}
    1 & \dot{z}_1 & \hdots& \dot{z}_1^n \\
    \vdots & \vdots& \vdots & \vdots \\
    1 & \dot{z}_i & \hdots& \dot{z}_i^n \\
  \end{pmatrix}
  \bm{\theta}\,,
\end{equation}
where $\dot{z}_i$ may denote the observable value of the $i$th bin (bin center) of the
template distribution.
The best estimators for the polynomial parameters from a distribution $\bm{y}_{(j)}$ are thus
\begin{equation}
  \bm{\hat\theta} = (P^{\text{T}}\mathcal{W}_{(j)}P)^{-1}\mathcal{W}_{(j)}P \bm{y}_{(j)} := B_{(j)}\bm{y}_{(j)}
  \label{eq:npol}
\end{equation}
and $\mathcal{W}_{(j)}=\mathcal{V}_{(j)}^{-1}$ is the inverse covariance matrix of
the uncertainties of the template values.
Since in the Linear Template Fit the templates are sufficiently
similar (cf.\ section~\ref{sec:templrange}, and $\dot\alpha_{(j+1)}\sim\dot\alpha_{(j)}+\epsilon$), it is appropriate to
assume that the uncertainties are (approximately) equivalent for all
templates
\begin{align}
  \mathcal{V} := \mathcal{V}_{(1)} \sim \mathcal{V}_{(2)} \sim \hdots
  \sim \mathcal{V}_{(j)} \sim c^2\mathcal{V}_{\text{data}}\,,
  \label{eq:Vsim}
\end{align}
and the uncertainties of the data scale by a constant factor $c^2$ only, probably
because of a different amount of statistics.
Therefore, the $g$-inverse $B$ of the $n$th degree polynomial fit (eq.~\eqref{eq:npol}) is
equivalent for any of the templates and also for the data distribution, $B_{(j)}=B$.
Thus, the Linear Template Fit when using the $n$+1 polynomial function
parameters becomes
\begin{align}
  \chisq = \left(B\dt-BY\mb-BY\mt \right)^{\text{T}} P^{\text{T}}WP \left(B\dt-BY\mb-BY\mt \right)\,,
\end{align}
which after re-sorting becomes equivalent to the \chisq~equation of the Linear Template Fit
\begin{align}
  \chisq = \left(\dt-Y\mb-Y\mt \right)^{\text{T}} \mathscr{W}  \left(\dt-Y\mb-Y\mt \right)\,,
\end{align}
but using
\begin{align}
  \mathscr{W} =
  \left[P(P^{\text{T}}\mathcal{W}P)^{-1}\mathcal{W}P\right]^{\text{T}} W \left[P (P^{\text{T}}\mathcal{W}P)^{-1}\mathcal{W}P\right]\,.
\end{align}
It is therefore observed that a preceded fit of a function to
the templates results only in a change of the Hesse
matrix $W\to\mathscr{W}$.
This observation is valid for all pre-fits that
behave linearly, like $B$, and if eq.~\eqref{eq:Vsim} holds.

\section{Nonlinear approximation of a multivariate model}
\label{sec:nonlinndim}
In section~\ref{sec:NonLinearModel}, the nonlinear approximation of
a univariate model using an $n$th degree regression was described; see
eq.~\eqref{eq:multtheta}. When applying a first-order
Taylor expansion to it, a linearized model is obtained, and
from the equations of the Linear Template Fit, a best estimator for the
fitting problem is found.
A similar ansatz for a nonlinear model approximation of a
multivariate model is straightforward.
However, in the nonlinear approximation of a multivariate model, it is also
of interest to consider interference terms
($\sim\alpha_{k_1}\alpha_{k_2}$), and this will be discussed in the following.  
It will be shown that the subsequent linear approximation can again become input to a
linear least-squares problem similar to the Linear Template Fit.

In each bin $i$, the parameter dependence of a multivariate model with
$k$ parameters $\bm\alpha$ can be approximated as
\begin{align}
  \lambda_i(\bm{\alpha}) &\approx 
  \mathrm{y}_i(\bm{\alpha};\hat\theta_{(0,i)},\bm{\hat\theta}_{(1,i)},\bm{\hat\theta}_{(2,i)},\bm{\hat\theta}_{(\text{Infrc},i)})
  =: \hat{\mathrm{y}}_i(\bm{\alpha})\nonumber\\
  &= \hat\theta_{(0,i)}          + \hat\theta_{1}^{(1,i)}\alpha_1 + \hdots + \hat\theta_{k}^{(1,i)}\alpha_k
  + \hat\theta_{1}^{(2,i)}\alpha^2_1 + \hdots + \hat\theta_{k}^{(2,i)}\alpha^2_k 
  +  \hat\theta_{1}^{(\text{Infrc},i)}\alpha_1\alpha_2 + \hdots 
  \label{eq:nonlinyndim}
\end{align}
The best estimators for the $\theta$~parameters are obtained from a
regression analysis using a Vandermonde-like design matrix (similar to~Ref.\cite{skikit-matrix})
\begin{align}
  M &=
  \begin{pmatrix}
    M_{(\text{linear fit})} & M_{(\text{squared})} & M_{(\text{Interference})}
  \end{pmatrix}
  \nonumber\\
  &=
  \begin{pmatrix}
    1 & \dot{\alpha}_{1,1} & \cdots  & \dot{\alpha}_{1,k} & \dot{\alpha}^2_{1,1} & \cdots  & \dot{\alpha}^2_{1,k} &\dot{\alpha}_{1,1}\dot{\alpha}_{1,2} & \cdots& \dot{\alpha}_{1,k_1}\dot{\alpha}_{1,k_2} \\
    \vdots & \vdots & \ddots &  \vdots & \vdots & \ddots & \vdots  & \vdots & \ddots & \vdots \\                                                                                                                                          1 &  \dot{\alpha}_{j,1} & \cdots  & \dot{\alpha}_{j,k} & \dot{\alpha}^2_{j,1} & \cdots  & \dot{\alpha}^2_{j,k} &\dot{\alpha}_{j,1}\dot{\alpha}_{j,2}& \cdots & \dot{\alpha}_{j,k_1}\dot{\alpha}_{j,k_2}
  \end{pmatrix}\,,
\end{align}
where the entries are calculated from the reference points of the templates $\dot\alpha_{j,k}$.
The first equation shows literally the three block matrices that form
that design matrix.
The best estimators of the $\theta$~parameters in each bin are
then obtained from its $g$-inverse as similar to eq.~\eqref{eq:theta}
\begin{equation}
   \begin{pmatrix}
     \hat\theta_{(0,i)} \\
     \bm{\hat\theta}_{(1,i)}\\
     \bm{\hat\theta}_{(2,i)}\\
     \bm{\hat\theta}_{(\text{Infrc},i)}
   \end{pmatrix}
   =
   M^+\begin{pmatrix} y_{(1),i} \\ \vdots \\ y_{(j),i} \end{pmatrix}\,.
   \label{eq:nonlinMndim}
\end{equation}

For reasons outlined in the main text, an unweighted regression is
applicable, and thus the regression matrix is equivalent for every bin
$i$.
When substituting the $\hat\theta$~estimators into eq.~\eqref{eq:nonlinyndim}, one
obtains the approximated model in every bin, and thus the full model
as a second-order function $\bm{\hat{\mathrm{y}}}(\bm{\alpha})$.

Next, a first-order Taylor expansion is applied to the model
$\bm{\hat{\mathrm{y}}}(\bm{\alpha})$.
The expansion point is identified\,\footnote{Of course, any other
  expansion point can be used in these equations, for instance
  the result from a previous measurement or the result from a
  previous nonlinear template fit. The latter defines naturally an
  iterative fitting algorithm.}
 with the best estimator of the
Linear Template Fit $\bm{\hat\alpha}$.
For that purpose it is useful to refer to the columns of the
transposed $g$-inverse matrix with vectors ($\bm{\bar{m}}$ and
multiple $\bm{\tilde{m}}$) like
\begin{equation}
  {M^+}^{\text{T}} =:
  \begin{pmatrix}
    \bm{\bar{m}}   & 
    \bm{\tilde{m}_{[1]}} &
    \cdots &
    \bm{\tilde{m}_{[k]}} &
    \bm{\tilde{m}_{[1^2]}} &
    \cdots &
    \bm{\tilde{m}_{[k^2]}} &
    \bm{\tilde{m}_{[1\, 2]}} &
    \cdots &
    \bm{\tilde{m}_{[k_1 k_2]}} 
  \end{pmatrix}\,.
\end{equation}
The resulting terms from the first-order Taylor expansion are
organized in the $k\times j$-matrix
\begin{equation}
  \tilde{\tilde{M}} =:
  \begin{pmatrix}
    \bm{\tilde{\tilde{m}}_{[1]}}^{\text{T}} \\
    \vdots \\
    \bm{\tilde{\tilde{m}}_{[k]}}^{\text{T}} \\
  \end{pmatrix}
~~~\text{with elements}~~~
  \bm{\tilde{\tilde{m}}_{[k]}} := 
  \bm{\tilde{m}_{[k]}}
  + 2 \bm{\tilde{m}_{[k^2]}}\hat\alpha_k
  + \sum_{k_2 \neq k} \bm{\tilde{m}_{[k k_2]}} \hat\alpha_{k}\hat\alpha_{k_2}
\end{equation}
and the vector
\begin{equation}
  \bm{\tilde{\bar{m}}} =
   \bm{\bar{m}}
   - \sum_k \bm{\tilde{m}_{[k^2]}} \hat\alpha^2_k
   - \sum_{k_1} \sum_{k_2\neq k_1} \bm{\tilde{m}_{[k_1 k_2]}} \hat\alpha_{k_1}\hat\alpha_{k_2}\,.
\end{equation}
Hence, the estimated model, similar to eq.~\eqref{eq:approxm}, is
found as
  \begin{equation}
    \bm{\tilde{\hat{\mathrm{y}}}}(\bm{\alpha}) =
    Y\bm{\tilde{\bar{m}}}  + Y\tilde{\tilde{M}}\bm{\alpha}\,.
    \label{eq:approxmnDim}
\end{equation}
This model, $\bm{\tilde{\hat{\mathrm{y}}}}(\bm{\alpha})\approx\bm{\lambda}(\bm\alpha)$,
can be used together with the equations of the linear
template fit to calculate the best estimator, since it is linear in all
parameters of interest.

\section{Calculation of the template fit matrix $\mathcal{F}$}
\label{sec:numerics}
The calculation of the template fit matrix $\mathcal{F}$
(eq.~\eqref{eq:full0}) involves the calculation of $M^+$ as well as the
inversions of \V\ and $\mathcal{D}$.
These calculations shall be briefly discussed in the following.

The calculation of $M^+$ includes the calculation and subsequent
inversion of the  sum of squares and cross products (SSCP) matrix of
an extended Vandermonde, $(M^{\text{T}}M)^{-1}$, which is widely studied in
literature (see e.g.\ Ref.~\cite{seber2008matrix}) and typically does
not involve numerical problems for $n=1$ (or 2).

The inverse covariance matrix $\W = \V^{-1}$ is commonly
calculable since $\V$ includes uncertainties without bin-to-bin
correlations and it is thus positive definite. In fact,
it needs to be invertible for a valid problem, or may even be a diagonal matrix.
An efficient algorithm to prove that \V\ is positive definite is a
Cholesky decomposition, and when successful, the inverse is 
almost already calculated~\cite{madsen}.

The inversion of the $(k+l)\times(k+l)$ matrix $\mathcal{D}$ may be
more challenging, but it can be expressed using three sub-matrices as
\begin{equation}
  \mathcal{D} =
  \begin{pmatrix}
    (\Y)^{\text{T}}\W\Y  & (\Y)^{\text{T}}\W S \\
    S^{\text{T}}\W\Y  &  \bm{1}_l + S^{\text{T}}\W S \\
  \end{pmatrix}
  =:
  \begin{pmatrix}
     D  & B \\
     B^{\text{T}}  &  \Sigma \\
  \end{pmatrix}\,,
  \label{eq:HMblocks}
\end{equation}
where the right side introduces some shorthand notations (for $D$,
compare also eq.~\eqref{eq:full0}).
The inverse of $\mathcal{D}$ is then given in two variants by~\cite{Henderson:10.2307/2029838}:
\begin{eqnarray}
  \mathcal{D}^{-1} =& 
  \begin{pmatrix}
    D^{-1}-D^{-1}B\mathcal{L} & & \mathcal{L}^{\text{T}} \\
     \mathcal{L} & & L^{-1} \\
  \end{pmatrix}
    ~~~&\text{using} ~~~\mathcal{L}=-L^{-1}B^{\text{T}}D^{-1}\,, \\
  =& 
  \begin{pmatrix}
     K^{-1}  & & \mathcal{K} \\
     \mathcal{K}^{\text{T}}  & & \Sigma^{-1} - \Sigma^{-1}B^{\text{T}}\mathcal{K} \\
  \end{pmatrix}
  ~~~&\text{using} ~~~\mathcal{K}=-K^{-1}B\Sigma^{-1}
  \,.
\end{eqnarray}
This requires one to calculate the inverse of only two matrices, $L$ and
$D$, or $K$ and $\Sigma$, which are defined as
\begin{equation}
  D = (\Y)^{\text{T}}\W\Y
  ~~~~\text{and}~~~
  L := \Sigma - B^{\text{T}}D^{-1}B
  \,,
\end{equation}
or
\begin{align}
  \Sigma = \bm{1}_l + S^{\text{T}}\W S
    ~~~~\text{and}~~~
  K :=& D - B\Sigma^{-1}B^{\text{T}} \\
     =&  (\Y)^{\text{T}}(\W-\W V_s\W+\W V_sV^{-1}V_s\W)\Y
  \,.
          \label{eq:K}
\end{align}
It is seen that these either can be inverted similarly to weighted
least-squares
matrices~\cite{Bjorck:1411947,Seber2003LinearRA,Bjorck:1968883},
or are just sums of two matrices, and the
Sherman--Morrison--Woodbury
formula can be
applied~\cite{Duncan:doi:10.1080/14786444408520897,woodbury-1950}
(see also Ref.~\cite{seber2008matrix}).

\vspace{2cm}
\small
\bibliography{LTF_impl}

\end{document}